\documentclass[useAMS,usenatbib,longauth]{aa}

\usepackage{txfonts}
\usepackage[T1]{fontenc}
\usepackage{ae,aecompl}
\usepackage{newtxtext,newtxmath}
\usepackage{graphicx}	
\usepackage{amsmath}	
\usepackage{euclid} 
\usepackage{silence} 
\usepackage{soul}
\usepackage[dvipsnames]{xcolor}
\usepackage[switch, modulo]{lineno} 
\usepackage{multirow} 
\usepackage{booktabs} 
\usepackage{rotating} 
\usepackage[normalem]{ulem} 

\usepackage[pdfpagelabels=false]{hyperref}	
\hypersetup{colorlinks=true,linkcolor=blue,citecolor=blue,filecolor=blue,urlcolor=blue,}

\makeatletter
\renewcommand*\aa@pageof{, page \thepage{} of \pageref*{LastPage}}
\makeatother
\def\NFW{\textsc{nfw}}
\def\BMO{\textsc{bmo}}
\defcitealias{giocoli24}{ECG}
\defcitealias{PaperIII}{ECS}
\defcitealias{PaperIV}{ECL}
\newcommand*{\COMBCL}{\texttt{COMB-CL}\xspace}
\newcommand*{\AMICO}{\texttt{AMICO}\xspace}
\newcommand*{\PZWav}{\texttt{PZWav}\xspace}
\newcommand{\splitcell}[2]{{\begin{tabular}{@{}c@{}}{#1} \\ {#2}\end{tabular}}}
\newcommand{\splitcellbis}[4]{{\begin{tabular}{l@{}l@{}l@{}l@{}}{#1} \\ {#2} \\ {#3} \\ {#4}\end{tabular}}}
\newcommand{\splitcelltwo}[2]{{\begin{tabular}{r@{}r@{}}{#1} \\ {#2}\end{tabular}}}
\newcommand{\splitcellthree}[3]{{\begin{tabular}{r@{}r@{}r@{}}{#1} \\ {#2} \\ {#3}\end{tabular}}}
\newcommand{\splitcellfour}[4]{{\begin{tabular}{r@{}r@{}r@{}r@{}}{#1} \\ {#2} \\ {#3} \\ {#4}\end{tabular}}}

\begin{document}
\label{firstpage}


\title{\Euclid preparation}
\subtitle{Determining the weak lensing mass accuracy and precision for galaxy clusters}    

\author{Euclid Collaboration: L.~Ingoglia\orcid{0000-0002-7587-0997}\thanks{\email{lorenzo.ingoglia@inaf.it}}\inst{\ref{aff1},\ref{aff2}}
\and M.~Sereno\orcid{0000-0003-0302-0325}\inst{\ref{aff3},\ref{aff4}}
\and S.~Farrens\orcid{0000-0002-9594-9387}\inst{\ref{aff5}}
\and C.~Giocoli\orcid{0000-0002-9590-7961}\inst{\ref{aff3},\ref{aff6}}
\and L.~Baumont\orcid{0000-0002-1518-0150}\inst{\ref{aff5}}
\and G.~F.~Lesci\orcid{0000-0002-4607-2830}\inst{\ref{aff1},\ref{aff3}}
\and L.~Moscardini\orcid{0000-0002-3473-6716}\inst{\ref{aff1},\ref{aff3},\ref{aff4}}
\and C.~Murray\inst{\ref{aff7}}
\and M.~Vannier\inst{\ref{aff8}}
\and A.~Biviano\orcid{0000-0002-0857-0732}\inst{\ref{aff9},\ref{aff10}}
\and C.~Carbone\orcid{0000-0003-0125-3563}\inst{\ref{aff11}}
\and G.~Covone\orcid{0000-0002-2553-096X}\inst{\ref{aff12},\ref{aff13},\ref{aff14}}
\and G.~Despali\orcid{0000-0001-6150-4112}\inst{\ref{aff1},\ref{aff3},\ref{aff4}}
\and M.~Maturi\orcid{0000-0002-3517-2422}\inst{\ref{aff15},\ref{aff16}}
\and S.~Maurogordato\inst{\ref{aff8}}
\and M.~Meneghetti\orcid{0000-0003-1225-7084}\inst{\ref{aff3},\ref{aff4}}
\and M.~Radovich\orcid{0000-0002-3585-866X}\inst{\ref{aff17}}
\and B.~Altieri\orcid{0000-0003-3936-0284}\inst{\ref{aff18}}
\and A.~Amara\inst{\ref{aff19}}
\and S.~Andreon\orcid{0000-0002-2041-8784}\inst{\ref{aff20}}
\and N.~Auricchio\orcid{0000-0003-4444-8651}\inst{\ref{aff3}}
\and C.~Baccigalupi\orcid{0000-0002-8211-1630}\inst{\ref{aff10},\ref{aff9},\ref{aff21},\ref{aff22}}
\and M.~Baldi\orcid{0000-0003-4145-1943}\inst{\ref{aff23},\ref{aff3},\ref{aff4}}
\and S.~Bardelli\orcid{0000-0002-8900-0298}\inst{\ref{aff3}}
\and F.~Bellagamba\inst{\ref{aff23},\ref{aff3}}
\and R.~Bender\orcid{0000-0001-7179-0626}\inst{\ref{aff24},\ref{aff25}}
\and F.~Bernardeau\inst{\ref{aff26},\ref{aff27}}
\and D.~Bonino\orcid{0000-0002-3336-9977}\inst{\ref{aff28}}
\and E.~Branchini\orcid{0000-0002-0808-6908}\inst{\ref{aff29},\ref{aff30},\ref{aff20}}
\and M.~Brescia\orcid{0000-0001-9506-5680}\inst{\ref{aff12},\ref{aff13},\ref{aff14}}
\and J.~Brinchmann\orcid{0000-0003-4359-8797}\inst{\ref{aff31},\ref{aff32}}
\and S.~Camera\orcid{0000-0003-3399-3574}\inst{\ref{aff33},\ref{aff34},\ref{aff28}}
\and V.~Capobianco\orcid{0000-0002-3309-7692}\inst{\ref{aff28}}
\and J.~Carretero\orcid{0000-0002-3130-0204}\inst{\ref{aff35},\ref{aff36}}
\and S.~Casas\orcid{0000-0002-4751-5138}\inst{\ref{aff37}}
\and M.~Castellano\orcid{0000-0001-9875-8263}\inst{\ref{aff38}}
\and G.~Castignani\orcid{0000-0001-6831-0687}\inst{\ref{aff3}}
\and S.~Cavuoti\orcid{0000-0002-3787-4196}\inst{\ref{aff13},\ref{aff14}}
\and A.~Cimatti\inst{\ref{aff39}}
\and C.~Colodro-Conde\inst{\ref{aff40}}
\and G.~Congedo\orcid{0000-0003-2508-0046}\inst{\ref{aff41}}
\and C.~J.~Conselice\orcid{0000-0003-1949-7638}\inst{\ref{aff42}}
\and L.~Conversi\orcid{0000-0002-6710-8476}\inst{\ref{aff43},\ref{aff18}}
\and Y.~Copin\orcid{0000-0002-5317-7518}\inst{\ref{aff44}}
\and F.~Courbin\orcid{0000-0003-0758-6510}\inst{\ref{aff45},\ref{aff46},\ref{aff47}}
\and H.~M.~Courtois\orcid{0000-0003-0509-1776}\inst{\ref{aff48}}
\and M.~Cropper\orcid{0000-0003-4571-9468}\inst{\ref{aff49}}
\and A.~Da~Silva\orcid{0000-0002-6385-1609}\inst{\ref{aff50},\ref{aff51}}
\and H.~Degaudenzi\orcid{0000-0002-5887-6799}\inst{\ref{aff52}}
\and G.~De~Lucia\orcid{0000-0002-6220-9104}\inst{\ref{aff9}}
\and J.~Dinis\orcid{0000-0001-5075-1601}\inst{\ref{aff50},\ref{aff51}}
\and F.~Dubath\orcid{0000-0002-6533-2810}\inst{\ref{aff52}}
\and C.~A.~J.~Duncan\inst{\ref{aff42}}
\and X.~Dupac\inst{\ref{aff18}}
\and S.~Dusini\orcid{0000-0002-1128-0664}\inst{\ref{aff53}}
\and A.~Ealet\orcid{0000-0003-3070-014X}\inst{\ref{aff44}}
\and M.~Farina\orcid{0000-0002-3089-7846}\inst{\ref{aff54}}
\and F.~Faustini\orcid{0000-0001-6274-5145}\inst{\ref{aff55},\ref{aff38}}
\and S.~Ferriol\inst{\ref{aff44}}
\and P.~Fosalba\orcid{0000-0002-1510-5214}\inst{\ref{aff56},\ref{aff57}}
\and M.~Frailis\orcid{0000-0002-7400-2135}\inst{\ref{aff9}}
\and E.~Franceschi\orcid{0000-0002-0585-6591}\inst{\ref{aff3}}
\and M.~Fumana\orcid{0000-0001-6787-5950}\inst{\ref{aff11}}
\and S.~Galeotta\orcid{0000-0002-3748-5115}\inst{\ref{aff9}}
\and W.~Gillard\orcid{0000-0003-4744-9748}\inst{\ref{aff58}}
\and B.~Gillis\orcid{0000-0002-4478-1270}\inst{\ref{aff41}}
\and P.~G\'omez-Alvarez\orcid{0000-0002-8594-5358}\inst{\ref{aff59},\ref{aff18}}
\and A.~Grazian\orcid{0000-0002-5688-0663}\inst{\ref{aff17}}
\and F.~Grupp\inst{\ref{aff24},\ref{aff25}}
\and L.~Guzzo\orcid{0000-0001-8264-5192}\inst{\ref{aff60},\ref{aff20}}
\and S.~V.~H.~Haugan\orcid{0000-0001-9648-7260}\inst{\ref{aff61}}
\and W.~Holmes\inst{\ref{aff62}}
\and F.~Hormuth\inst{\ref{aff63}}
\and A.~Hornstrup\orcid{0000-0002-3363-0936}\inst{\ref{aff64},\ref{aff65}}
\and P.~Hudelot\inst{\ref{aff27}}
\and S.~Ili\'c\orcid{0000-0003-4285-9086}\inst{\ref{aff66},\ref{aff67}}
\and K.~Jahnke\orcid{0000-0003-3804-2137}\inst{\ref{aff68}}
\and M.~Jhabvala\inst{\ref{aff69}}
\and B.~Joachimi\orcid{0000-0001-7494-1303}\inst{\ref{aff70}}
\and E.~Keih\"anen\orcid{0000-0003-1804-7715}\inst{\ref{aff71}}
\and S.~Kermiche\orcid{0000-0002-0302-5735}\inst{\ref{aff58}}
\and A.~Kiessling\orcid{0000-0002-2590-1273}\inst{\ref{aff62}}
\and M.~Kilbinger\orcid{0000-0001-9513-7138}\inst{\ref{aff5}}
\and B.~Kubik\orcid{0009-0006-5823-4880}\inst{\ref{aff44}}
\and M.~K\"ummel\orcid{0000-0003-2791-2117}\inst{\ref{aff25}}
\and M.~Kunz\orcid{0000-0002-3052-7394}\inst{\ref{aff72}}
\and H.~Kurki-Suonio\orcid{0000-0002-4618-3063}\inst{\ref{aff73},\ref{aff74}}
\and S.~Ligori\orcid{0000-0003-4172-4606}\inst{\ref{aff28}}
\and P.~B.~Lilje\orcid{0000-0003-4324-7794}\inst{\ref{aff61}}
\and V.~Lindholm\orcid{0000-0003-2317-5471}\inst{\ref{aff73},\ref{aff74}}
\and I.~Lloro\inst{\ref{aff75}}
\and G.~Mainetti\orcid{0000-0003-2384-2377}\inst{\ref{aff76}}
\and E.~Maiorano\orcid{0000-0003-2593-4355}\inst{\ref{aff3}}
\and O.~Mansutti\orcid{0000-0001-5758-4658}\inst{\ref{aff9}}
\and S.~Marcin\inst{\ref{aff77}}
\and O.~Marggraf\orcid{0000-0001-7242-3852}\inst{\ref{aff78}}
\and K.~Markovic\orcid{0000-0001-6764-073X}\inst{\ref{aff62}}
\and M.~Martinelli\orcid{0000-0002-6943-7732}\inst{\ref{aff38},\ref{aff79}}
\and N.~Martinet\orcid{0000-0003-2786-7790}\inst{\ref{aff80}}
\and F.~Marulli\orcid{0000-0002-8850-0303}\inst{\ref{aff1},\ref{aff3},\ref{aff4}}
\and R.~Massey\orcid{0000-0002-6085-3780}\inst{\ref{aff81}}
\and E.~Medinaceli\orcid{0000-0002-4040-7783}\inst{\ref{aff3}}
\and S.~Mei\orcid{0000-0002-2849-559X}\inst{\ref{aff7}}
\and M.~Melchior\inst{\ref{aff77}}
\and Y.~Mellier\inst{\ref{aff82},\ref{aff27}}
\and E.~Merlin\orcid{0000-0001-6870-8900}\inst{\ref{aff38}}
\and G.~Meylan\inst{\ref{aff45}}
\and M.~Moresco\orcid{0000-0002-7616-7136}\inst{\ref{aff1},\ref{aff3}}
\and E.~Munari\orcid{0000-0002-1751-5946}\inst{\ref{aff9},\ref{aff10}}
\and S.-M.~Niemi\inst{\ref{aff83}}
\and C.~Padilla\orcid{0000-0001-7951-0166}\inst{\ref{aff84}}
\and K.~Paech\orcid{0000-0003-0625-2367}\inst{\ref{aff24}}
\and S.~Paltani\orcid{0000-0002-8108-9179}\inst{\ref{aff52}}
\and F.~Pasian\orcid{0000-0002-4869-3227}\inst{\ref{aff9}}
\and K.~Pedersen\inst{\ref{aff85}}
\and W.~J.~Percival\orcid{0000-0002-0644-5727}\inst{\ref{aff86},\ref{aff87},\ref{aff88}}
\and V.~Pettorino\inst{\ref{aff83}}
\and S.~Pires\orcid{0000-0002-0249-2104}\inst{\ref{aff5}}
\and G.~Polenta\orcid{0000-0003-4067-9196}\inst{\ref{aff55}}
\and M.~Poncet\inst{\ref{aff89}}
\and L.~A.~Popa\inst{\ref{aff90}}
\and L.~Pozzetti\orcid{0000-0001-7085-0412}\inst{\ref{aff3}}
\and F.~Raison\orcid{0000-0002-7819-6918}\inst{\ref{aff24}}
\and A.~Renzi\orcid{0000-0001-9856-1970}\inst{\ref{aff91},\ref{aff53}}
\and J.~Rhodes\orcid{0000-0002-4485-8549}\inst{\ref{aff62}}
\and G.~Riccio\inst{\ref{aff13}}
\and E.~Romelli\orcid{0000-0003-3069-9222}\inst{\ref{aff9}}
\and M.~Roncarelli\orcid{0000-0001-9587-7822}\inst{\ref{aff3}}
\and E.~Rossetti\orcid{0000-0003-0238-4047}\inst{\ref{aff23}}
\and R.~Saglia\orcid{0000-0003-0378-7032}\inst{\ref{aff25},\ref{aff24}}
\and Z.~Sakr\orcid{0000-0002-4823-3757}\inst{\ref{aff15},\ref{aff67},\ref{aff92}}
\and J.-C.~Salvignol\inst{\ref{aff83}}
\and A.~G.~S\'anchez\orcid{0000-0003-1198-831X}\inst{\ref{aff24}}
\and D.~Sapone\orcid{0000-0001-7089-4503}\inst{\ref{aff93}}
\and B.~Sartoris\orcid{0000-0003-1337-5269}\inst{\ref{aff25},\ref{aff9}}
\and M.~Schirmer\orcid{0000-0003-2568-9994}\inst{\ref{aff68}}
\and P.~Schneider\orcid{0000-0001-8561-2679}\inst{\ref{aff78}}
\and A.~Secroun\orcid{0000-0003-0505-3710}\inst{\ref{aff58}}
\and G.~Seidel\orcid{0000-0003-2907-353X}\inst{\ref{aff68}}
\and S.~Serrano\orcid{0000-0002-0211-2861}\inst{\ref{aff56},\ref{aff94},\ref{aff57}}
\and C.~Sirignano\orcid{0000-0002-0995-7146}\inst{\ref{aff91},\ref{aff53}}
\and G.~Sirri\orcid{0000-0003-2626-2853}\inst{\ref{aff4}}
\and L.~Stanco\orcid{0000-0002-9706-5104}\inst{\ref{aff53}}
\and J.~Steinwagner\orcid{0000-0001-7443-1047}\inst{\ref{aff24}}
\and P.~Tallada-Cresp\'{i}\orcid{0000-0002-1336-8328}\inst{\ref{aff35},\ref{aff36}}
\and D.~Tavagnacco\orcid{0000-0001-7475-9894}\inst{\ref{aff9}}
\and A.~N.~Taylor\inst{\ref{aff41}}
\and I.~Tereno\inst{\ref{aff50},\ref{aff95}}
\and R.~Toledo-Moreo\orcid{0000-0002-2997-4859}\inst{\ref{aff96}}
\and F.~Torradeflot\orcid{0000-0003-1160-1517}\inst{\ref{aff36},\ref{aff35}}
\and I.~Tutusaus\orcid{0000-0002-3199-0399}\inst{\ref{aff67}}
\and L.~Valenziano\orcid{0000-0002-1170-0104}\inst{\ref{aff3},\ref{aff97}}
\and T.~Vassallo\orcid{0000-0001-6512-6358}\inst{\ref{aff25},\ref{aff9}}
\and G.~Verdoes~Kleijn\orcid{0000-0001-5803-2580}\inst{\ref{aff98}}
\and A.~Veropalumbo\orcid{0000-0003-2387-1194}\inst{\ref{aff20},\ref{aff30},\ref{aff99}}
\and Y.~Wang\orcid{0000-0002-4749-2984}\inst{\ref{aff100}}
\and J.~Weller\orcid{0000-0002-8282-2010}\inst{\ref{aff25},\ref{aff24}}
\and G.~Zamorani\orcid{0000-0002-2318-301X}\inst{\ref{aff3}}
\and E.~Zucca\orcid{0000-0002-5845-8132}\inst{\ref{aff3}}
\and M.~Bolzonella\orcid{0000-0003-3278-4607}\inst{\ref{aff3}}
\and E.~Bozzo\orcid{0000-0002-8201-1525}\inst{\ref{aff52}}
\and C.~Burigana\orcid{0000-0002-3005-5796}\inst{\ref{aff2},\ref{aff97}}
\and M.~Calabrese\orcid{0000-0002-2637-2422}\inst{\ref{aff101},\ref{aff11}}
\and D.~Di~Ferdinando\inst{\ref{aff4}}
\and J.~A.~Escartin~Vigo\inst{\ref{aff24}}
\and R.~Farinelli\inst{\ref{aff3}}
\and F.~Finelli\orcid{0000-0002-6694-3269}\inst{\ref{aff3},\ref{aff97}}
\and J.~Gracia-Carpio\inst{\ref{aff24}}
\and S.~Matthew\orcid{0000-0001-8448-1697}\inst{\ref{aff41}}
\and A.~Pezzotta\orcid{0000-0003-0726-2268}\inst{\ref{aff24}}
\and M.~P\"ontinen\orcid{0000-0001-5442-2530}\inst{\ref{aff73}}
\and V.~Scottez\inst{\ref{aff82},\ref{aff102}}
\and M.~Tenti\orcid{0000-0002-4254-5901}\inst{\ref{aff4}}
\and M.~Viel\orcid{0000-0002-2642-5707}\inst{\ref{aff10},\ref{aff9},\ref{aff22},\ref{aff21},\ref{aff103}}
\and M.~Wiesmann\orcid{0009-0000-8199-5860}\inst{\ref{aff61}}
\and Y.~Akrami\orcid{0000-0002-2407-7956}\inst{\ref{aff104},\ref{aff105}}
\and V.~Allevato\orcid{0000-0001-7232-5152}\inst{\ref{aff13}}
\and S.~Anselmi\orcid{0000-0002-3579-9583}\inst{\ref{aff53},\ref{aff91},\ref{aff106}}
\and M.~Archidiacono\orcid{0000-0003-4952-9012}\inst{\ref{aff60},\ref{aff107}}
\and F.~Atrio-Barandela\orcid{0000-0002-2130-2513}\inst{\ref{aff108}}
\and M.~Ballardini\orcid{0000-0003-4481-3559}\inst{\ref{aff109},\ref{aff3},\ref{aff110}}
\and D.~Bertacca\orcid{0000-0002-2490-7139}\inst{\ref{aff91},\ref{aff17},\ref{aff53}}
\and M.~Bethermin\orcid{0000-0002-3915-2015}\inst{\ref{aff111},\ref{aff80}}
\and A.~Blanchard\orcid{0000-0001-8555-9003}\inst{\ref{aff67}}
\and L.~Blot\orcid{0000-0002-9622-7167}\inst{\ref{aff112},\ref{aff106}}
\and H.~B\"ohringer\orcid{0000-0001-8241-4204}\inst{\ref{aff24},\ref{aff113},\ref{aff114}}
\and S.~Borgani\orcid{0000-0001-6151-6439}\inst{\ref{aff115},\ref{aff10},\ref{aff9},\ref{aff21}}
\and S.~Bruton\orcid{0000-0002-6503-5218}\inst{\ref{aff116}}
\and R.~Cabanac\orcid{0000-0001-6679-2600}\inst{\ref{aff67}}
\and A.~Calabro\orcid{0000-0003-2536-1614}\inst{\ref{aff38}}
\and G.~Ca\~nas-Herrera\orcid{0000-0003-2796-2149}\inst{\ref{aff83},\ref{aff117}}
\and A.~Cappi\inst{\ref{aff3},\ref{aff8}}
\and F.~Caro\inst{\ref{aff38}}
\and C.~S.~Carvalho\inst{\ref{aff95}}
\and T.~Castro\orcid{0000-0002-6292-3228}\inst{\ref{aff9},\ref{aff21},\ref{aff10},\ref{aff103}}
\and K.~C.~Chambers\orcid{0000-0001-6965-7789}\inst{\ref{aff118}}
\and S.~Contarini\orcid{0000-0002-9843-723X}\inst{\ref{aff24}}
\and A.~R.~Cooray\orcid{0000-0002-3892-0190}\inst{\ref{aff119}}
\and M.~Costanzi\orcid{0000-0001-8158-1449}\inst{\ref{aff115},\ref{aff9},\ref{aff10}}
\and O.~Cucciati\orcid{0000-0002-9336-7551}\inst{\ref{aff3}}
\and G.~Desprez\orcid{0000-0001-8325-1742}\inst{\ref{aff120}}
\and A.~D\'iaz-S\'anchez\orcid{0000-0003-0748-4768}\inst{\ref{aff121}}
\and J.~J.~Diaz\inst{\ref{aff122}}
\and S.~Di~Domizio\orcid{0000-0003-2863-5895}\inst{\ref{aff29},\ref{aff30}}
\and H.~Dole\orcid{0000-0002-9767-3839}\inst{\ref{aff123}}
\and S.~Escoffier\orcid{0000-0002-2847-7498}\inst{\ref{aff58}}
\and M.~Ezziati\orcid{0009-0003-6065-1585}\inst{\ref{aff80}}
\and A.~G.~Ferrari\orcid{0009-0005-5266-4110}\inst{\ref{aff39},\ref{aff4}}
\and P.~G.~Ferreira\orcid{0000-0002-3021-2851}\inst{\ref{aff124}}
\and I.~Ferrero\orcid{0000-0002-1295-1132}\inst{\ref{aff61}}
\and A.~Finoguenov\orcid{0000-0002-4606-5403}\inst{\ref{aff73}}
\and A.~Fontana\orcid{0000-0003-3820-2823}\inst{\ref{aff38}}
\and F.~Fornari\orcid{0000-0003-2979-6738}\inst{\ref{aff97}}
\and L.~Gabarra\orcid{0000-0002-8486-8856}\inst{\ref{aff124}}
\and K.~Ganga\orcid{0000-0001-8159-8208}\inst{\ref{aff7}}
\and J.~Garc\'ia-Bellido\orcid{0000-0002-9370-8360}\inst{\ref{aff104}}
\and T.~Gasparetto\orcid{0000-0002-7913-4866}\inst{\ref{aff9}}
\and V.~Gautard\inst{\ref{aff125}}
\and E.~Gaztanaga\orcid{0000-0001-9632-0815}\inst{\ref{aff57},\ref{aff56},\ref{aff126}}
\and F.~Giacomini\orcid{0000-0002-3129-2814}\inst{\ref{aff4}}
\and F.~Gianotti\orcid{0000-0003-4666-119X}\inst{\ref{aff3}}
\and G.~Gozaliasl\orcid{0000-0002-0236-919X}\inst{\ref{aff127},\ref{aff73}}
\and C.~M.~Gutierrez\orcid{0000-0001-7854-783X}\inst{\ref{aff128}}
\and A.~Hall\orcid{0000-0002-3139-8651}\inst{\ref{aff41}}
\and H.~Hildebrandt\orcid{0000-0002-9814-3338}\inst{\ref{aff129}}
\and J.~Hjorth\orcid{0000-0002-4571-2306}\inst{\ref{aff85}}
\and A.~Jimenez~Mu\~noz\orcid{0009-0004-5252-185X}\inst{\ref{aff130}}
\and J.~J.~E.~Kajava\orcid{0000-0002-3010-8333}\inst{\ref{aff131},\ref{aff132}}
\and V.~Kansal\orcid{0000-0002-4008-6078}\inst{\ref{aff133},\ref{aff134}}
\and D.~Karagiannis\orcid{0000-0002-4927-0816}\inst{\ref{aff135},\ref{aff136}}
\and C.~C.~Kirkpatrick\inst{\ref{aff71}}
\and A.~M.~C.~Le~Brun\orcid{0000-0002-0936-4594}\inst{\ref{aff106}}
\and J.~Le~Graet\orcid{0000-0001-6523-7971}\inst{\ref{aff58}}
\and L.~Legrand\orcid{0000-0003-0610-5252}\inst{\ref{aff137}}
\and J.~Lesgourgues\orcid{0000-0001-7627-353X}\inst{\ref{aff37}}
\and T.~I.~Liaudat\orcid{0000-0002-9104-314X}\inst{\ref{aff138}}
\and A.~Loureiro\orcid{0000-0002-4371-0876}\inst{\ref{aff139},\ref{aff140}}
\and J.~Macias-Perez\orcid{0000-0002-5385-2763}\inst{\ref{aff130}}
\and G.~Maggio\orcid{0000-0003-4020-4836}\inst{\ref{aff9}}
\and M.~Magliocchetti\orcid{0000-0001-9158-4838}\inst{\ref{aff54}}
\and F.~Mannucci\orcid{0000-0002-4803-2381}\inst{\ref{aff141}}
\and R.~Maoli\orcid{0000-0002-6065-3025}\inst{\ref{aff142},\ref{aff38}}
\and J.~Mart\'{i}n-Fleitas\orcid{0000-0002-8594-569X}\inst{\ref{aff143}}
\and C.~J.~A.~P.~Martins\orcid{0000-0002-4886-9261}\inst{\ref{aff144},\ref{aff31}}
\and L.~Maurin\orcid{0000-0002-8406-0857}\inst{\ref{aff123}}
\and R.~B.~Metcalf\orcid{0000-0003-3167-2574}\inst{\ref{aff1},\ref{aff3}}
\and M.~Miluzio\inst{\ref{aff18},\ref{aff145}}
\and P.~Monaco\orcid{0000-0003-2083-7564}\inst{\ref{aff115},\ref{aff9},\ref{aff21},\ref{aff10}}
\and A.~Montoro\orcid{0000-0003-4730-8590}\inst{\ref{aff57},\ref{aff56}}
\and A.~Mora\orcid{0000-0002-1922-8529}\inst{\ref{aff143}}
\and C.~Moretti\orcid{0000-0003-3314-8936}\inst{\ref{aff22},\ref{aff103},\ref{aff9},\ref{aff10},\ref{aff21}}
\and G.~Morgante\inst{\ref{aff3}}
\and S.~Nadathur\orcid{0000-0001-9070-3102}\inst{\ref{aff126}}
\and Nicholas~A.~Walton\orcid{0000-0003-3983-8778}\inst{\ref{aff146}}
\and L.~Pagano\orcid{0000-0003-1820-5998}\inst{\ref{aff109},\ref{aff110}}
\and L.~Patrizii\inst{\ref{aff4}}
\and V.~Popa\orcid{0000-0002-9118-8330}\inst{\ref{aff90}}
\and D.~Potter\orcid{0000-0002-0757-5195}\inst{\ref{aff147}}
\and I.~Risso\orcid{0000-0003-2525-7761}\inst{\ref{aff148}}
\and P.-F.~Rocci\inst{\ref{aff123}}
\and M.~Sahl\'en\orcid{0000-0003-0973-4804}\inst{\ref{aff149}}
\and E.~Sarpa\orcid{0000-0002-1256-655X}\inst{\ref{aff22},\ref{aff103},\ref{aff21}}
\and A.~Schneider\orcid{0000-0001-7055-8104}\inst{\ref{aff147}}
\and M.~Schultheis\inst{\ref{aff8}}
\and P.~Simon\inst{\ref{aff78}}
\and A.~Spurio~Mancini\orcid{0000-0001-5698-0990}\inst{\ref{aff150},\ref{aff49}}
\and J.~Stadel\orcid{0000-0001-7565-8622}\inst{\ref{aff147}}
\and S.~A.~Stanford\orcid{0000-0003-0122-0841}\inst{\ref{aff151}}
\and K.~Tanidis\inst{\ref{aff124}}
\and C.~Tao\orcid{0000-0001-7961-8177}\inst{\ref{aff58}}
\and G.~Testera\inst{\ref{aff30}}
\and R.~Teyssier\orcid{0000-0001-7689-0933}\inst{\ref{aff152}}
\and S.~Toft\orcid{0000-0003-3631-7176}\inst{\ref{aff153},\ref{aff154}}
\and S.~Tosi\orcid{0000-0002-7275-9193}\inst{\ref{aff29},\ref{aff30}}
\and A.~Troja\orcid{0000-0003-0239-4595}\inst{\ref{aff91},\ref{aff53}}
\and M.~Tucci\inst{\ref{aff52}}
\and C.~Valieri\inst{\ref{aff4}}
\and J.~Valiviita\orcid{0000-0001-6225-3693}\inst{\ref{aff73},\ref{aff74}}
\and D.~Vergani\orcid{0000-0003-0898-2216}\inst{\ref{aff3}}
\and G.~Verza\orcid{0000-0002-1886-8348}\inst{\ref{aff155},\ref{aff156}}
\and P.~Vielzeuf\orcid{0000-0003-2035-9339}\inst{\ref{aff58}}}
										   
\institute{Dipartimento di Fisica e Astronomia "Augusto Righi" - Alma Mater Studiorum Universit\`a di Bologna, via Piero Gobetti 93/2, 40129 Bologna, Italy\label{aff1}
\and
INAF, Istituto di Radioastronomia, Via Piero Gobetti 101, 40129 Bologna, Italy\label{aff2}
\and
INAF-Osservatorio di Astrofisica e Scienza dello Spazio di Bologna, Via Piero Gobetti 93/3, 40129 Bologna, Italy\label{aff3}
\and
INFN-Sezione di Bologna, Viale Berti Pichat 6/2, 40127 Bologna, Italy\label{aff4}
\and
Universit\'e Paris-Saclay, Universit\'e Paris Cit\'e, CEA, CNRS, AIM, 91191, Gif-sur-Yvette, France\label{aff5}
\and
Istituto Nazionale di Fisica Nucleare, Sezione di Bologna, Via Irnerio 46, 40126 Bologna, Italy\label{aff6}
\and
Universit\'e Paris Cit\'e, CNRS, Astroparticule et Cosmologie, 75013 Paris, France\label{aff7}
\and
Universit\'e C\^{o}te d'Azur, Observatoire de la C\^{o}te d'Azur, CNRS, Laboratoire Lagrange, Bd de l'Observatoire, CS 34229, 06304 Nice cedex 4, France\label{aff8}
\and
INAF-Osservatorio Astronomico di Trieste, Via G. B. Tiepolo 11, 34143 Trieste, Italy\label{aff9}
\and
IFPU, Institute for Fundamental Physics of the Universe, via Beirut 2, 34151 Trieste, Italy\label{aff10}
\and
INAF-IASF Milano, Via Alfonso Corti 12, 20133 Milano, Italy\label{aff11}
\and
Department of Physics "E. Pancini", University Federico II, Via Cinthia 6, 80126, Napoli, Italy\label{aff12}
\and
INAF-Osservatorio Astronomico di Capodimonte, Via Moiariello 16, 80131 Napoli, Italy\label{aff13}
\and
INFN section of Naples, Via Cinthia 6, 80126, Napoli, Italy\label{aff14}
\and
Institut f\"ur Theoretische Physik, University of Heidelberg, Philosophenweg 16, 69120 Heidelberg, Germany\label{aff15}
\and
Zentrum f\"ur Astronomie, Universit\"at Heidelberg, Philosophenweg 12, 69120 Heidelberg, Germany\label{aff16}
\and
INAF-Osservatorio Astronomico di Padova, Via dell'Osservatorio 5, 35122 Padova, Italy\label{aff17}
\and
ESAC/ESA, Camino Bajo del Castillo, s/n., Urb. Villafranca del Castillo, 28692 Villanueva de la Ca\~nada, Madrid, Spain\label{aff18}
\and
School of Mathematics and Physics, University of Surrey, Guildford, Surrey, GU2 7XH, UK\label{aff19}
\and
INAF-Osservatorio Astronomico di Brera, Via Brera 28, 20122 Milano, Italy\label{aff20}
\and
INFN, Sezione di Trieste, Via Valerio 2, 34127 Trieste TS, Italy\label{aff21}
\and
SISSA, International School for Advanced Studies, Via Bonomea 265, 34136 Trieste TS, Italy\label{aff22}
\and
Dipartimento di Fisica e Astronomia, Universit\`a di Bologna, Via Gobetti 93/2, 40129 Bologna, Italy\label{aff23}
\and
Max Planck Institute for Extraterrestrial Physics, Giessenbachstr. 1, 85748 Garching, Germany\label{aff24}
\and
Universit\"ats-Sternwarte M\"unchen, Fakult\"at f\"ur Physik, Ludwig-Maximilians-Universit\"at M\"unchen, Scheinerstrasse 1, 81679 M\"unchen, Germany\label{aff25}
\and
Institut de Physique Th\'eorique, CEA, CNRS, Universit\'e Paris-Saclay 91191 Gif-sur-Yvette Cedex, France\label{aff26}
\and
Institut d'Astrophysique de Paris, UMR 7095, CNRS, and Sorbonne Universit\'e, 98 bis boulevard Arago, 75014 Paris, France\label{aff27}
\and
INAF-Osservatorio Astrofisico di Torino, Via Osservatorio 20, 10025 Pino Torinese (TO), Italy\label{aff28}
\and
Dipartimento di Fisica, Universit\`a di Genova, Via Dodecaneso 33, 16146, Genova, Italy\label{aff29}
\and
INFN-Sezione di Genova, Via Dodecaneso 33, 16146, Genova, Italy\label{aff30}
\and
Instituto de Astrof\'isica e Ci\^encias do Espa\c{c}o, Universidade do Porto, CAUP, Rua das Estrelas, PT4150-762 Porto, Portugal\label{aff31}
\and
Faculdade de Ci\^encias da Universidade do Porto, Rua do Campo de Alegre, 4150-007 Porto, Portugal\label{aff32}
\and
Dipartimento di Fisica, Universit\`a degli Studi di Torino, Via P. Giuria 1, 10125 Torino, Italy\label{aff33}
\and
INFN-Sezione di Torino, Via P. Giuria 1, 10125 Torino, Italy\label{aff34}
\and
Centro de Investigaciones Energ\'eticas, Medioambientales y Tecnol\'ogicas (CIEMAT), Avenida Complutense 40, 28040 Madrid, Spain\label{aff35}
\and
Port d'Informaci\'{o} Cient\'{i}fica, Campus UAB, C. Albareda s/n, 08193 Bellaterra (Barcelona), Spain\label{aff36}
\and
Institute for Theoretical Particle Physics and Cosmology (TTK), RWTH Aachen University, 52056 Aachen, Germany\label{aff37}
\and
INAF-Osservatorio Astronomico di Roma, Via Frascati 33, 00078 Monteporzio Catone, Italy\label{aff38}
\and
Dipartimento di Fisica e Astronomia "Augusto Righi" - Alma Mater Studiorum Universit\`a di Bologna, Viale Berti Pichat 6/2, 40127 Bologna, Italy\label{aff39}
\and
Instituto de Astrof\'isica de Canarias, Calle V\'ia L\'actea s/n, 38204, San Crist\'obal de La Laguna, Tenerife, Spain\label{aff40}
\and
Institute for Astronomy, University of Edinburgh, Royal Observatory, Blackford Hill, Edinburgh EH9 3HJ, UK\label{aff41}
\and
Jodrell Bank Centre for Astrophysics, Department of Physics and Astronomy, University of Manchester, Oxford Road, Manchester M13 9PL, UK\label{aff42}
\and
European Space Agency/ESRIN, Largo Galileo Galilei 1, 00044 Frascati, Roma, Italy\label{aff43}
\and
Universit\'e Claude Bernard Lyon 1, CNRS/IN2P3, IP2I Lyon, UMR 5822, Villeurbanne, F-69100, France\label{aff44}
\and
Institute of Physics, Laboratory of Astrophysics, Ecole Polytechnique F\'ed\'erale de Lausanne (EPFL), Observatoire de Sauverny, 1290 Versoix, Switzerland\label{aff45}
\and
Institut de Ci\`{e}ncies del Cosmos (ICCUB), Universitat de Barcelona (IEEC-UB), Mart\'{i} i Franqu\`{e}s 1, 08028 Barcelona, Spain\label{aff46}
\and
Instituci\'o Catalana de Recerca i Estudis Avan\c{c}ats (ICREA), Passeig de Llu\'{\i}s Companys 23, 08010 Barcelona, Spain\label{aff47}
\and
UCB Lyon 1, CNRS/IN2P3, IUF, IP2I Lyon, 4 rue Enrico Fermi, 69622 Villeurbanne, France\label{aff48}
\and
Mullard Space Science Laboratory, University College London, Holmbury St Mary, Dorking, Surrey RH5 6NT, UK\label{aff49}
\and
Departamento de F\'isica, Faculdade de Ci\^encias, Universidade de Lisboa, Edif\'icio C8, Campo Grande, PT1749-016 Lisboa, Portugal\label{aff50}
\and
Instituto de Astrof\'isica e Ci\^encias do Espa\c{c}o, Faculdade de Ci\^encias, Universidade de Lisboa, Campo Grande, 1749-016 Lisboa, Portugal\label{aff51}
\and
Department of Astronomy, University of Geneva, ch. d'Ecogia 16, 1290 Versoix, Switzerland\label{aff52}
\and
INFN-Padova, Via Marzolo 8, 35131 Padova, Italy\label{aff53}
\and
INAF-Istituto di Astrofisica e Planetologia Spaziali, via del Fosso del Cavaliere, 100, 00100 Roma, Italy\label{aff54}
\and
Space Science Data Center, Italian Space Agency, via del Politecnico snc, 00133 Roma, Italy\label{aff55}
\and
Institut d'Estudis Espacials de Catalunya (IEEC),  Edifici RDIT, Campus UPC, 08860 Castelldefels, Barcelona, Spain\label{aff56}
\and
Institute of Space Sciences (ICE, CSIC), Campus UAB, Carrer de Can Magrans, s/n, 08193 Barcelona, Spain\label{aff57}
\and
Aix-Marseille Universit\'e, CNRS/IN2P3, CPPM, Marseille, France\label{aff58}
\and
FRACTAL S.L.N.E., calle Tulip\'an 2, Portal 13 1A, 28231, Las Rozas de Madrid, Spain\label{aff59}
\and
Dipartimento di Fisica "Aldo Pontremoli", Universit\`a degli Studi di Milano, Via Celoria 16, 20133 Milano, Italy\label{aff60}
\and
Institute of Theoretical Astrophysics, University of Oslo, P.O. Box 1029 Blindern, 0315 Oslo, Norway\label{aff61}
\and
Jet Propulsion Laboratory, California Institute of Technology, 4800 Oak Grove Drive, Pasadena, CA, 91109, USA\label{aff62}
\and
Felix Hormuth Engineering, Goethestr. 17, 69181 Leimen, Germany\label{aff63}
\and
Technical University of Denmark, Elektrovej 327, 2800 Kgs. Lyngby, Denmark\label{aff64}
\and
Cosmic Dawn Center (DAWN), Denmark\label{aff65}
\and
Universit\'e Paris-Saclay, CNRS/IN2P3, IJCLab, 91405 Orsay, France\label{aff66}
\and
Institut de Recherche en Astrophysique et Plan\'etologie (IRAP), Universit\'e de Toulouse, CNRS, UPS, CNES, 14 Av. Edouard Belin, 31400 Toulouse, France\label{aff67}
\and
Max-Planck-Institut f\"ur Astronomie, K\"onigstuhl 17, 69117 Heidelberg, Germany\label{aff68}
\and
NASA Goddard Space Flight Center, Greenbelt, MD 20771, USA\label{aff69}
\and
Department of Physics and Astronomy, University College London, Gower Street, London WC1E 6BT, UK\label{aff70}
\and
Department of Physics and Helsinki Institute of Physics, Gustaf H\"allstr\"omin katu 2, 00014 University of Helsinki, Finland\label{aff71}
\and
Universit\'e de Gen\`eve, D\'epartement de Physique Th\'eorique and Centre for Astroparticle Physics, 24 quai Ernest-Ansermet, CH-1211 Gen\`eve 4, Switzerland\label{aff72}
\and
Department of Physics, P.O. Box 64, 00014 University of Helsinki, Finland\label{aff73}
\and
Helsinki Institute of Physics, Gustaf H{\"a}llstr{\"o}min katu 2, University of Helsinki, Helsinki, Finland\label{aff74}
\and
NOVA optical infrared instrumentation group at ASTRON, Oude Hoogeveensedijk 4, 7991PD, Dwingeloo, The Netherlands\label{aff75}
\and
Centre de Calcul de l'IN2P3/CNRS, 21 avenue Pierre de Coubertin 69627 Villeurbanne Cedex, France\label{aff76}
\and
University of Applied Sciences and Arts of Northwestern Switzerland, School of Engineering, 5210 Windisch, Switzerland\label{aff77}
\and
Universit\"at Bonn, Argelander-Institut f\"ur Astronomie, Auf dem H\"ugel 71, 53121 Bonn, Germany\label{aff78}
\and
INFN-Sezione di Roma, Piazzale Aldo Moro, 2 - c/o Dipartimento di Fisica, Edificio G. Marconi, 00185 Roma, Italy\label{aff79}
\and
Aix-Marseille Universit\'e, CNRS, CNES, LAM, Marseille, France\label{aff80}
\and
Department of Physics, Institute for Computational Cosmology, Durham University, South Road, Durham, DH1 3LE, UK\label{aff81}
\and
Institut d'Astrophysique de Paris, 98bis Boulevard Arago, 75014, Paris, France\label{aff82}
\and
European Space Agency/ESTEC, Keplerlaan 1, 2201 AZ Noordwijk, The Netherlands\label{aff83}
\and
Institut de F\'{i}sica d'Altes Energies (IFAE), The Barcelona Institute of Science and Technology, Campus UAB, 08193 Bellaterra (Barcelona), Spain\label{aff84}
\and
DARK, Niels Bohr Institute, University of Copenhagen, Jagtvej 155, 2200 Copenhagen, Denmark\label{aff85}
\and
Waterloo Centre for Astrophysics, University of Waterloo, Waterloo, Ontario N2L 3G1, Canada\label{aff86}
\and
Department of Physics and Astronomy, University of Waterloo, Waterloo, Ontario N2L 3G1, Canada\label{aff87}
\and
Perimeter Institute for Theoretical Physics, Waterloo, Ontario N2L 2Y5, Canada\label{aff88}
\and
Centre National d'Etudes Spatiales -- Centre spatial de Toulouse, 18 avenue Edouard Belin, 31401 Toulouse Cedex 9, France\label{aff89}
\and
Institute of Space Science, Str. Atomistilor, nr. 409 M\u{a}gurele, Ilfov, 077125, Romania\label{aff90}
\and
Dipartimento di Fisica e Astronomia "G. Galilei", Universit\`a di Padova, Via Marzolo 8, 35131 Padova, Italy\label{aff91}
\and
Universit\'e St Joseph; Faculty of Sciences, Beirut, Lebanon\label{aff92}
\and
Departamento de F\'isica, FCFM, Universidad de Chile, Blanco Encalada 2008, Santiago, Chile\label{aff93}
\and
Satlantis, University Science Park, Sede Bld 48940, Leioa-Bilbao, Spain\label{aff94}
\and
Instituto de Astrof\'isica e Ci\^encias do Espa\c{c}o, Faculdade de Ci\^encias, Universidade de Lisboa, Tapada da Ajuda, 1349-018 Lisboa, Portugal\label{aff95}
\and
Universidad Polit\'ecnica de Cartagena, Departamento de Electr\'onica y Tecnolog\'ia de Computadoras,  Plaza del Hospital 1, 30202 Cartagena, Spain\label{aff96}
\and
INFN-Bologna, Via Irnerio 46, 40126 Bologna, Italy\label{aff97}
\and
Kapteyn Astronomical Institute, University of Groningen, PO Box 800, 9700 AV Groningen, The Netherlands\label{aff98}
\and
Dipartimento di Fisica, Universit\`a degli studi di Genova, and INFN-Sezione di Genova, via Dodecaneso 33, 16146, Genova, Italy\label{aff99}
\and
Infrared Processing and Analysis Center, California Institute of Technology, Pasadena, CA 91125, USA\label{aff100}
\and
Astronomical Observatory of the Autonomous Region of the Aosta Valley (OAVdA), Loc. Lignan 39, I-11020, Nus (Aosta Valley), Italy\label{aff101}
\and
ICL, Junia, Universit\'e Catholique de Lille, LITL, 59000 Lille, France\label{aff102}
\and
ICSC - Centro Nazionale di Ricerca in High Performance Computing, Big Data e Quantum Computing, Via Magnanelli 2, Bologna, Italy\label{aff103}
\and
Instituto de F\'isica Te\'orica UAM-CSIC, Campus de Cantoblanco, 28049 Madrid, Spain\label{aff104}
\and
CERCA/ISO, Department of Physics, Case Western Reserve University, 10900 Euclid Avenue, Cleveland, OH 44106, USA\label{aff105}
\and
Laboratoire Univers et Th\'eorie, Observatoire de Paris, Universit\'e PSL, Universit\'e Paris Cit\'e, CNRS, 92190 Meudon, France\label{aff106}
\and
INFN-Sezione di Milano, Via Celoria 16, 20133 Milano, Italy\label{aff107}
\and
Departamento de F{\'\i}sica Fundamental. Universidad de Salamanca. Plaza de la Merced s/n. 37008 Salamanca, Spain\label{aff108}
\and
Dipartimento di Fisica e Scienze della Terra, Universit\`a degli Studi di Ferrara, Via Giuseppe Saragat 1, 44122 Ferrara, Italy\label{aff109}
\and
Istituto Nazionale di Fisica Nucleare, Sezione di Ferrara, Via Giuseppe Saragat 1, 44122 Ferrara, Italy\label{aff110}
\and
Universit\'e de Strasbourg, CNRS, Observatoire astronomique de Strasbourg, UMR 7550, 67000 Strasbourg, France\label{aff111}
\and
Center for Data-Driven Discovery, Kavli IPMU (WPI), UTIAS, The University of Tokyo, Kashiwa, Chiba 277-8583, Japan\label{aff112}
\and
Ludwig-Maximilians-University, Schellingstrasse 4, 80799 Munich, Germany\label{aff113}
\and
Max-Planck-Institut f\"ur Physik, Boltzmannstr. 8, 85748 Garching, Germany\label{aff114}
\and
Dipartimento di Fisica - Sezione di Astronomia, Universit\`a di Trieste, Via Tiepolo 11, 34131 Trieste, Italy\label{aff115}
\and
Minnesota Institute for Astrophysics, University of Minnesota, 116 Church St SE, Minneapolis, MN 55455, USA\label{aff116}
\and
Institute Lorentz, Leiden University, Niels Bohrweg 2, 2333 CA Leiden, The Netherlands\label{aff117}
\and
Institute for Astronomy, University of Hawaii, 2680 Woodlawn Drive, Honolulu, HI 96822, USA\label{aff118}
\and
Department of Physics \& Astronomy, University of California Irvine, Irvine CA 92697, USA\label{aff119}
\and
Department of Astronomy \& Physics and Institute for Computational Astrophysics, Saint Mary's University, 923 Robie Street, Halifax, Nova Scotia, B3H 3C3, Canada\label{aff120}
\and
Departamento F\'isica Aplicada, Universidad Polit\'ecnica de Cartagena, Campus Muralla del Mar, 30202 Cartagena, Murcia, Spain\label{aff121}
\and
Instituto de Astrof\'isica de Canarias (IAC); Departamento de Astrof\'isica, Universidad de La Laguna (ULL), 38200, La Laguna, Tenerife, Spain\label{aff122}
\and
Universit\'e Paris-Saclay, CNRS, Institut d'astrophysique spatiale, 91405, Orsay, France\label{aff123}
\and
Department of Physics, Oxford University, Keble Road, Oxford OX1 3RH, UK\label{aff124}
\and
CEA Saclay, DFR/IRFU, Service d'Astrophysique, Bat. 709, 91191 Gif-sur-Yvette, France\label{aff125}
\and
Institute of Cosmology and Gravitation, University of Portsmouth, Portsmouth PO1 3FX, UK\label{aff126}
\and
Department of Computer Science, Aalto University, PO Box 15400, Espoo, FI-00 076, Finland\label{aff127}
\and
Instituto de Astrof\'\i sica de Canarias, c/ Via Lactea s/n, La Laguna E-38200, Spain. Departamento de Astrof\'\i sica de la Universidad de La Laguna, Avda. Francisco Sanchez, La Laguna, E-38200, Spain\label{aff128}
\and
Ruhr University Bochum, Faculty of Physics and Astronomy, Astronomical Institute (AIRUB), German Centre for Cosmological Lensing (GCCL), 44780 Bochum, Germany\label{aff129}
\and
Univ. Grenoble Alpes, CNRS, Grenoble INP, LPSC-IN2P3, 53, Avenue des Martyrs, 38000, Grenoble, France\label{aff130}
\and
Department of Physics and Astronomy, Vesilinnantie 5, 20014 University of Turku, Finland\label{aff131}
\and
Serco for European Space Agency (ESA), Camino bajo del Castillo, s/n, Urbanizacion Villafranca del Castillo, Villanueva de la Ca\~nada, 28692 Madrid, Spain\label{aff132}
\and
ARC Centre of Excellence for Dark Matter Particle Physics, Melbourne, Australia\label{aff133}
\and
Centre for Astrophysics \& Supercomputing, Swinburne University of Technology,  Hawthorn, Victoria 3122, Australia\label{aff134}
\and
School of Physics and Astronomy, Queen Mary University of London, Mile End Road, London E1 4NS, UK\label{aff135}
\and
Department of Physics and Astronomy, University of the Western Cape, Bellville, Cape Town, 7535, South Africa\label{aff136}
\and
ICTP South American Institute for Fundamental Research, Instituto de F\'{\i}sica Te\'orica, Universidade Estadual Paulista, S\~ao Paulo, Brazil\label{aff137}
\and
IRFU, CEA, Universit\'e Paris-Saclay 91191 Gif-sur-Yvette Cedex, France\label{aff138}
\and
Oskar Klein Centre for Cosmoparticle Physics, Department of Physics, Stockholm University, Stockholm, SE-106 91, Sweden\label{aff139}
\and
Astrophysics Group, Blackett Laboratory, Imperial College London, London SW7 2AZ, UK\label{aff140}
\and
INAF-Osservatorio Astrofisico di Arcetri, Largo E. Fermi 5, 50125, Firenze, Italy\label{aff141}
\and
Dipartimento di Fisica, Sapienza Universit\`a di Roma, Piazzale Aldo Moro 2, 00185 Roma, Italy\label{aff142}
\and
Aurora Technology for European Space Agency (ESA), Camino bajo del Castillo, s/n, Urbanizacion Villafranca del Castillo, Villanueva de la Ca\~nada, 28692 Madrid, Spain\label{aff143}
\and
Centro de Astrof\'{\i}sica da Universidade do Porto, Rua das Estrelas, 4150-762 Porto, Portugal\label{aff144}
\and
HE Space for European Space Agency (ESA), Camino bajo del Castillo, s/n, Urbanizacion Villafranca del Castillo, Villanueva de la Ca\~nada, 28692 Madrid, Spain\label{aff145}
\and
Institute of Astronomy, University of Cambridge, Madingley Road, Cambridge CB3 0HA, UK\label{aff146}
\and
Department of Astrophysics, University of Zurich, Winterthurerstrasse 190, 8057 Zurich, Switzerland\label{aff147}
\and
INAF-Osservatorio Astronomico di Brera, Via Brera 28, 20122 Milano, Italy, and INFN-Sezione di Genova, Via Dodecaneso 33, 16146, Genova, Italy\label{aff148}
\and
Theoretical astrophysics, Department of Physics and Astronomy, Uppsala University, Box 515, 751 20 Uppsala, Sweden\label{aff149}
\and
Department of Physics, Royal Holloway, University of London, TW20 0EX, UK\label{aff150}
\and
Department of Physics and Astronomy, University of California, Davis, CA 95616, USA\label{aff151}
\and
Department of Astrophysical Sciences, Peyton Hall, Princeton University, Princeton, NJ 08544, USA\label{aff152}
\and
Cosmic Dawn Center (DAWN)\label{aff153}
\and
Niels Bohr Institute, University of Copenhagen, Jagtvej 128, 2200 Copenhagen, Denmark\label{aff154}
\and
Center for Cosmology and Particle Physics, Department of Physics, New York University, New York, NY 10003, USA\label{aff155}
\and
Center for Computational Astrophysics, Flatiron Institute, 162 5th Avenue, 10010, New York, NY, USA\label{aff156}}    



 
\abstract
{
The ability to measure unbiased weak-lensing (WL) masses is a key ingredient to exploit galaxy clusters as a competitive cosmological probe with the ESA \Euclid survey or future missions. 
We investigate the level of accuracy and precision of cluster masses measured with the \Euclid data processing pipeline. We use the DEMNUni-Cov $N$-body simulations to assess how well the WL mass probes the true halo mass, and, then, how well WL masses can be recovered in the presence of measurement uncertainties. We consider different halo mass density models, priors, and mass point estimates, i.e., the biweight, mean, and median of the marginalised posterior distribution and the maximum likelihood parameter. 
WL mass differs from true mass due to, e.g., the intrinsic ellipticity of sources, correlated or uncorrelated matter and large-scale structure, halo triaxiality and orientation, and merging or irregular morphology. In an ideal scenario without observational or measurement errors, the maximum likelihood estimator is the most accurate, with WL masses biased low by $\langle b_M \rangle = -14.6 \pm 1.7 \, \%$ on average over the full range $M_\text{200c} > 5 \times 10^{13} \, M_\odot$ and $z < 1$. Due to the stabilising effect of the prior, the biweight, mean, and median estimates are more precise, i.e., they have a smaller intrinsic scatter. The scatter decreases with increasing mass and informative priors can significantly reduce the scatter. Halo mass density profiles with a truncation provide better fits to the lensing signal, while the accuracy and precision are not significantly affected.
We further investigate the impact of various additional sources of systematic uncertainty on the WL mass estimates, namely the impact of photometric redshift uncertainties and source selection, the expected performance of \Euclid cluster detection algorithms, and the presence of masks. Taken in isolation, we find that the largest effect is induced by non-conservative source selection with $\langle b_M \rangle = - 33.4 \pm 1.6 \, \%$. This effect can be mostly removed with a robust selection. As a final \Euclid-like test, we combine systematic effects in a realistic observational setting and find $\langle b_M \rangle = - 15.5 \pm 2.4 \, \%$ under a robust selection. This is very similar to the ideal case, though with a slightly larger scatter mostly due to cluster redshift uncertainty and miscentering.
}



\authorrunning{Euclid Collaboration: L. Ingoglia et al.}
\titlerunning{Cluster weak lensing mass precision and accuracy}
\maketitle

\section{Introduction} \label{sec: Introduction}

Galaxy clusters are robust tracers of the matter density field \citep[e.g.,][]{2011ARA&A..49..409A}. They are hosted in dark matter haloes, which are seeded by initial perturbations of the matter overdensity field and subsequently form via the hierarchical growth of structure. At the present time, many clusters have reached virial equilibrium and are among the most massive gravitationally bound systems in the Universe \citep{2005RvMP...77..207V, 2008LNP...740..287B}. 

Clusters probe the distribution and evolution of cosmic structures through their abundance, spatial distribution, and as a function of their redshift and mass. The high-mass end of the halo-mass function has been measured by cluster surveys and is particularly sensitive to dark energy \citep{2009ApJ...692.1060V, 2015MNRAS.446.2205M, 2016A&A...594A..22P, 2019MNRAS.488.4779C, 2019ApJ...878...55B}. Surveys measure baryonic mass proxies, such as richness, the Sunyaev--Zeldovich effect, and X-ray luminosity, from intracluster gas or the distribution of galaxies. A mass-observable relation is additionally needed to anchor these survey measurements to the halo mass function. Therefore, unbiased cluster mass measurements are essential to accurately constrain cosmological parameters \citep{2021A&A...653A..19G, 2022MNRAS.511.1484I}.

One reliable method to measure galaxy cluster masses is via weak gravitational lensing \citep[henceforth WL, for a review see][]{2001PhR...340..291B, 2006glsw.conf..269S, 2015RPPh...78h6901K}, an effect by which the images of background galaxies are distorted due to a foreground mass. The inference of cluster masses from WL is close to unbiased \citep{2011ApJ...740...25B, 2011MNRAS.414.1851O, 2012MNRAS.421.1073B}, independent of the dynamical state of the cluster, and probes the entire halo mass, which is dominated by dark matter \citep{2012MNRAS.427.1298H, 2015MNRAS.449..685H, 2014ApJ...795..163U, 2016ApJ...821..116U,2018NatAs...2..744S}. Individual WL cluster masses have successfully been used in cluster count measurements to constrain cosmology \citep{2019ApJ...878...55B, 2021PhRvD.103d3522C}.

In the past decade, the emergence of large and deep photometric surveys, such as the Canada-France-Hawaii Telescope Legacy Survey \citep[CFHTLS;][]{2012MNRAS.427..146H}, the Kilo-Degree Survey \citep[KiDS;][]{2013ExA....35...25D}, the Hyper Suprime-Cam Subaru Strategic Program \citep[HSC-SSP;][]{2018PASJ...70S...4A, 2018PASJ...70S...1M}, and the Dark Energy Survey \citep[DES;][]{2018ApJS..239...18A}, has enabled precise WL mass measurements of large samples of individual galaxy clusters \citep[e.g.,][]{2017MNRAS.472.1946S, 2018NatAs...2..744S, ume+20, 2022MNRAS.512.4785M, PaperIII}. These measurements have laid the groundwork for the next generation of surveys.

The ESA \Euclid Survey \citep{2011arXiv1110.3193L, 2022A&A...662A.112E, EuclidSkyOverview} will observe galaxies over a significant fraction of the sky (the nominal area of the wide survey is $14 \, 000 \, \rm deg^2$) in wide optical and near-infrared bands. From the photometric galaxy catalogue, galaxy clusters will be identified using two detection algorithms, \AMICO \citep{2011MNRAS.413.1145B, 2018MNRAS.473.5221B, 2019A&A...627A..23E} and \PZWav \citep{2014becs.confE...7G, 2019A&A...627A..23E, 2024ApJ...967..123T}, and their masses will be estimated with the \Euclid combined clusters and WL pipeline \COMBCL\footnote{\url{https://gitlab.euclid-sgs.uk/PF-LE3-CL/LE3_COMB_CL}. The access is restricted to members of the \Euclid Consortium. A public release is expected with Euclid Collaboration: Farrens et al. (in prep.).} using the shape catalogue constructed from the high-resolution images taken with the VIS instrument \citep{2012SPIE.8442E..0VC, EuclidSkyVIS}. While a comprehensive description of the code structure and methods employed by \COMBCL will be presented in a forthcoming paper (Euclid Collaboration: Farrens et al. in prep.), a brief overview of the pipeline can be found in \citet[][]{PaperIII}.

Measuring individual galaxy cluster masses using the WL signal is challenging due to large intrinsic scatter in the shapes of galaxies (shape noise), contamination from foreground or cluster member galaxies in the source catalogue, and uncertainty in the photometric redshifts (henceforth photo-$z$) of source galaxies \citep{2019MNRAS.482.1352M, PaperIV}. Other issues, such as miscentring \citep{2022MNRAS.509.1127S}, triaxiality \citep{2011ApJ...740...25B}, or merging \citep{2023ApJ...945...71L}, can also create a bias with respect to the true cluster mass.
Numerical simulations are powerful tools to measure these various effects as they provide a true halo mass and allow us to test how individual systematic effects influence the lensing signal. Recent works on hydrodynamical simulations have made significant progress in determining bias and uncertainties in WL mass calibration \citep{2019MNRAS.488.2041G, 2021MNRAS.507.5671G}. 

\citet[][referred hereafter as \citetalias{giocoli24}]{giocoli24} performed a systematic study of the cluster mass bias based on the Three Hundred Project \citep{2018MNRAS.480.2898C, 2022MNRAS.514..977C} simulations, a sample of 324 high massive clusters resimulated with hydrodynamical physics. They examined the impact of cluster projection effects, different halo profile modelling, and free, fixed, or mass-dependent concentration on the measured WL mass in the Euclid Wide Survey. 

In this paper, we present an analysis, complementary to \citetalias{giocoli24}, of the bias affecting the WL cluster mass estimates obtained with \COMBCL in a \Euclid-like setting. 
Using a large $N$-body simulation data set, we measure the mass bias for different statistical point estimates over the broad mass range $M_{\rm 200c} > 5 \times 10^{13} \, M_\odot$, where $M_{\rm 200c}$ is the mass enclosed by a spherical overdensity 200 times the critical density of the Universe at the cluster redshift. Our analysis differs from the Three Hundred study in several ways. Firstly, the distribution in mass and redshift of the simulations we use is more representative of the clusters that \Euclid is expected to detect. Secondly, while the Three Hundred study focuses on the most massive clusters, we here perform a WL mass bias analysis on a larger sample of clusters that includes relatively low-mass clusters. Furthermore, we study the impact of individual and combined systematic effects on both lens and source catalogues. In addition to shape noise, we assess the impact of uncertain source redshift estimates, different algorithms used for cluster detection in \Euclid (\AMICO or \PZWav), cluster miscentring, masks, models of the halo density profile, and priors.

In this work, we assume a spatially flat $\Lambda$CDM model consistent with the simulation data adopted.
All references to `$\ln$' and `$\log$' stand for the natural and decimal logarithms, respectively. 
Masses expressed in logarithms are $\log (M / M_\odot)$ or $\ln (M / M_\odot)$.

This paper is part of a series presenting and discussing WL mass measurements of clusters exploiting the \COMBCL pipeline.  Euclid Collaboration: Farrens et al. (in prep.) describes the algorithms and code, \citet{PaperIII} tests the robustness of \COMBCL through the reanalysis of precursor photometric surveys, and \citet{PaperIV} introduces a novel method for colour selections of background galaxies. 

The present paper discusses the simulation data used for the analysis (Sect.~\ref{sec: Simulated data}), methods for lensing mass measurements (Sect.~\ref{sec: Lensing Measurements}), accuracy and precision of WL mass estimates (Sect.~\ref{sec: Accuracy and Precision}), impacts of various systematic effects on the mass bias (Sect.~\ref{sec: Systematics Assessment}), and the role of halo profile modelling and priors (Sect.~\ref{sec: Model and prior}). All systematic effects are considered together in Sect.~\ref{sec: Combined systematics}, and a conclusive summary and discussion are presented in Sects.~\ref{sec: Summary}, and ~\ref{sec: Discussion}, respectively.



\section{Simulated data} \label{sec: Simulated data}

In this work, we are interested in simulated data of massive dark matter haloes, typical of those that host galaxy clusters. 
The following section presents the simulations used in the paper for a consistent test of individual cluster WL mass measurements. 

\subsection{DEMNUni-Cov} \label{sec: DEMNUni-Cov}

The Dark Energy and Massive Neutrino Universe \citep[DEMNUni,][]{2016JCAP...07..034C, 2022JCAP...11..041P} is a set of cosmological $N$-body simulations that follows the redshift evolution of the large-scale structure (LSS) of the Universe with and without massive neutrinos. These simulations were designed to study covariance matrices of various cosmological observables, such as galaxy clustering, WL and complementary CMB data. Their large volume and high resolution make them ideal for cluster WL studies. 

In this work, we use one of the DEMNUni-Cov independent $N$-body simulations \citep{2021JCAP...01..009P, 2023A&A...673A...1B}. It consists of the gravitational evolution of $1024^3$ CDM (Cold Dark Matter) particles with mass resolution $m_{\rm p} \sim 8 \times 10^{10} \, h^{-1} M_\odot$ in a box of comoving size equal to $1 \, h^{-1} \rm Gpc$ on a side. Initial conditions were generated at $z=99$, using a theoretical linear power spectrum calculated using \texttt{CAMB} \citep{camb}. The considered cosmological parameters are consistent with \citet{planck18}; specifically, $\Omega_{\rm cdm} = 0.27$, $\Omega_{\rm b} = 0.05$, Hubble constant $H_0=67 \, \rm km \, s^{-1} \, Mpc^{-1}$, initial scalar amplitude $A_{\rm s} = 2.1265 \times 10^{-9}$, and primordial spectra index $n_{\rm s} = 0.96$. 63 snapshots were stored while running the simulations from redshift $z = 99$ to $z = 0$. This is sufficient to construct continuous past-light cones up to high redshifts.

Lensing past-light cone simulations are constructed with the \texttt{MapSim} pipeline routines \citep{2015MNRAS.452.2757G}.
The size of the simulation box is sufficient to design a pyramidal past-light cone with a square base up to $z=4$ and an aperture of 10 degrees on a side. A total of 43 lens planes are constructed by reading 40 stored snapshots from $z=0$ to $z=4$, replicating the boxes five times along the light-cone, and stacking the boxes \citep{2015JCAP...10..036T, 2017MNRAS.470.3574G, 2018MNRAS.478.1305C, 2018MNRAS.481.2813G, 2018MNRAS.478.5436G, 2018A&A...619A..38P, 2020MNRAS.493..305H, 2021MNRAS.505.2886B}. Rays are shot through these lens planes in the Born approximation regime from various sources redshifts, located at the upper bound of each lens plane, down to the observer placed at the vertex of the pyramid. The convergence and shear maps are computed with the \texttt{MOKA} library pipeline \citep{giocoli12a} and resolved with 4096 pixels, which correspond to a pixel resolution of $\ang{;;8.8}$, adequate for WL cluster studies. 

The Born approximation is reasonable given the resolution and the volume of DEMNUni-Cov. It is worth underlining that this assumption is an excellent approximation on small angular scales in the WL regime and remains valid in the cluster lensing regime, as long as we avoid the core region of the cluster \citep{2012MNRAS.420..455S}. 
This method has been used and tested on a variety of cosmological simulations \citep{2015JCAP...10..036T, 2018MNRAS.478.1305C, 2018MNRAS.478.5436G, 2023A&A...675A.120E} and recently compared with other algorithms \citep{2020MNRAS.493..305H}.
From the calibration of shear measured in \Euclid-like surveys, we expect a sub-percent level of accuracy \citep{2013MNRAS.431.3103C}. We extract the corresponding shear catalogue, as done in \citet{2023A&A...675A.120E}, by populating the past-light cone with the expected \Euclid source redshift distribution for the wide survey \citep{blanchard20}.

\subsection{Source and lens specifics} \label{sec: Source and lens specifics}

\subsubsection{Sources}

An unbiased shear catalogue of unclustered sources is derived from the DEMNUni-Cov past-light cones with information about source position $\rm (RA, Dec)$, redshift, shear components $\gamma_1$, $\gamma_2$, and convergence $\kappa$.
The sources are uniformly spatially distributed with a number density of $30 \, \rm arcmin^{-2}$ up to $z= 3$. 

We simulate observed ellipticities, including both intrinsic ellipticity and shear distortion, based on the simulated shear and convergence. In the WL regime, where $\gamma \ll 1$ and $\kappa \ll 1$, the average intrinsic shape of randomly oriented sources has zero ellipticity, and the ensemble average observed ellipticity of the sources is equivalent to the reduced shear $g \equiv \gamma / (1 - \kappa)$. 

Due to the weak deformation of the source shape, the shear components are dominated by shot noise. We simulate observed ellipticities by adding shape noise to the reduced shear as \citep{1997A&A...318..687S}
\begin{equation}
g_{\rm obs} = \frac{\epsilon_{\rm s} + g}{1 + g^{\ast} \epsilon_{\rm s}} \, ,
\label{eq: g_obs}
\end{equation}
where the intrinsic ellipticity of the source $\epsilon_{\rm s}$ is normally generated assuming a shape dispersion of $\sigma_{\epsilon} = 0.26$. as expected for \Euclid \citep{2019A&A...627A..59E, 2023A&A...675A.120E}.

For our analysis, we consider either true simulated source redshifts, or redshifts scattered to mimic the process of a photometric redshift measurement, see Sect.~\ref{sec: Reshifts uncertainty}. We discuss two sets of simulated photo-$z$s with appropriate selections: one accounting for outliers and using a non-conservative cut, and another with more reliable photo-$z$s and a robust cut. We do not consider uncertainties in the source position, which are negligible in the WL regime. 
The astrometric uncertainty for \Euclid sources is lower than $15 \arcsecond$ \citep{2022A&A...658A.126E}, and the effect on the signal is negligible in WL analyses.

\subsubsection{Lenses}

The lens halo catalogue extracted from the DM haloes provides the following information: position $\rm (RA, Dec)$; redshift of the deflector $z_{\rm d} < 1$; 
mass $M_{\rm 200c} > 5 \times 10^{13} \, M_\odot$.

\begin{figure}
    \includegraphics[width=\columnwidth]{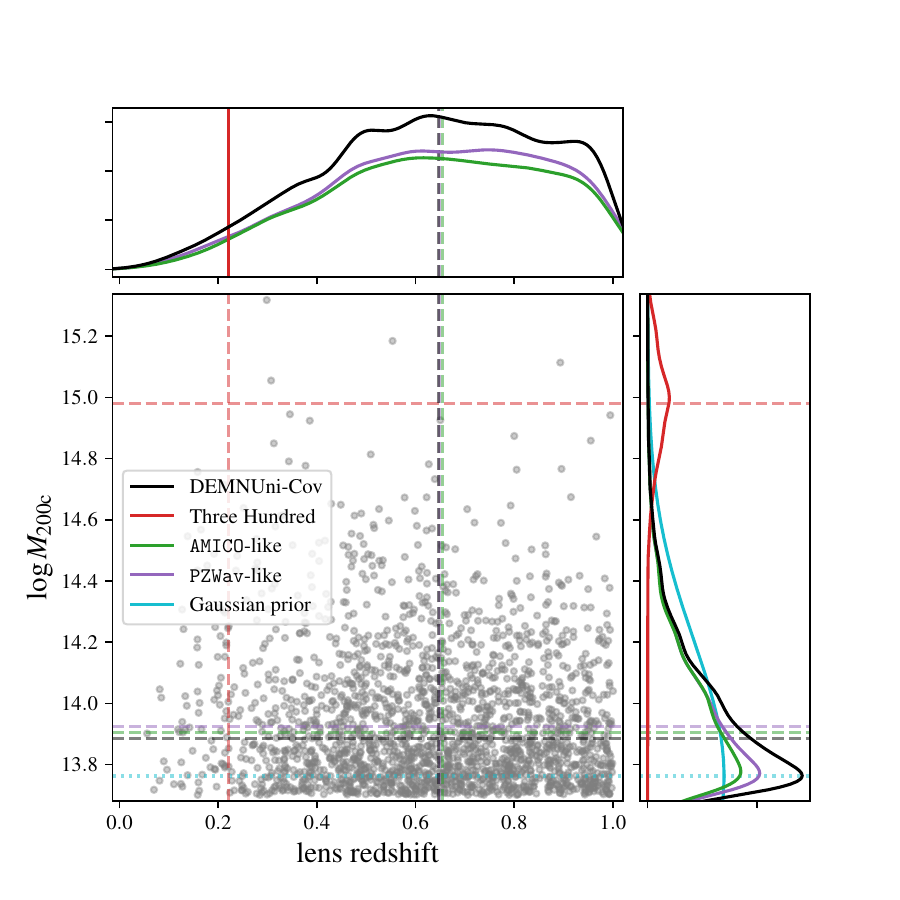}
    \caption{Distribution of the DEMNUni-Cov clusters and of the Three Hundred clusters \citepalias{giocoli24} in the mass-redshift plane. The marginalised distribution of the redshift and mass of the clusters is shown in the top and right panels, respectively. Distributions are normalised to the full DEMNUni-Cov sample. Distributions of \AMICO and \PZWav-like selected objects (see Sect.~\ref{sec: Euclid-like clusters}) are also shown. The median is shown with dashed lines. The mean of the Gaussian prior on mass (see Sect.~\ref{sec: Gaussian mass prior}) is shown on the right panel as a dotted line.}
    \label{fig: clusters_distribution}
\end{figure}

The DEMNUni-Cov sample consists of 6155 clusters. Cuts on the halo catalogue are made at low mass, $M_{\rm 200c} > 5 \times 10^{13} \, M_\odot$, and high redshift, $z_{\rm d} < 1$, to focus on the broad mass-redshift region of \Euclid-detected clusters for which the WL mass can be obtained with a large signal-to-noise ratio \citep{2019A&A...627A..23E, PaperIII}. 
Figure~\ref{fig: clusters_distribution} depicts the distribution of selected clusters. The mass distribution peaks at the lower bound of the sample while the cluster redshifts are more uniformly distributed above $z_{\rm d} > 0.4$.

For our analysis, we consider either the full sample with true simulated lens position and redshift or subsamples with scattered values to mimic the detection process, see Sect.~\ref{sec: Euclid-like clusters}. Two \Euclid-like cluster catalogues are generated, simulating the centre positions and cluster distributions using the selection functions of the detection algorithms \AMICO \citep{2011MNRAS.413.1145B, 2018MNRAS.473.5221B, 2019A&A...627A..23E} and \PZWav \citep{2014becs.confE...7G, 2019A&A...627A..23E, 2024ApJ...967..123T}.

\subsection{Three Hundred} \label{sec: Differences with Three Hundred}

We compare our findings with the recent results presented by \citetalias{giocoli24} and based on the Three Hundred simulations. These runs consist of 324 hydrodynamical simulated regions by the Three Hundred Collaboration \citep{2018MNRAS.480.2898C, 2022MNRAS.514..977C} centred on the most massive clusters ($M_{\rm 200c} \ga 8 \times 10^{14} \, h^{-1} M_\odot$) identified at $z_{\rm d} = 0$ in a box size of $1 \, \rm Gpc$ on a side, where haloes have been identified using \texttt{AHF} \citep[Amiga's Halo Finder,][]{2009ApJS..182..608K}.

The WL mass measurements from the Three Hundred simulations are discussed in Sect.~\ref{sec: Comparison with the Three Hundred}. In \citetalias{giocoli24}, the analyses of cluster WL mass bias are performed using the projections of lensing signals along different orientations of the mass tensor ellipsoid. The authors consider the haloes at $z_{\rm d} = 0.22$ and three random projections, which raises the effective number of cluster mass measurements to 972.

The authors fitted a Baltz--Marshall--Oguri model \citep{2009JCAP...01..015B}, as implemented in the \texttt{CosmoBolognaLib}\footnote{\url{https://gitlab.com/federicomarulli/CosmoBolognaLib}} \citep{marulli16} libraries, with fixed, free, or mass-constrained values of the concentration parameter and various values of the truncated radius, finding that the results are sensitive to the model parameter choices.

The mass and redshift range of DEMNUni-Cov clusters more closely resemble what is expected from \Euclid observations than the mass range of the Three Hundred Project. In particular, the DEMNUni-Cov catalogue focuses on a lower cluster mass range; thus, the cluster sample is much larger than that of the Three Hundred. 

For the Three Hundred simulations, the WL signal was derived by projecting the particles in a slice of depth $\pm 5 \, \rm Mpc$ in front and behind the cluster onto a single lens plane to focus on the lensing effect of the main halo. On the contrary, a full multiplane ray-tracing was performed in DEMNUni-Cov simulations, and we can account for the full projected mass density distribution from the source redshift to the observer located at $z=0$, including the effects of LSS and correlated matter around the main halo.

The large DEMNUni-Cov sample allows us to measure WL mass bias with systematics similar to those we expect from \Euclid data in the low-mass range. We also look at the precision of the lensing signal in different redshift bins as DEMNUni-Cov haloes are distributed up to redshift $z=1$. It is worth highlighting that hydrodynamical effects primarily impact the lensing signal in the cluster core, $R \lesssim 100 \, \rm kpc$ \citep[e.g.,][]{2008MNRAS.391.1685S}. In order to avoid the impact of these effects, we only model the WL mass outside of this region.



\section{Lensing measurements} \label{sec: Lensing Measurements}

In this section, we detail the methodology we use to estimate WL cluster masses from the data presented in Sect.~\ref{sec: Simulated data}. In particular, we describe how we measure the WL signal in radial bins and fit the data to a fiducial model.

\subsection{Shear profiles} \label{sec: Shear profiles}

\subsubsection{Lensing properties} \label{sec: Lensing properties}

The intrinsic shapes of galaxies are distorted by the matter inhomogeneities along the line of sight, which include galaxy clusters. This yields isotropic and anisotropic deformations of the intrinsic ellipticity of the galaxies: the convergence $\kappa$ and the shear $\gamma$, respectively.
The lensing information of an intervening axially symmetric lens on a single source plane is encoded by the convergence \citep{2001PhR...340..291B, 2006glsw.conf..269S, 2015RPPh...78h6901K}
\begin{equation}
\Sigma = \kappa \Sigma_{\rm cr} \, ,
\label{eq: Sigma}
\end{equation}
and by excess surface density, which can be expressed in terms of the tangential component of the shear,
\begin{equation}
\Delta \Sigma = \gamma_{\rm t} \Sigma_{\rm cr} \, .
\label{eq: Delta_Sigma}
\end{equation}
Here, the surface mass density $\Sigma$ is the projected matter density, and the excess surface mass density $\Delta \Sigma$ is the difference between the mean value of the surface mass density calculated over a disc with a radius that comprises the projected distance between the lens centre and the source, and its local value. The critical surface mass density $\Sigma_{\rm cr}$ is defined as
\begin{equation}
\Sigma_{\rm cr} \equiv \frac{\rm c^2}{\rm 4 \pi G} \frac{D_{\rm s}}{D_{\rm d} D_{\rm ds}} \, ,
\label{eq: Sigma_cr}
\end{equation}
where $c$ is the speed of light in vacuum, $G$ is the gravitational constant, and $D_{\rm s}$, $D_{\rm d}$ and $D_{\rm ds}$ are the angular diameter distances from the observer to the source, from the observer to the lens, and from the lens to the source, respectively.

We also introduce the observed quantity $\Delta\Sigma_{g_{\rm t}}$ analogous to Eq.~\eqref{eq: Delta_Sigma},
\begin{equation}
\Delta \Sigma_{g_{\rm t}} = g_{\rm t} \Sigma_{\rm cr} \, .
\label{eq: Delta_Sigma_g}
\end{equation} 
In the following, we will use the excess surface mass density as in Eq.~\eqref{eq: Delta_Sigma_g}.

\subsubsection{Profile measurement} \label{sec: Profile measurement}

Because the intrinsic shape noise dominates the lensing signal of individual sources, we derive density profiles by measuring the mean source shear in fixed radial bins. The averaged lensing observable is calculated over the $j$-th radial annulus as
\begin{equation}
\langle \Delta \Sigma_{g_{\rm t}} \rangle_j = \frac{\sum_{i \in j} w_{\Delta\Sigma, i} g_{{\rm t}, i} \Sigma_{{\rm cr}, i}}{\sum_{i \in j} w_{\Delta\Sigma, i}} \, , 
\label{eq: Delta_Sigma_av}
\end{equation}
where the weight of the $i$-th source is $w_{\Delta\Sigma, i} = w_{\rm s} \Sigma_{{\rm cr}, i}^{-2}$. We consider a uniform lensing weight related to shape estimate uncertainties, $w_{\rm s} = 1 / \sigma^2_{\epsilon}$. In this study, the statistical uncertainty of the lensing estimate accounts for the shape noise of the background sources, and it is computed as the uncertainty of the weighted mean,
\begin{equation}
\delta_{\Delta\Sigma_{g_{\rm t}}, j} = \biggl( \sum_{i \in j} w_{\Delta\Sigma, i} \biggr) ^{-1/2} \, .
\label{eq: Delta_Sigma_err}
\end{equation}
Here, we do not account for bias in ellipticity measurements and shear calibration in the lensing signal, unlike methods applied to real survey data \citep[e.g.,][]{2019MNRAS.482.1352M}.

We select background sources with the following criterion for the source photo-$z$
\begin{equation}
z_{\rm s} > z_{\rm d} + \Delta z \, ,
\label{eq: basic selection}
\end{equation}
where $\Delta z$ is a secure interval that minimises the contamination. We set $\Delta z = 0.1$ to be twice the threshold value defined in \citet{2018PASJ...70...30M}.

Shears are measured at the shear-weighted radial position \citep{2017MNRAS.472.1946S}
\begin{equation}
\langle R \rangle_j = \left( \frac{\sum_{i \in j} w_{\Delta\Sigma, i} R_i^{-\alpha} }{\sum_{i \in j} w_{\Delta\Sigma, i} } \right)^{-1 / \alpha} \, .
\label{eq: R_av}
\end{equation}
The effective radius is computed with $\alpha=1$. 

To calculate $\Sigma_{\rm cr}$ in a radial bin with $N$ background sources, we derive the effective surface critical density as \citep{2017MNRAS.472.1946S}
\begin{equation}
\langle \Sigma_{\rm cr}^{-1} \rangle_j = \frac{\sum_{i \in j} \Sigma_{{\rm cr}, i}^{-1}}{N} \, .
\label{eq: Sigma_cr_back}
\end{equation}
In addition, we define the WL signal-to-noise ratio per halo as \citep{2017MNRAS.472.1946S}
\begin{equation}
{\rm (S/N)_\textsc{wl}} = \frac{\langle \Delta\Sigma_{g_{\rm t}} \rangle}{\delta_t} \, .
\label{eq: SNR}
\end{equation}
Here, $\langle \Delta\Sigma_{g_{\rm t}} \rangle$ is the average lensing observable measured in the full radial range of the lens, while the error budget $\delta_t$ accounts for the contribution of statistical uncertainties and cosmic noise, see Sect.~\ref{sec: Fitting configuration}. 

In the following, we consider shears averaged in 8 logarithmically equispaced radial bins covering the range $[0.4, 4.0] \, \rm Mpc$ from the cluster centre, similar to the binning scheme set in \citet{PaperIII}.

\subsection{Mass inference} \label{sec: Mass inference}

\subsubsection{Density models} \label{sec: Density models}

We derive WL masses by constraining a fiducial model of the halo density profile with the measured shear density profiles. Modelling the mass density distribution of the haloes is challenging as it results from various physical effects, e.g., miscentering \citep{2006MNRAS.373.1159Y, 2007ApJ...656...27J} or the contribution of correlated matter \citep{2014ApJ...784L..25C, 2022MNRAS.511.1484I}. In this study, as a reference model, we assume the simple but effective Navarro--Frenk--White (NFW) density profile \citep{1996ApJ...462..563N, 1997ApJ...490..493N}
\begin{equation}
\rho_{\NFW} (r) = \frac{\rho_{\rm s}}{(r/r_{\rm s})(1+r/r_{\rm s})^2} \, ,
\label{eq: NFW}
\end{equation}
where $\rho_{\rm s}$ is the scale density, and $r_{\rm s}$ the scale radius. 

We also consider a truncated version of the NFW profile \citep{{2009JCAP...01..015B}}, known as the Baltz--Marshall--Oguri (BMO) profile,
\begin{equation}
\rho_{\BMO} (r) = \rho_{\NFW} (r) \left( \frac{r_{\rm t}^2}{r^2 + r_{\rm t}^2} \right)^2 \, ,
\label{eq: BMO}
\end{equation}
where $r_{\rm t}$ is the truncation radius set to $r_{\rm t} = 3 r_{\rm 200c}$ \citep{2011MNRAS.414.1851O, 2019MNRAS.484.1598B}.

The mass density and the excess surface mass density can be expressed as a function of mass, $M_{\rm 200c}$, and concentration $c_{\rm 200c}$.
For the fitting parameters, we consider the logarithm (base 10) of mass and concentration. 

\subsubsection{Fitting procedure} \label{sec: Fitting configuration}

Following a Bayesian approach, we derive the posterior probability density function $p$ of parameters $\vec{\rm p} = [\log M_{\rm 200c}, \log c_{\rm 200c}]$ given the likelihood function $\mathcal{L}$ and the prior $p_{\rm prior}$ as
\begin{equation}
p (\vec{\rm p} \vert \langle \Delta\Sigma_{g_{\rm t}} \rangle) \propto \mathcal{L} ( \langle \Delta\Sigma_{g_{\rm t}} \rangle \vert \vec{\rm p}) p_{\rm prior} (\vec{\rm p}) \, ,
\label{eq: posterior}
\end{equation}
where $\langle \Delta\Sigma_{g_{\rm t}} \rangle$ are the data, $\mathcal{L} \propto \exp ( - \chi^2 / 2 )$, and 
\begin{equation}
\resizebox{\hsize}{!}{%
$\chi^2 = \sum\limits_{i, j} \left[ \langle \Delta\Sigma_{g_{\rm t}} \rangle_i - \Delta\Sigma_{g_{\rm t}} (\langle R \rangle_i, \vec{\rm p}) \right]^{-1} \tens{C}_{ij}^{-1} \left[ \langle \Delta\Sigma_{g_{\rm t}} \rangle_j - \Delta\Sigma_{g_{\rm t}} (\langle R \rangle_j, \vec{\rm p}) \right] \, ,$ %
}
\label{eq: chi_square}
\end{equation}
where the sum runs over the radial bins $i, j$. We measure the reduced $\chi^2$ as 
\begin{equation}
\chi^2_{\rm red} \equiv \frac{\chi^2}{N_{\rm dof}},
\label{eq: reduced chi_square}
\end{equation}
where $N_{\rm dof}$ is the number of degrees of freedom.

In the present analysis, the covariance matrix $\tens{C}$ accounts for shape noise, $\delta \Delta\Sigma_{g_{\rm t}}$, and cosmic noise, $\Delta\Sigma_{\rm LSS}$, as \citep{2015MNRAS.449.4264G,2018NatAs...2..744S}
\begin{equation}
\tens{C}= \tens{C}^{\rm stat} + \tens{C}^{\rm LSS} \, .
\label{eq: covariance}
\end{equation}
In the above equation, $\tens{C}^{\rm stat}$ is a diagonal matrix with terms $C^{\rm stat}_{ii} = (\delta \Delta\Sigma_{{g_{\rm t}}, i})^2$, and $\tens{C}^{\rm LSS}_{ij} \equiv \langle \Delta\Sigma_{\rm LSS}(\Delta \theta_i) \Delta \Sigma_{\rm LSS}(\Delta \theta_j) \rangle$ characterises the effects of uncorrelated LSS in each pair of annular bins $(\Delta \theta_i, \Delta \theta_j)$ \citep{1998MNRAS.296..873S, 2003MNRAS.339.1155H}
\begin{equation} 
\resizebox{\hsize}{!}{%
$\langle \Delta \Sigma_{\rm LSS}(\Delta \theta_i) \Delta \Sigma_{\rm LSS}(\Delta \theta_j) \rangle = 2 \pi \Sigma_{\rm cr}^2 \int_0^{\infty} P_k(\ell)g(\ell, \Delta \theta_i) g(\ell, \Delta \theta_j) \ \ell \ \diff \ell \, ,$
}
\label{eq: C_LSS}
\end{equation}
where $P_k(l)$ is the effective projected lensing power spectrum and the function $g$ is the filter.
 
We adopt a uniform prior with ranges $\log M_{\rm 200c} \in [13, 16]$ and $\log c_{\rm 200c} \in [0, 1]$. In Sect.~\ref{sec: Gaussian mass prior}, we show the impact of using a Gaussian prior on the mass inference. 

We sample the posterior distribution using an affine invariant Markov Chain Monte Carlo (MCMC) approach \citep{2013PASP..125..306F}. Each Markov chain runs for 3200 steps starting from initial values randomly taken from a bivariate normal distribution of mean mass $\log M_{\rm 200c} = 14$ and mean concentration $\log c_{\rm 200c} = 0.6$. The posterior is estimated after removing a burn-in phase, which is assumed as four times the autocorrelation time of the chain.

\subsubsection{Mass point estimators} \label{sec: Statistical estimators}

The MCMC chains sample the posterior distribution of the model parameters, and can be summarised with a point estimate of the WL mass. Cluster masses measured using individual shear density profiles may vary depending on the statistical estimator employed. In the following sections, we compare WL masses recovered using several different point estimators. We look at the mean, and the related standard deviation, or the median of the marginalised posterior mass distribution, for which the associated uncertainties are the standard deviation, and the $16^{\rm th}$ and $84^{\rm th}$ percentiles of the sample distribution. We also consider the biweight location and scale, hereafter referred to as CBI and SBI, respectively, which are robust statistics for summarising a distribution \citep{1990AJ....100...32B}. Finally, we look at the maximum likelihood ${\rm max}(\mathcal{L})$, hereafter referred to as ML.

\subsection{Cluster ensemble average} \label{sec: Clusters average}

By averaging the shear signal across an ensemble of clusters, the precision of lensing profile measurements improves in comparison to those obtained from a single lens. Therefore, the uncertainties on the lensing mass measurements are also reduced. This provides a complementary method to single lens mass bias analyses for quantifying the impact of systematic effects.

We measure the surface mass density in the $j$-th radial annulus similarly to Eq.~\eqref{eq: lensing_average} with the average quantity
\begin{equation}
\langle \Delta\Sigma_{g_{\rm t}} \rangle_{W, j} = \frac{\sum_{n \in N} W_{\Delta\Sigma, j, n} \langle \Delta\Sigma_{g_{\rm t}} \rangle_{j, n}}{\sum_{n \in N} W_{\Delta\Sigma, j, n}} \, , 
\label{eq: Delta_Sigma_av_stack}
\end{equation}
where $\langle \Delta\Sigma_{g_{\rm t}} \rangle_{j, n}$ is the surface mass density in the $j$-th radial bin, see Eq.~(\ref{eq: Delta_Sigma_av}), of the $n$-th cluster, and the weight can be expressed in terms of the uncertainty on the weighted mean, see Eq.~(\ref{eq: Delta_Sigma_err}), as $W_{\Delta\Sigma, j, n} = \delta^{-2}_{\Delta\Sigma_{g_{\rm t}}, j, n}$. 

The associated uncertainty on the averaged cluster lensing profile accounts for the total shape noise of the sources. For this analysis, contributions from correlated or uncorrelated large-scale structure are not considered in the covariance of the averaged lensing profiles.

In the following, any observable $O$ measured by fitting the averaged lensing profile, e.g., mass or concentration, is quoted as $O_{\langle \Delta\Sigma \rangle}$.

\begin{figure}
	\includegraphics[width=\columnwidth]{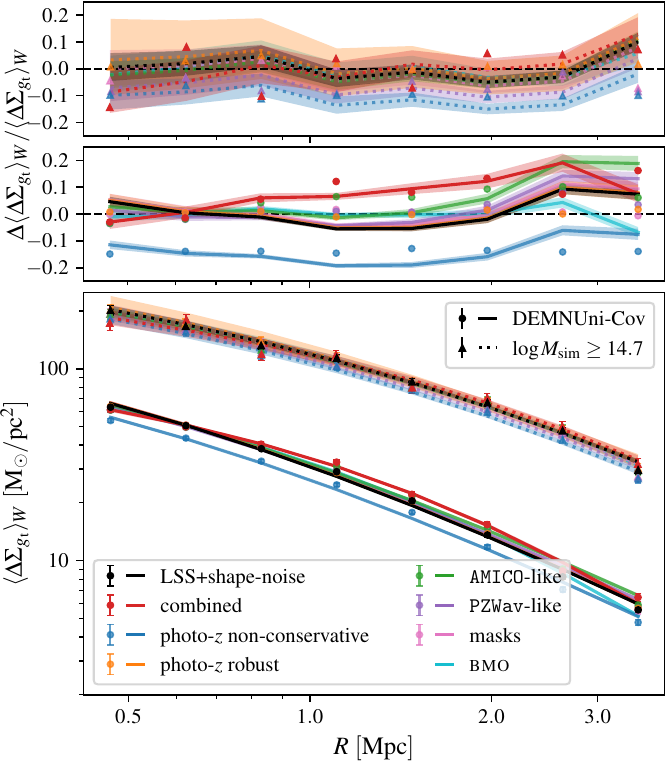}
    \caption{Average excess surface density and maximum likelihood NFW / BMO density fitted models. The figure shows the density profiles for the LSS plus shape noise (Sect.~\ref{sec: Accuracy and Precision}), non-conservative photo-$z$ selection (Sect.~\ref{sec: Basic selection}), robust photo-$z$ selection (Sect.~\ref{sec: Robust selection}), \AMICO-like sample (Sect.~\ref{sec: Euclid-like clusters}), \PZWav-like sample (Sect.~\ref{sec: Euclid-like clusters}), masks (Sect.\ref{sec: Masks effects}), BMO model (Sect.\ref{sec: BMO profile}), and combined effects (Sect.\ref{sec: Combined systematics}). Signals are shown either for the full sample, or for the massive subsample at $\log M_{\rm sim} \geq 14.7$. The top panels show the relative change with respect to the fiducial shape and LSS noise shear data (points), and with respect to their models (lines).}
    \label{fig: stacked_shears}
\end{figure}

Figure~\ref{fig: stacked_shears} displays the average shear profiles for the samples and cases discussed in Sects.~\ref{sec: Accuracy and Precision}, and \ref{sec: Systematics Assessment}, and the density model for the ML point estimate. Mass and concentration are estimated at the mean lens redshift of the cluster sample considered, i.e. as measured in Eq.~\eqref{eq: lensing_average}, and given in Table~\ref{tab: stacked_mass} along with the corresponding reduced $\chi^2$.

For a given population of $N$ galaxy clusters, we compute the ensemble average of the observable $O$, e.g., the mass bias or the mass change, see Eqs.~(\ref{eq: mass_bias}, \ref{eq: relative_change}), as a lensing-weighted mean \citep[e.g.,][]{2014ApJ...795..163U}
\begin{equation}
\langle O \rangle_W = \frac{\sum_{n \in N} W_{\Delta\Sigma, n} O_n }{\sum_{n \in N} W_{\Delta\Sigma, n}} \, ,
\label{eq: lensing_average}
\end{equation}
where $O_n$ is the observable measured for the $n$-th cluster, and $W_{\Delta\Sigma, n} = \sum_j W_{\Delta\Sigma, j, n}$ is the total weight of the $n$-th cluster, see Eq.~(\ref{eq: Delta_Sigma_err}). The associated scale of the weighted mean in Eq.~(\ref{eq: lensing_average}) is calculated as
\begin{equation}
\sigma_O = \biggl( \frac{\sum_{n \in N} W_{\Delta\Sigma, n} (\langle O \rangle_W - O_n)^2}{\sum_{n \in N} W_{\Delta\Sigma, n}} \biggr) ^{1/2} \, .
\label{eq: lensing_scatter}
\end{equation}

\begin{table}
    \centering
    \caption{ML logarithmic mass and concentration of the NFW fit to average cluster lensing profiles, and associated reduced $\chi^2$.}
    \resizebox{\hsize}{!}{
    \begin{tabular}{lccc}
        \toprule
        Data Set & $\log M_{\langle \Delta\Sigma \rangle}$ & $\log c_{\langle \Delta\Sigma \rangle}$ & $\chi^2_{\rm red}$ \\
        \midrule
        LSS + shape noise (\ref{sec: Accuracy and Precision}) & \splitcell{$13.992 \pm 0.004$}{$14.927 \pm 0.012$} & \splitcell{$0.509 \pm 0.016$}{$0.437 \pm 0.025$} & \splitcell{6.51}{1.36} \\
        \addlinespace
        Three Hundred \citepalias{giocoli24} & $14.914 \pm 0.026$ & $0.671 \pm 0.033$ & 0.43 \\
        \addlinespace
        photo-$z$ non-conservative (\ref{sec: Basic selection}) & \splitcell{$13.895 \pm 0.005$}{$14.858 \pm 0.012$} & \splitcell{$0.467 \pm 0.017$}{$0.423 \pm 0.027$} & \splitcell{6.30}{1.24} \\
        \addlinespace
        photo-$z$ robust (\ref{sec: Robust selection}) & \splitcell{$13.994 \pm 0.005$}{$14.935 \pm 0.014$} & \splitcell{$0.509 \pm 0.020$}{$0.445 \pm 0.032$} & \splitcell{3.82}{0.80} \\
        \addlinespace
        \AMICO-like (\ref{sec: Euclid-like clusters}) & \splitcell{$14.015 \pm 0.005$}{$14.927 \pm 0.014$} & \splitcell{$0.436 \pm 0.016$}{$0.420 \pm 0.027$} & \splitcell{8.49}{2.74} \\
        \addlinespace
        \PZWav-like (\ref{sec: Euclid-like clusters}) & \splitcell{$13.996 \pm 0.005$}{$14.892 \pm 0.013$} & \splitcell{$0.454 \pm 0.016$}{$0.425 \pm 0.027$} & \splitcell{4.49}{2.05} \\
        \addlinespace
        masks (\ref{sec: Masks effects}) & \splitcell{$13.997 \pm 0.005$}{$14.919 \pm 0.013$} & \splitcell{$0.502 \pm 0.018$}{$0.445 \pm 0.029$} & \splitcell{5.64}{1.80} \\
        \addlinespace
        BMO (\ref{sec: BMO profile}) & \splitcell{$14.013 \pm 0.004$}{$14.927 \pm 0.012$} & \splitcell{$0.309 \pm 0.018$}{$0.369 \pm 0.031$} & \splitcell{1.43}{0.68} \\
        \addlinespace
        \splitcellbis{photo-$z$ robust +}{\AMICO-like +}{masks +}{BMO + prior} (\ref{sec: Combined systematics}) & \splitcell{$14.048 \pm 0.007$}{$14.950 \pm 0.019$} & \splitcell{$0.210 \pm 0.024$}{$0.303 \pm 0.042$} & \splitcell{1.94}{2.11} \\
        \bottomrule
    \end{tabular}
    }
    \tablefoot{Reported uncertainties are the standard deviation of the marginalised distribution. For each data set, lensing average quantities are shown for the full mass range (top row), or $\log M_{\rm sim} \geq 14.7$ (bottom row).}
    \label{tab: stacked_mass}
\end{table}



\section{Accuracy and precision for WL masses for intrinsic scatter and noise} \label{sec: Accuracy and Precision}

Simulated data can be used to probe the accuracy and precision of WL cluster mass measurements, as the true halo masses are known a priori. In this section, we assess the WL mass estimates we obtained for DEMNUni-Cov clusters measured from unbiased catalogues.
By unbiased catalogues, we mean that the catalogues are not affected by measurement errors, and the WL mass differs from the true mass only for intrinsic noise and bias. This can be due to a number of effects. For example, due to intrinsic ellipticity, the measured shape of a source may differ from the underlying reduced shear. LSS distorts the source shape. Mass derived assuming the spherical approximation may be scattered due to traxiality and cluster orientation, or irregular morphology \citep{2010A&A...514A..93M, 2011ApJ...740...25B, 2014MNRAS.440.1899G, 2014ApJ...797...34M, 2022MNRAS.513.2178H, giocoli24}. These effects are all accounted for in simulated $N$-body samples. We refer to this case as `LSS + shape noise'.

\subsection{Precision of the lensing signal} \label{sec: Quality assessment}

\begin{figure}
	\includegraphics[width=\columnwidth]{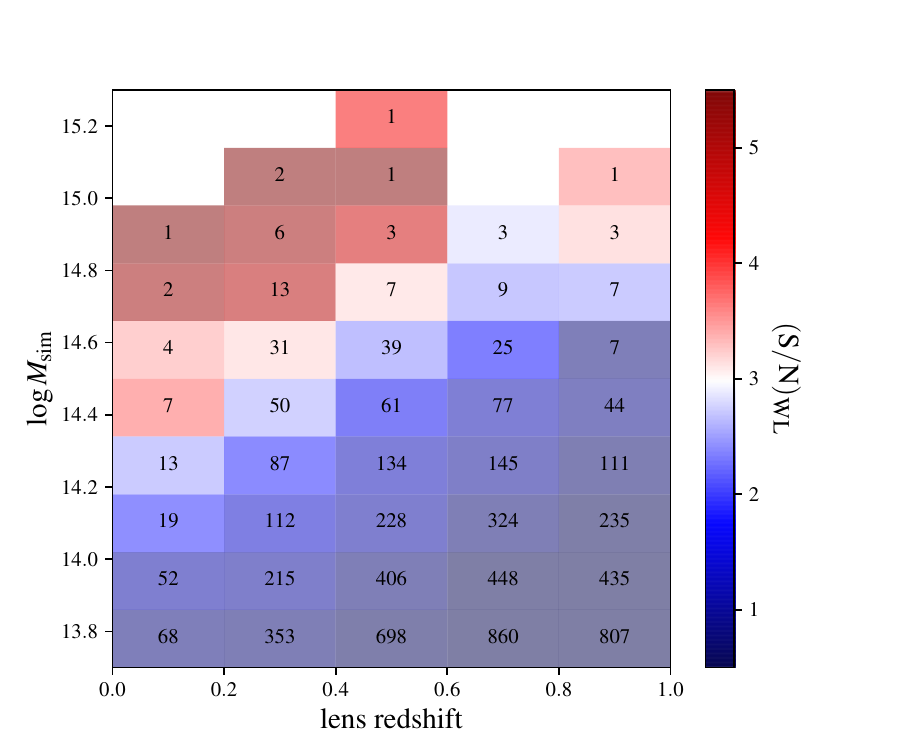}
    \caption{$\rm (S/N)_\textsc{wl}$ of DEMNUni-Cov clusters in bins of mass and redshift, computed as the CBI of the values in the bin, for the `LSS + shape noise' case. The number of clusters is displayed in each bin.}
    \label{fig: SNR_distribution}
\end{figure}

To assess the precision to which the WL signal can be measured, we first calculate $\rm (S/N)_\textsc{wl}$ for each lens as in Eq.~\eqref{eq: SNR}. In Fig.~\ref{fig: SNR_distribution}, we plot the distribution of $\rm (S/N)_\textsc{wl}$ for clusters binned in mass and redshift. For each bin, we show the CBI value. We find the highest $\rm (S/N)_\textsc{wl}$ values for the few high-mass and low-redshift haloes, while the more numerous distant or low-mass objects have lensing signals largely dominated by noise. 

Table~\ref{tab: SNR threshold} gives the number of clusters with $\rm (S/N)_\textsc{wl}$ larger than a certain threshold. Only 226 ($3.7 \, \%$) clusters have lensing signals with $\rm (S/N)_\textsc{wl} > 3$. This number depends on the radial aperture of the lensing profile, the number density of the background source population, and the cosmological framework of the analysis.

Our results are consistent with the semi-analytical forecasting presented in \citet{PaperIII}. Applying the methodology presented in \citet{PaperIII} to haloes with the same mass and redshift distribution as the DEMNUni-Cov simulations, and assuming ideal observational conditions with only shape and LSS noise, the expected number of clusters with $\rm (S/N)_\textsc{wl} > 3$ is 66. This estimate can be interpreted as the number of massive clusters that individually have a probability in excess of $50 \, \%$ of having $\rm (S/N)_\textsc{wl} > 3$ for different noise realisations. 

In the presence of noise, the signal is scattered. Due to the steepness of the halo mass function, a large number of low-mass clusters are expected to have their signal boosted to $\rm (S/N)_\textsc{wl} > 3$. The semi-analytical modelling of \citet{PaperIII} predicts $206 \pm 12$ haloes with $\rm (S/N)_\textsc{wl} > 3$, in good agreement with the result from the DEMNUni-Cov $N$-body simulations. However, the semi-analytical prediction could be slightly biased low due to effects not taken into account in the modelling, such as projection effects or triaxiality.

\begin{table}
	\centering
	\caption{Number (percentage) of clusters with $\rm (S/N)_\textsc{wl}$ larger than a given threshold.}
    \resizebox{\hsize}{!}{
	\begin{tabular}{lccc}
		\toprule
		data set & $\rm (S/N)_\textsc{wl} \geq 3$ & $\rm (S/N)_\textsc{wl} \geq 5$ & $\rm (S/N)_\textsc{wl} \geq 10$ \\
		\midrule
        LSS + shape noise (\ref{sec: Accuracy and Precision}) & $226 \, (3.67)$ & $28 \, (0.45)$ & $1 \, (0.02)$ \\
        \addlinespace
        \addlinespace
		photo-$z$ non-conservative (\ref{sec: Basic selection}) & $167 \, (2.71)$ & $22 \, (0.36)$ & $1 \, (0.02)$ \\
        \addlinespace
		photo-$z$ robust (\ref{sec: Robust selection}) & $221 \, (3.59)$ & $23 \, (0.37)$ & $1 \, (0.02)$ \\
        \addlinespace
		\AMICO-like (\ref{sec: Euclid-like clusters}) & $184 \, (4.04)$ & $26 \, (0.57)$ & $1 \, (0.02)$ \\
        \addlinespace
		\PZWav-like (\ref{sec: Euclid-like clusters}) & $201 \, (4.06)$ & $22 \, (0.44)$ & $2 \, (0.04)$ \\
        \addlinespace
		masks (\ref{sec: Masks effects}) & $183 \, (3.85)$ & $26 \, (0.55)$ & $1 \, (0.02)$ \\
        \addlinespace
		\splitcellbis{photo-$z$ robust + }{\AMICO-like + }{masks + }{BMO + prior} (\ref{sec: Combined systematics}) & $140 \, (3.94)$ & $19 \, (0.54)$ & $1 \, (0.03)$ \\
        \bottomrule
	\end{tabular}
    }
	\label{tab: SNR threshold}
\end{table}

\subsection{Accuracy and precision of the lensing mass}

We measure WL masses by fitting an NFW model to the shear profile for each lens in the DEMNUni-Cov simulations, in the `LSS + shape noise' case. Other observational uncertainties are discussed in Sect.~\ref{sec: Systematics Assessment}. Therefore, we compare the true mass to the WL mass determined when only intrinsic noise and bias affect the estimate. 

The WL mass accuracy and precision is assessed using two different approaches. In Sect.~\ref{sec: Linear regression and scatter}, we compare WL and true mass with a linear regression, while in Sect.~\ref{sec: Mass bias} we analyse the weighted mass bias, i.e., the ratio between the WL mass and the true mass. The two analyses are complementary. On the one hand, we consider the linear relation between the logarithm of masses, whose uncertainty is proportional to the relative uncertainty on the mass and close to the $\rm (S/N)_\textsc{wl}$. High-signal, more massive clusters weigh more since the noise is (nearly) uniform at a given lens redshit. On the other hand, the mass bias in Sect.~\ref{sec: Mass bias} is weighted by the (inverse of the squared) noise. This weight is (nearly) uniform for clusters of different masses. 

\subsubsection{Linear regression} \label{sec: Linear regression and scatter}

\begin{figure*}
    \centering
	\includegraphics[width=\textwidth]{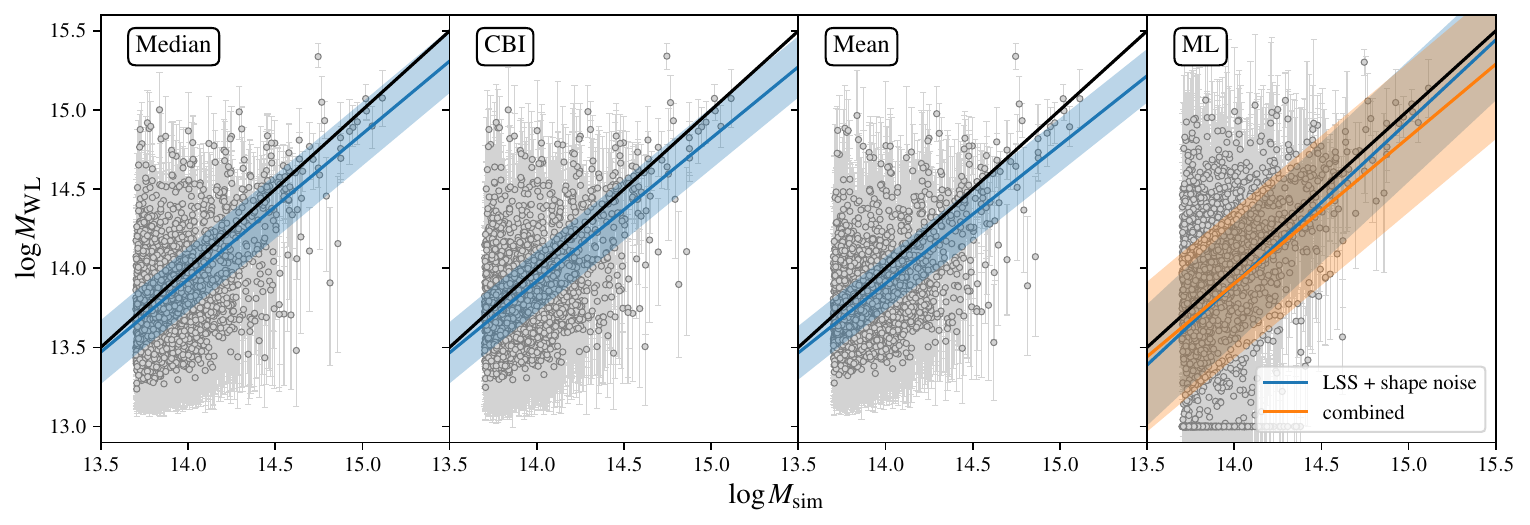}
    \caption{Scatter plot for WL mass point estimates of DEMNUni-Cov clusters. Error bars correspond to the standard deviation of the mean (in the mean/ML panels), $16^{\rm th}$ and $84^{\rm th}$ percentiles (in the median panel), or SBI (in the CBI pannel). The one-to-one line is shown in black, while blue lines give results for the linear regression over the data points for the `LSS + shape noise' case. The orange line shows the linear regression for ML WL mass measured on DEMNUni-Cov data with the combination of systematic effects discussed in Sect.~\ref{sec: Combined systematics}. The shaded region corresponds to the scatter.}
    \label{fig: scatter_point_estimate}
\end{figure*}

\begin{table*}[!ht]
    \centering
    \caption{Fitted parameters of the $\log M_\textsc{wl}$--$\log M_{\rm sim}$ relation.}
    \begin{tabular}{lrcccc}
        \toprule
        Data Set & & Median & CBI & Mean & ML \\
        \midrule
        LSS + shape noise (\ref{sec: Accuracy and Precision}) & $\left\{ \splitcellthree{$\alpha = $}{$\beta = $}{$\sigma = $} \right.$ & \splitcellthree{$-0.070 \pm 0.006$}{$0.920 \pm 0.022$}{$0.202 \pm 0.005$} & \splitcellthree{$-0.082 \pm 0.006$}{$0.904 \pm 0.021$}{$0.194 \pm 0.005$} & \splitcellthree{$-0.097 \pm 0.005$}{$0.875 \pm 0.021$}{$0.169 \pm 0.005$} & \splitcellthree{$-0.098 \pm 0.007$}{$1.027 \pm 0.030$}{$0.382 \pm 0.006$} \\
        \addlinespace
        Three Hundred \citepalias{giocoli24} & $\left\{ \splitcellthree{$\alpha = $}{$\beta = $}{$\sigma = $} \right.$ & \splitcellthree{$-0.114 \pm 0.046$}{$1.081 \pm 0.044$}{$0.064 \pm 0.003$} & - & - & - \\
        \addlinespace
        photo-$z$ non-conservative (\ref{sec: Basic selection}) & $\left\{ \splitcellthree{$\alpha = $}{$\beta = $}{$\sigma = $} \right.$ & \splitcellthree{$-0.150 \pm 0.005$}{$0.855 \pm 0.020$}{$0.170 \pm 0.005$} & \splitcellthree{$-0.158 \pm 0.005$}{$0.835 \pm 0.021$}{$0.160 \pm 0.005$} & \splitcellthree{$-0.167 \pm 0.005$}{$0.807 \pm 0.020$}{$0.135 \pm 0.005$} & \splitcellthree{$-0.216 \pm 0.007$}{$0.977 \pm 0.029$}{$0.365 \pm 0.006$} \\
        \addlinespace
        photo-$z$ robust (\ref{sec: Robust selection}) & $\left\{ \splitcellthree{$\alpha = $}{$\beta = $}{$\sigma = $} \right.$ & \splitcellthree{$-0.066 \pm 0.006$}{$0.881 \pm 0.023$}{$0.220 \pm 0.005$} & \splitcellthree{$-0.094 \pm 0.006$}{$0.843 \pm 0.021$}{$0.178 \pm 0.005$} & \splitcellthree{$-0.094 \pm 0.006$}{$0.843 \pm 0.021$}{$0.178 \pm 0.005$} & \splitcellthree{$-0.099 \pm 0.008$}{$1.008 \pm 0.031$}{$0.418 \pm 0.007$} \\
        \addlinespace
        \AMICO-like (\ref{sec: Euclid-like clusters}) & $\left\{ \splitcellthree{$\alpha = $}{$\beta = $}{$\sigma = $} \right.$ & \splitcellthree{$-0.086 \pm 0.007$}{$0.897 \pm 0.024$}{$0.206 \pm 0.006$} & \splitcellthree{$-0.098 \pm 0.006$}{$0.883 \pm 0.024$}{$0.199 \pm 0.006$} & \splitcellthree{$-0.111 \pm 0.006$}{$0.864 \pm 0.023$}{$0.172 \pm 0.006$} & \splitcellthree{$-0.118 \pm 0.008$}{$0.988 \pm 0.034$}{$0.394 \pm 0.007$} \\
        \addlinespace
        \PZWav-like (\ref{sec: Euclid-like clusters}) & $\left\{ \splitcellthree{$\alpha = $}{$\beta = $}{$\sigma = $} \right.$ & \splitcellthree{$-0.091 \pm 0.006$}{$0.877 \pm 0.025$}{$0.213 \pm 0.006$} & \splitcellthree{$-0.102 \pm 0.006$}{$0.859 \pm 0.024$}{$0.205 \pm 0.005$} & \splitcellthree{$-0.102 \pm 0.006$}{$0.861 \pm 0.024$}{$0.205 \pm 0.006$} & \splitcellthree{$-0.127 \pm 0.008$}{$0.983 \pm 0.033$}{$0.397 \pm 0.007$} \\
        \addlinespace
        masks (\ref{sec: Masks effects}) & $\left\{ \splitcellthree{$\alpha = $}{$\beta = $}{$\sigma = $} \right.$ & \splitcellthree{$-0.070 \pm 0.007$}{$0.906 \pm 0.025$}{$0.212 \pm 0.006$} & \splitcellthree{$-0.081 \pm 0.006$}{$0.887 \pm 0.026$}{$0.203 \pm 0.006$} & \splitcellthree{$-0.094 \pm 0.006$}{$0.860 \pm 0.024$}{$0.177 \pm 0.005$} & \splitcellthree{$-0.095 \pm 0.008$}{$1.020 \pm 0.033$}{$0.393 \pm 0.008$} \\
        \addlinespace
        BMO + prior (\ref{sec: Model and prior}) & $\left\{ \splitcellthree{$\alpha = $}{$\beta = $}{$\sigma = $} \right.$ & \splitcellthree{$-0.116 \pm 0.005$}{$0.732 \pm 0.017$}{$0.094 \pm 0.005$} & \splitcellthree{$-0.124 \pm 0.004$}{$0.727 \pm 0.017$}{$0.092 \pm 0.005$} & \splitcellthree{$-0.135 \pm 0.004$}{$0.714 \pm 0.016$}{$0.082 \pm 0.006$} & \splitcellthree{$-0.091 \pm 0.007$}{$0.953 \pm 0.030$}{$0.414 \pm 0.006$} \\
        
        \splitcellbis{photo-$z$ robust +}{\AMICO-like +}{masks +}{BMO + prior} (\ref{sec: Combined systematics}) & $\left\{ \splitcellthree{$\alpha = $}{$\beta = $}{$\sigma = $} \right.$ & \splitcellthree{$-0.135 \pm 0.006$}{$0.696 \pm 0.022$}{$0.108 \pm 0.006$} & \splitcellthree{$-0.154 \pm 0.005$}{$0.673 \pm 0.023$}{$0.099 \pm 0.006$} & \splitcellthree{$-0.154 \pm 0.006$}{$0.672 \pm 0.023$}{$0.099 \pm 0.006$} & \splitcellthree{$-0.097 \pm 0.010$}{$0.924 \pm 0.043$}{$0.473 \pm 0.008$} \\
        \bottomrule
    \end{tabular}
    \tablefoot{The intercept $\alpha$, slope $\beta$, and intrinsic scatter $\sigma$ of the regression line (see Eq.~\eqref{eq: linear_regression}) are reported for the median, CBI, mean, or ML estimators. Uncertainties are reported as the standard deviation of the marginalised parameter distribution.}
    \label{tab: linear_regression}
\end{table*}

In Fig.~\ref{fig: scatter_point_estimate}, we show the scatter between the WL mass and the true mass of DEMNUni-Cov lenses. We display masses as median, CBI, mean, or ML point estimates obtained with the \COMBCL pipeline. As a first attempt to assess the accuracy of the measurements, we look at the fraction of clusters with WL mass lower than the true halo mass. The median, CBI, and mean point estimates show a similar fraction of underestimated WL masses ($65.5 \, \%$, $66.9 \, \%$, and $68.5 \, \%$), while almost half of the clusters have a ML WL mass lower than the true mass ($49.2 \, \%$).

To better evaluate the level of WL mass accuracy and precision, we perform a linear regression of the true mass versus the WL mass point estimate using the \texttt{LIRA} \citep[LInear Regression in Astronomy,][]{2016MNRAS.455.2149S} package. Intercept $\alpha$, slope $\beta$, and intrinsic scatter $\sigma$ are calculated from the linear regression of the logarithmic masses as
\begin{equation}
\log (M_\textsc{wl} / M_{\rm piv}) = \alpha + \beta \log (M_{\rm sim} / M_{\rm piv}) \pm \sigma \, ,
\label{eq: linear_regression}
\end{equation}
where the pivot mass is $M_{\rm piv} = 10^{14} M_\odot$. Results are presented in Table~\ref{tab: linear_regression}. The slope and intercept values indicate how close the regression line is to one-to-one relationship, and thus account for the accuracy. The scatter quantifies how close the mass dependence is to the linear relation in Eq.~\eqref{eq: linear_regression}, and thus accounts for the precision. 

The ML is the most accurate point estimate for the overall mass range ($\beta \sim 1$). Given the adopted priors, the bias is mass dependent for the mean, median, or CBI estimators, ($\beta \sim 0.9$). In the low-mass regime, the mean, median, or CBI estimators are accurate, but significantly deviate from the one-to-one relation at higher cluster masses. The scatter of the ML masses is about twice the size of that from the other statistical estimators, which makes the ML masses less precise. 

The level of precision of the different estimators can be strongly impacted by the mass prior. The ML estimator is by design not affected by the shape of the prior but it is still affected by the lower limit of the fitted parameter space. At low masses, the prior can play a stabilising role. For low or negative $\rm (S/N)_\textsc{wl}$, the ML estimator is very close to the minimum considered mass, $\log M_\textsc{wl} = 13$. In our analysis, $18.9 \, \%$ of the ML estimates reach the lower limit of the mass range. On the other hand, even when the peak of the mass posterior probability is at very low masses, the median, CBI, or mean estimators better account for the tail of the distribution at larger masses, and hence have a significantly lower scatter.


\subsubsection{Mass bias} \label{sec: Mass bias}

To quantify the WL mass accuracy, we can look at the mass bias, i.e., the ratio between the WL mass and the true halo mass,
\begin{equation}
b_M := \frac{M_\textsc{wl}}{M_{\rm sim}} - 1 \, .
\label{eq: mass_bias}
\end{equation}
To first-order in the Taylor expansion, the mass bias can be written as
\begin{equation}
b_M \simeq \ln (M_\textsc{wl} / M_{\rm sim}) \, .
\label{eq: mass_bias_taylor}
\end{equation}
Hereafter, all references to the mass bias correspond to the definition in Eq.~\eqref{eq: mass_bias_taylor}. 

The ensemble average mass bias $\langle b_M \rangle$, which measures the accuracy, and the related scatter $\sigma_b$, which measures the precision, are computed as in Eqs.~(\ref{eq: lensing_average}, \ref{eq: lensing_scatter}), respectively. We calculate the uncertainty of mass bias and scatter estimates as the standard deviation of 1000 bootstrap realizations.

\begin{figure}
	\includegraphics[width=\columnwidth]{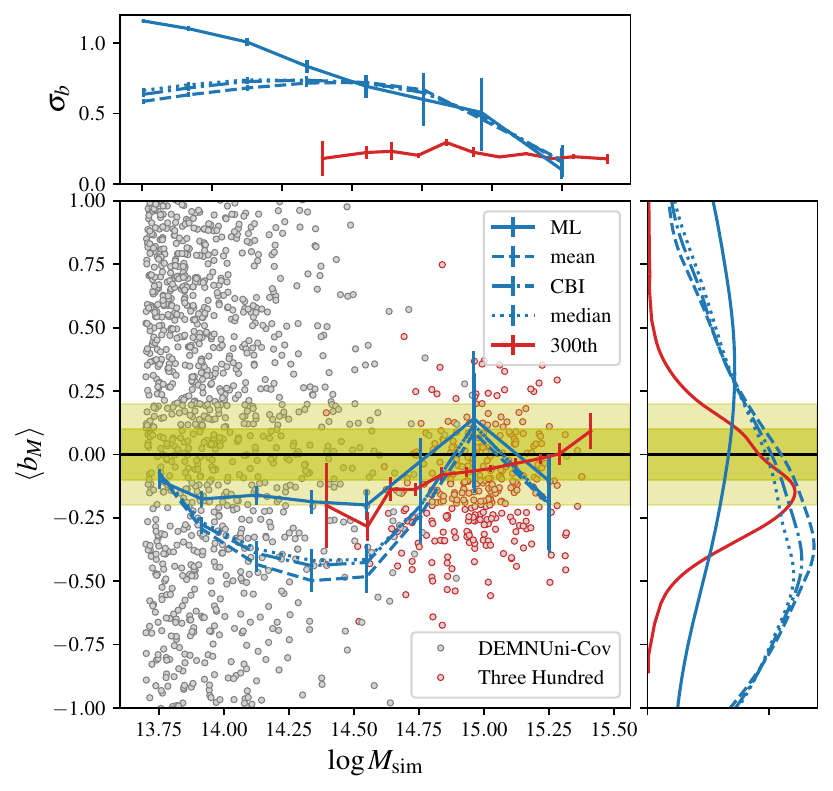}
    \caption{Weighted average (plotted as lines) of the WL mass bias for median, CBI, mean, and ML point estimates as a function of the true mass of the DEMNUni-Cov clusters. The grey points show the WL mass bias for ML estimates. Results from \citetalias{giocoli24} are presented with red lines and points. The light and dark yellow bands mark $\pm 10 \, \%$ and $20 \, \%$ thresholds, respectively. Top and right panels give the mass bias scatter and the distribution of the mass bias points for each estimator, respectively. Error bars are the standard deviation of the bootstrap sample distribution of weighted means.}
    \label{fig: mass_bias_point_estimate}
\end{figure}

Figure~\ref{fig: mass_bias_point_estimate} shows the weighted average of the mass bias in 8 equispaced logarithmic mass bins.
The WL mass for low-mass clusters is underestimated. This is particularly severe for median, CBI, and mean point estimates, for which the WL mass, in the mass range $14 \la \log M_{\rm sim} \la 14.5$, is about $50 \, \%$ of the true mass. On the other hand, the ML mass suffers from a less severe bias, and, for the same mass bins, it is no less than $80 \, \%$ of the true mass on average. For massive clusters, the bias is smaller. However, in the high-mass range, the number of simulated clusters is low and the bias measurement is strongly affected by statistical fluctuations. 

The intrinsic scatter decreases as the cluster mass increases. In low-mass regime, the scatter of the ML masses is larger than the scatter of the median, CBI, or mean measurements. This difference vanishes for higher masses, where the $\rm (S/N)_\textsc{wl}$ can become significant. 

Table~\ref{tab: mass_bias} provides the weighted mean of the WL mass bias for each point estimate. Median, CBI, and mean data points underestimate the true mass by approximately $23$--$27 \, \%$, while ML masses are on average more accurate, $\langle b_M \rangle = -14.6 \, \%$. For the overall mass range, the scatter is mainly driven by the numerous, low-mass haloes, and it is larger for the ML mass, which can often coincide with the lower limit of the parameter space. The scatter reduces from 1.07 to 0.83 when only considering objects with ML mass strictly larger than the lower limit of the mass prior. On the other hand, the scatter of high-mass clusters is lower and nearly independent from the point estimate of the mass, see Fig.~\ref{fig: mass_bias_point_estimate}. 

In summary, for the prior we considered, the ML mass is more accurate but less precise than median, CBI, or mean masses. The precision of the ML is similar to other point estimates in the high-mass regime, $\log M_{\rm sim} \ga 14.5$. The results may depend on priors, which should be optimised for each analysis based on the the scientific goal.

\begin{table*}[!ht]
    \centering
    \caption{Ensemble weighted average mass bias and scatter of median, CBI, mean, and ML point estimates.}
    \begin{tabular}{lrcccc}
        \toprule
        Data Set & & Median & CBI & Mean & ML \\
        \midrule
        LSS + shape noise (\ref{sec: Accuracy and Precision}) & $\left\{ \splitcell{$\langle b_M \rangle = $}{$\sigma_b = $} \right.$ & \splitcell{$-0.235 \pm 0.013$}{$0.712 \pm 0.010$} & \splitcell{$-0.245 \pm 0.012$}{$0.695 \pm 0.010$} & \splitcell{$-0.267 \pm 0.012$}{$0.657 \pm 0.011$} & \splitcell{$-0.146 \pm 0.017$}{$1.071 \pm 0.011$} \\
        \addlinespace
        Three Hundred \citepalias{giocoli24} & $\left\{ \splitcell{$\langle b_M \rangle = $}{$\sigma_b = $} \right.$ & \splitcell{$-0.053 \pm 0.008$}{$0.221 \pm 0.009$} & - & - & - \\
        \addlinespace
        photo-$z$ non-conservative (\ref{sec: Basic selection}) & $\left\{ \splitcell{$\langle b_M \rangle = $}{$\sigma_b = $} \right.$ & \splitcell{$-0.354 \pm 0.012$}{$0.681 \pm 0.009$} & \splitcell{$-0.360 \pm 0.011$}{$0.662 \pm 0.010$} & \splitcell{$-0.375 \pm 0.010$}{$0.626 \pm 0.010$} & \splitcell{$-0.334 \pm 0.016$}{$1.066 \pm 0.010$} \\
        \addlinespace
        photo-$z$ robust (\ref{sec: Robust selection}) & $\left\{ \splitcell{$\langle b_M \rangle = $}{$\sigma_b = $} \right.$ & \splitcell{$-0.220 \pm 0.012$}{$0.743 \pm 0.010$} & \splitcell{$-0.230 \pm 0.013$}{$0.724 \pm 0.010$} & \splitcell{$-0.255 \pm 0.012$}{$0.681 \pm 0.011$} & \splitcell{$-0.146 \pm 0.019$}{$1.137 \pm 0.011$} \\
        \addlinespace
        \AMICO-like (\ref{sec: Euclid-like clusters}) & $\left\{ \splitcell{$\langle b_M \rangle = $}{$\sigma_b = $} \right.$ & \splitcell{$-0.273 \pm 0.014$}{$0.728 \pm 0.011$} & \splitcell{$-0.282 \pm 0.015$}{$0.712 \pm 0.011$} & \splitcell{$-0.305 \pm 0.014$}{$0.673 \pm 0.012$} & \splitcell{$-0.182 \pm 0.020$}{$1.085 \pm 0.013$} \\
        \addlinespace
        \PZWav-like (\ref{sec: Euclid-like clusters}) & $\left\{ \splitcell{$\langle b_M \rangle = $}{$\sigma_b = $} \right.$ & \splitcell{$-0.284 \pm 0.014$}{$0.731 \pm 0.012$} & \splitcell{$-0.294 \pm 0.014$}{$0.714 \pm 0.012$} & \splitcell{$-0.316 \pm 0.013$}{$0.676 \pm 0.012$} & \splitcell{$-0.210 \pm 0.020$}{$1.099 \pm 0.013$} \\
        \addlinespace
        masks (\ref{sec: Masks effects}) & $\left\{ \splitcell{$\langle b_M \rangle = $}{$\sigma_b = $} \right.$ & \splitcell{$-0.225 \pm 0.014$}{$0.732 \pm 0.011$} & \splitcell{$-0.235 \pm 0.014$}{$0.714 \pm 0.012$} & \splitcell{$-0.255 \pm 0.013$}{$0.674 \pm 0.012$} & \splitcell{$-0.139 \pm 0.021$}{$1.101 \pm 0.013$} \\
        \addlinespace
        BMO + prior (\ref{sec: Model and prior}) & $\left\{ \splitcell{$\langle b_M \rangle = $}{$\sigma_b = $} \right.$ & \splitcell{$-0.271 \pm 0.010$}{$0.548 \pm 0.010$} & \splitcell{$-0.284 \pm 0.010$}{$0.544 \pm 0.010$} & \splitcell{$-0.306 \pm 0.009$}{$0.532 \pm 0.009$} & \splitcell{$-0.138 \pm 0.017$}{$1.069 \pm 0.011$} \\
        \addlinespace
        \splitcellbis{photo-$z$ robust +}{\AMICO-like +}{masks +}{BMO + prior} (\ref{sec: Combined systematics}) & $\left\{ \splitcell{$\langle b_M \rangle = $}{$\sigma_b = $} \right.$ & \splitcell{$-0.322 \pm 0.014$}{$0.578 \pm 0.011$} & \splitcell{$-0.334 \pm 0.013$}{$0.574 \pm 0.011$} & \splitcell{$-0.355 \pm 0.013$}{$0.560 \pm 0.011$} & \splitcell{$-0.154 \pm 0.025$}{$1.180 \pm 0.014$} \\
        \bottomrule
    \end{tabular}
    \tablefoot{Uncertainties are the standard deviation of the bootstrap sample distribution.}
    \label{tab: mass_bias}
\end{table*}

\subsection{Comparison with the Three Hundred} \label{sec: Comparison with the Three Hundred}

\begin{figure}
	\includegraphics[width=\columnwidth]{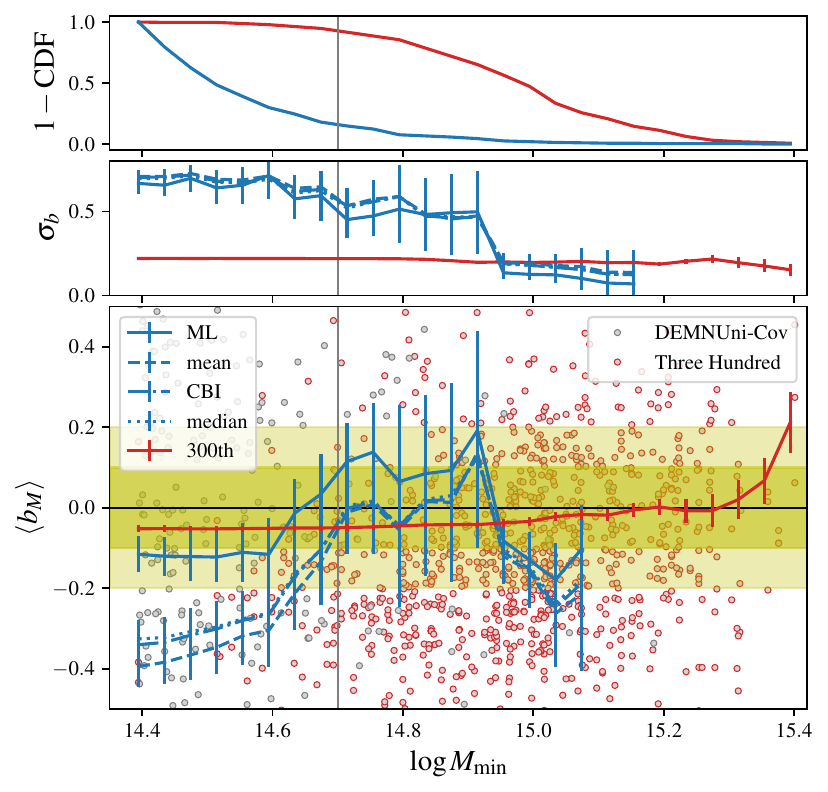}
    \caption{Weighted average of the WL mass bias of DEMNUni-Cov clusters with $M_{\rm sim} > M_{\rm min}$ as a function of $M_{\rm min}$. Results are compared with measurements performed on the Three Hundred cluster sample from \citetalias{giocoli24}. Top panels give the fraction of clusters in the considered mass range (top) and the mass bias scatter (middle). The vertical line is at $\log M_{\rm min} = 14.7$ lower limit. Error bars are the standard deviation of the bootstrap sample distribution.}
    \label{fig: high_mass_total_mass_bias}
\end{figure}

\begin{figure}
    \centering
	\includegraphics[width=\columnwidth]{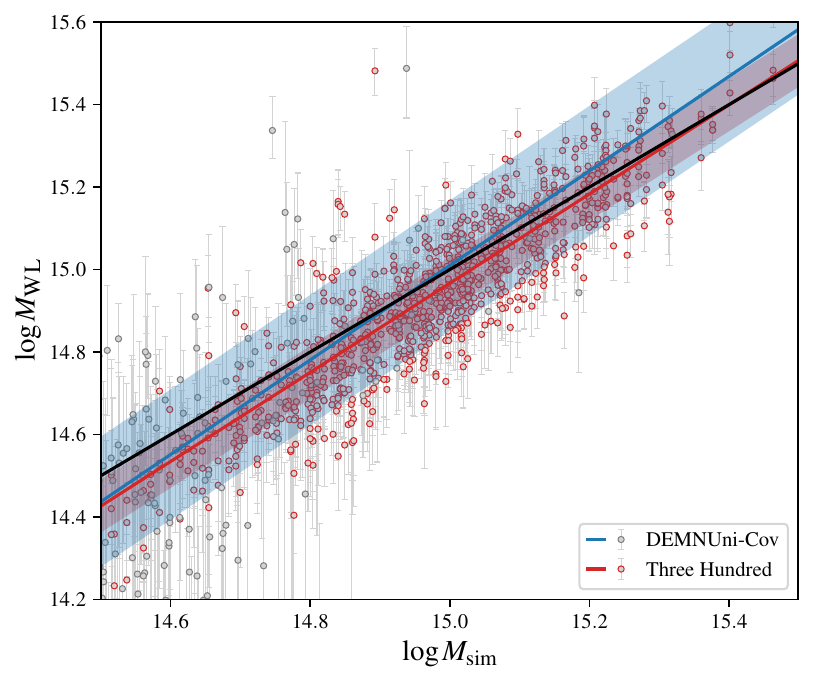}
    \caption{Scatter plot for median WL mass point estimates of DEMNUni-Cov clusters in the \citetalias{giocoli24} mass range. Error bars correspond to the $16^{\rm th}$ and $84^{\rm th}$ percentiles. The red line shows the linear regression for median WL mass measured in \citetalias{giocoli24}. The shaded region corresponds to the scatter.}
    \label{fig: scatter_point_estimate_high_mass}
\end{figure}

The mass bias of the Three Hundred clusters is smaller than the bias of the DEMNUni-Cov clusters, which cover a larger mass and redshift range, see Fig.~\ref{fig: clusters_distribution}. A fair comparison requires one to consider the mass bias of the 318 DEMNUni-Cov clusters within the mass range of the $3 \times 324$ Three Hundred clusters. 

In Fig.~\ref{fig: high_mass_total_mass_bias}, we show the weighted mean of the mass bias and associated scatter for clusters with true mass $M_{\rm sim} > M_{\rm min}$, where $\log M_{\rm min} \geq 14.4$ is the lower limit of the cluster mass range. The WL mass bias for the ML estimator is in agreement regardless of the lower limit considered, while other statistical estimators agree for clusters with $\log M_{\rm sim} \ga 14.7$. Our mass bias scatter is about twice the scatter for the Three Hundred clusters, but the bias scatter values agree at the high-mass end. 
The difference can be explained in terms of the simulation settings. The distribution of the Three Hundred clusters better samples the high-mass end of the halo mass function, whereas the DEMNUni-Cov cluster mass distribution follows the halo mass function, thus predominantly populating the low-mass range. There are numerous DEMNUni-Cov objects with $\log M_{\rm sim} \leq 14.7$, 269 out of 318, which increase the mass bias when calculating the mean of the full sample. Conversely. the contribution of the 66 Three Hundred clusters (out of 972) have less impact on the mass bias.
Furthermore, \citetalias{giocoli24} focused on the main halo and only considered particles in a slice of depth $\pm 5 \, \rm Mpc$ in front of and behind the cluster. On the other hand, the DEMNUni-Cov settings fully accounts for correlated matter around the halo and uncorrelated matter from LSS, which is very significant source of scatter at the \Euclid lensing depth \citep{PaperIII}.

Averaged density profiles and measured masses for $\log M_{\rm sim} \geq 14.7$ are shown in Fig.~\ref{fig: stacked_shears} and reported in Table~\ref{tab: stacked_mass}, respectively.

The agreement with the Three Hundred results is also found in the analysis of the linear regression of the true mass versus the median WL mass, see Fig.~\ref{fig: scatter_point_estimate_high_mass}. We use the median to be consistent with \citetalias{giocoli24}, which measured the $16^{\rm th}$, $50^{\rm th}$, and $84^{\rm th}$ percentiles of the mass. The parameters fitted to the DEMNUni-Cov clusters are $\alpha = -0.136 \pm 0.064$ and $\beta = 1.146 \pm 0.099$, while those for the Three Hundred are $\alpha = -0.114 \pm 0.046$ and $\beta = 1.081 \pm 0.044$. However, consistently with the scatter of the weighted average, our measured scatter from the linear regression is about two to three times larger than that of the Three Hundred, $\sigma = 0.158 \pm 0.018$ vs. $\sigma = 0.064 \pm 0.003$. We attribute this to the different mass distribution of objects, and to the uncorrelated LSS noise in the shear.


\section{Assessment of systematic effects in lens or source catalogues} \label{sec: Systematics Assessment}

\begin{figure}
	\includegraphics[width=\columnwidth]{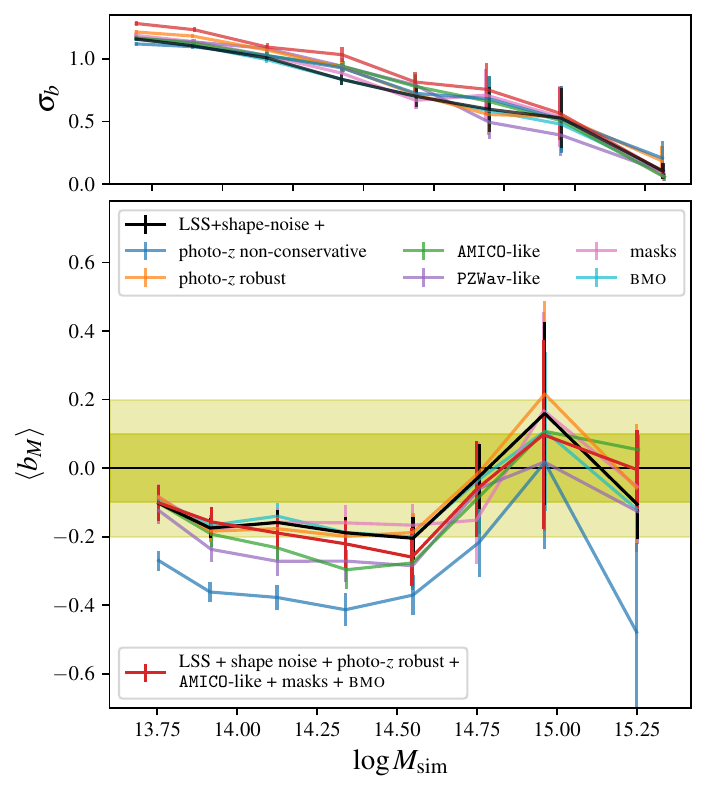}
    \caption{Weighted average of the WL mass bias for the ML mass estimator. Data sets considered are `LSS + shape noise'  (Sect.~\ref{sec: Accuracy and Precision}), photo-$z$ non-conservative selection (Sect.~\ref{sec: Basic selection}), photo-$z$ robust selection (Sect.~\ref{sec: Robust selection}), \AMICO-like sample (Sect.~\ref{sec: Euclid-like clusters}), \PZWav-like sample (Sect.~\ref{sec: Euclid-like clusters}), masks (Sect.\ref{sec: Masks effects}), BMO model (Sect.\ref{sec: BMO profile}), and combined effects (Sect.\ref{sec: Combined systematics}). The top panel shows the mass bias scatter. Error bars are the standard deviation of the bootstrap sample distribution.}
    \label{fig: mass_bias_systematics}
\end{figure}

\begin{figure}
	\includegraphics[width=\columnwidth]{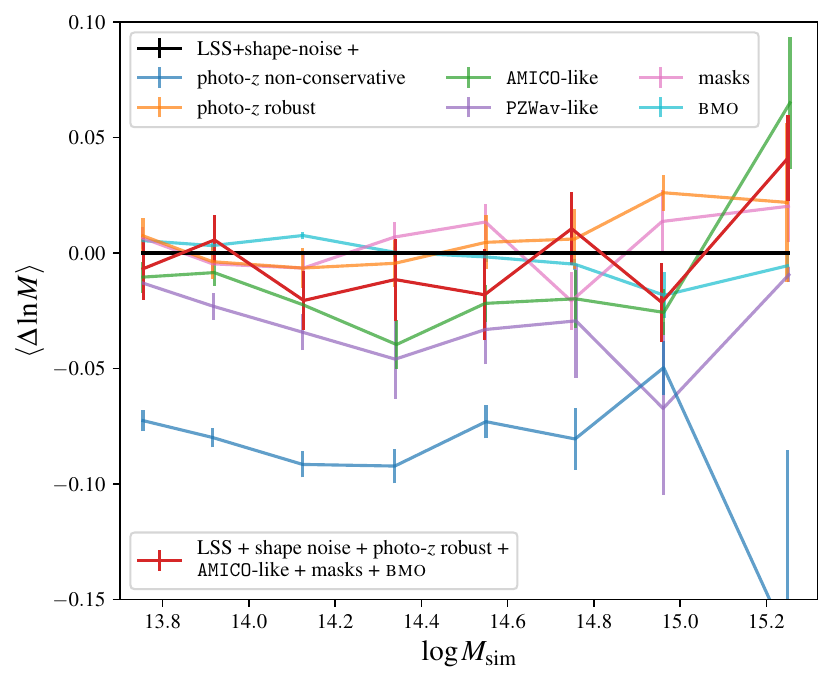}
    \caption{Relative change of the WL mass with respect to the `LSS plus shape noise' case due to systematic effects. We consider the ML estimator. Data sets considered are photo-$z$ non-conservative selection (Sect.~\ref{sec: Basic selection}), photo-$z$ robust selection (Sect.~\ref{sec: Robust selection}), \AMICO-like sample (Sect.~\ref{sec: Euclid-like clusters}), \PZWav-like sample (Sect.~\ref{sec: Euclid-like clusters}), masks (Sect.\ref{sec: Masks effects}), BMO model (Sect.\ref{sec: BMO profile}), and combined effects (Sect.\ref{sec: Combined systematics}). Error bars are the standard deviation of the bootstrap sample distribution.}
    \label{fig: mass_change_systematics}
\end{figure}

\begin{table}
	\centering
	\caption{Relative change of the WL mass with respect to the `LSS plus shape noise' case due to systematic effects.}
    \resizebox{\hsize}{!}{
	\begin{tabular}{lcc}
		\toprule
		data set & $\langle \Delta \ln M \rangle$ & $\Delta \ln M_{\langle \Delta\Sigma \rangle}$ \\
		\midrule
		photo-$z$ non-conservative (\ref{sec: Basic selection})& $-0.080 \pm 0.002$ & $-0.223 \pm 0.015$ \\
        \addlinespace
		photo-$z$ robust (\ref{sec: Robust selection})& $0.000 \pm 0.004$ & $0.007 \pm 0.017$ \\
        \addlinespace
		\AMICO-like (\ref{sec: Euclid-like clusters}) & $-0.015 \pm 0.003$ & $0.053 \pm 0.017$ \\
        \addlinespace
		\PZWav-like (\ref{sec: Euclid-like clusters}) & $-0.025 \pm 0.003$ & $0.009 \pm 0.015$ \\
        \addlinespace
		masks (\ref{sec: Masks effects}) & $0.000 \pm 0.002$ & $0.013 \pm 0.015$ \\
        \addlinespace
		BMO (\ref{sec: BMO profile}) & $0.004 \pm 0.001$ & $0.051 \pm 0.015$ \\
  
		\splitcellbis{photo-$z$ robust + }{\AMICO-like + }{masks + }{BMO + prior} (\ref{sec: Combined systematics}) & $-0.006 \pm 0.006$ & $0.129 \pm 0.019$ \\
        \bottomrule
	\end{tabular}
    }
    \tablefoot{We consider the ML estimator. Measured quantities are derived from the cluster mass average ($\langle \Delta \ln M \rangle$), or the cluster lensing profile average ($\Delta \ln M_{\langle \Delta\Sigma \rangle}$). Errors are computed as the standard deviation of the bootstrap sample distribution (mass average) and the standard deviation of the marginalised mass posterior distribution, added in quadrature (lensing profile).}
	\label{tab: relative_change}
\end{table}

The analysis in Sect.~\ref{sec: Accuracy and Precision} presents cluster mass measurements in an idealised setting where lens and source catalogues are unbiased and where only sources of intrinsic scatter and noise, e.g., LSS and shape noise, triaxiality, or cluster orientation, play a role. In this section, we test how well WL mass can be measured and how observational and measurement effects impact WL mass measurements.
In particular, we examine the effects of redshift uncertainty, selection effects from optical cluster detection algorithms, cluster centroid offsets, and masked data.

To quantify the impact of each systematic effect, we define the relative mass change
\begin{equation}
\Delta \ln M = \ln (M_{\rm sys} / M_{\rm ref}) \, ,
\label{eq: relative_change}
\end{equation}
where $M_{\rm ref}$ is the reference WL mass measured in the unbiased `LSS + shape noise' case as in the previous section, and $M_{\rm sys}$ is the WL mass measured on catalogues with additional systematic effects. As in Eq.~\eqref{eq: mass_bias_taylor}, we express the relative mass change in terms of the natural logarithm. We measure the ensemble average mass change, $\langle \Delta \ln M \rangle$, as given in Eq.~\eqref{eq: lensing_average} for clusters in common in reference and comparison samples. The error for this measurement is the standard deviation of the bootstrap sample distribution.

The mass shift can be further assessed with the WL mass measured by fitting the cluster ensemble averaged lensing profile, see Eq.~\eqref{eq: Delta_Sigma_av_stack}, $\Delta \ln M_{\langle \Delta\Sigma \rangle}$. All clusters of the samples are accounted in the averaged lensing profile. The uncertainty for $\Delta \ln M_{\langle \Delta\Sigma \rangle}$ is the sum in quadrature of the standard deviations of the posterior distribution for both reference and comparison samples. 

Table~\ref{tab: relative_change} lists the relative mass variation for the cluster mass ensemble average and the cluster lensing profile ensemble average. Figure~\ref{fig: mass_change_systematics} shows mass change for the cluster mass ensemble average in different mass bins.

Hereafter, we primarily present results on WL mass, mass bias, and relative mass change using ML point estimates. 
The ML point estimate is often considered in lensing analyses of Stage-III surveys or precursors \citep{PaperIII}. Properties of each sample presented in this section are summarised in Table~\ref{tab: specifics_samples}.

\subsection{Source redshift uncertainty} \label{sec: Reshifts uncertainty}

Measuring unbiased WL masses requires a robust identification of background galaxies. Contamination from foreground or cluster member galaxies substantially dilutes the lensing signal \citep{2017MNRAS.469.4899M}, thus increasing the WL mass bias towards negative values. We define the contamination fraction following \citet{PaperIV} as
\begin{equation}
f(z_{\rm d}\,|\,\vec{p}) := \frac{N_{\rm sel}(z_{\rm g} < z_{\rm d} \, | \, \vec{p})}{N_{\rm sel}(z_{\rm g} \geq 0 \, | \, \vec{p})} \, ,
\label{eq: contamination}
\end{equation}
where $\vec{p}$ is the photo-$z$ selection criterion, defined via Eq.~\eqref{eq: basic selection} or Eq.~\eqref{eq: robust selection}, and $N_{\rm sel}$ is the number of galaxies selected with the condition $\vec{p}$.

Contamination by cluster members could be corrected with a so-called `boost factor', applied either directly to the shear data \citep[e.g.,][]{2018ApJ...854..120M} or added to the mass model \citep[e.g.,][]{2019MNRAS.482.1352M}. In this work, we do not include a boost factor to correct for signal dilution. In fact, we find, in agreement with \citet{PaperIV}, that a boost correction is not necessary if one applies a sufficiently conservative source selection. However, boosting the WL signal allows for effective correction of the lensing signal, resulting in accurate mass calibration \citep[e.g.,][]{2019MNRAS.489.2511V}.

Primary \Euclid WL probes will divide galaxies into tomographic bins and perform a redshift calibration a posteriori to correct the galaxy distribution $n(z)$ in each bin. Cluster WL masses can be measured considering tomographic bins for background source redshifts with the ensemble $n(z)$ calibration \citep[e.g.,][]{2023arXiv231012213B, 2024A&A...687A.178G}. Using tomographic bins can result in significant foreground contamination in the background galaxy sample. However, this can be corrected using a well-known and calibrated source redshift distribution.

To study the effects of source selection on cluster WL masses, we simulate observed photometric source redshifts, $z_{\rm obs}$, from the true source redshifts, $z_{\rm true}$, 
\begin{equation}
\tilde{\Delta} z = \frac{z_{\rm obs} - z_{\rm true}}{1 + z_{\rm true}} \, .
\label{eq: Delta_z}
\end{equation}
We consider two source populations. The first population accounts for well-behaved photo-$z$s, whose deviations $\tilde{\Delta} z$ follow a Gaussian distribution centred on a redshift bias $\mu_{\tilde{\Delta} z}$ and with scatter $\sigma_{\tilde{\Delta} z}$.

The second population accounts for catastrophic outliers with $|\tilde{\Delta} z | > 0.15$, for which $\tilde{\Delta} z$ is uniformly distributed in the $|\tilde{\Delta} z | > 0.15$ region up to $z_s = 3$. The distribution of simulated photo-$z$s is shown in Fig.~\ref{fig: generated_redshift_distribution}.

The parameters of the distributions we use were informed by \citet{2023MNRAS.520.3529E}, who measured photo-$z$ bias and scatter from simulated \Euclid data \citep{2020MNRAS.494.2337B} using various codes. Measurements of the photo-$z$ bias are in agreement to those found for cosmic shear analyses \citep{2021A&A...647A.117E}. The photometry was simulated for \Euclid and Rubin/LSST bands with Gaussian noise. The photometric noise distributions depends on the survey depth and was fixed to one tenth of the flux corresponding to a $\rm S/N > 10$.

\begin{figure}
	\includegraphics[width=\columnwidth]{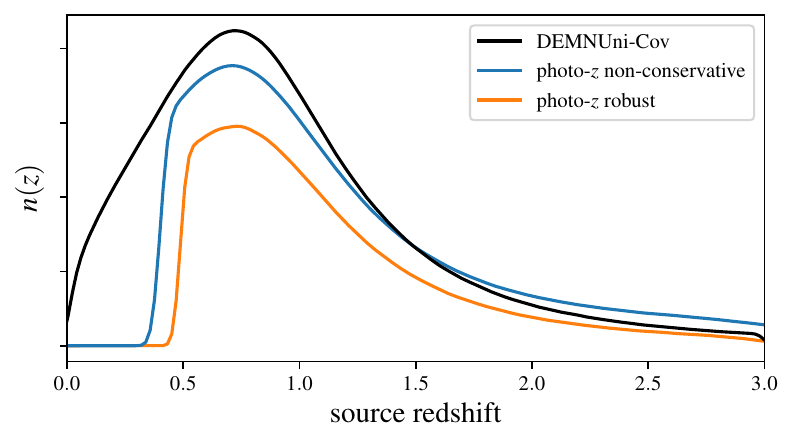}
    \caption{Redshift distribution of DEMNUni-Cov sources, and of non-conservative and robust photo-$z$-selected sources beyond $z_{\rm d} = 0.3$.}
    \label{fig: redshift_distribution}
\end{figure}

Below, we describe our simulations of two different photo-$z$ selected source samples, a non-conservative or a robust galaxy selection, and we examine the precision in lensing signal and mass in both scenarios. The distribution $n(z)$ beyond the cluster redshift $z_{\rm d}=0.3$, as well as the distribution $n(z)$ of the original DEMUni-Cov sample, for both photo-$z$ selections, are shown in Fig.~\ref{fig: redshift_distribution}.

\subsubsection{Non-conservative photo-\texorpdfstring{$z$}{z} selection} \label{sec: Basic selection}

Here, we consider the systematic uncertainty on cluster WL masses derived using a non-conservative source selection that selects background galaxies based only on the photo-$z$ point estimate, without any further photo-$z$ quality cuts. This is analogous to cluster WL with tomographic bins.

For this non-conservative selection, we expect the distribution of observed photo-$z$s to comprise a fraction of well-behaved galaxies with Gaussian scatter around the true redshift as well as a fraction of catastrophic outliers. Based on models of \Euclid plus ground-based complementary photometry from \citet{2023MNRAS.520.3529E}, and their analysis of $\HE$-band detected galaxies with $\rm S/N > 3$, we expect an outlier fraction of $12.7 \, \%$. For the distribution of non-catastrophic, well-behaved photo-$z$s, we consider a normal distribution with mean $\mu_{\tilde{\Delta} z} = -0.002$ and dispersion $\sigma_{\tilde{\Delta} z} = 0.045$. Figure~\ref{fig: generated_redshift_distribution} shows the distribution of simulated galaxy redshifts. 

Sources are selected to be background galaxies following Eq.~\eqref{eq: basic selection}. As shown in Fig.~\ref{fig: photoz_contamination}, this simple selection criterion results in significant foreground source contamination. The contamination, as defined in Eq.~\eqref{eq: contamination}, increases with the cluster redshift. This is because the total number of background sources reduces while the number of contaminated sources increases with increasing redshift.

\subsubsection{Robust photo-\texorpdfstring{$z$}{z} selection} \label{sec: Robust selection}

To reduce foreground contamination, we can make additional quality cuts in the galaxy source sample, e.g., requiring that the photo-$z$ probability distribution is well behaved with a prominent peak, or that a significant fraction of the distribution is above the lens redshift \citep{2017MNRAS.472.1946S}. These conservative cuts can significantly reduce the systematic contamination at the expense of a slightly smaller sample of selected galaxies and a slightly higher noise level \citep{PaperIV}.

We model the properties of the robustly selected galaxies as follows. Firstly, we conservatively assume that reliable shapes and nearly unbiased photo-$z$s can be measured for about $70 \, \%$ of the full galaxy population \citep[e.g.,][]{2012MNRAS.422..553B, 2019MNRAS.484.1598B, 2022MNRAS.511.1484I}, and, therefore, we randomly discard $30 \, \%$ of the galaxies from the sample. \Euclid and Stage-IV surveys should perform even better in terms of completeness \citep{PaperIV}.

\begin{figure}
	\includegraphics[width=\columnwidth]{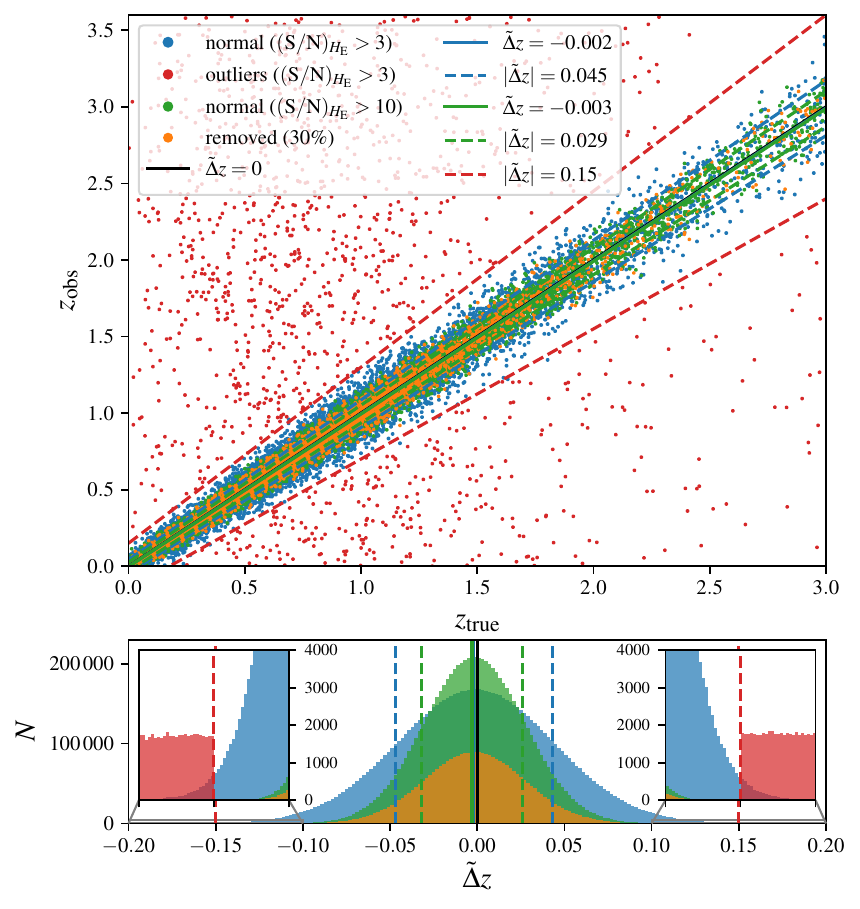}
    \caption{Scatter plot of simulated photo-$z$ vs. true redshift for non-conservative (${\rm red} + {\rm blue}$) or robust (${\rm green} - {\rm orange}$) selection. The bottom panel shows the marginalised 1D distribution of the normalised redshift deviation.}
    \label{fig: generated_redshift_distribution}
\end{figure}

Next, we assume that after applying redshift quality cuts, we are left with galaxies with a well-behaved, single peaked photo-$z$ probability density distribution whose signal is likely better detected than for the full sample. Consequently, we model the observed photo-$z$ distribution of the selected sources as galaxies expected to be observed in $\HE$-band images with $\rm S/N > 10$ \citep{2023MNRAS.520.3529E}. We expect $4 \, \%$ of outliers, but we do not include these objects in the sample as we assume that their fraction is significantly reduced to the sub-percent level thanks to the robust photo-$z$ selection, possibly coupled with thorough colour selections as discussed in \citet{PaperIV}. Deviations from the true redshifts of these photo-$z$s are normally distributed with mean $\mu_{\tilde{\Delta} z} = -0.003$ and dispersion $\sigma_{\tilde{\Delta} z} = 0.029$. Figure~\ref{fig: generated_redshift_distribution} shows the redshift distribution of the robustly selected galaxies.

\begin{figure}
	\includegraphics[width=\columnwidth]{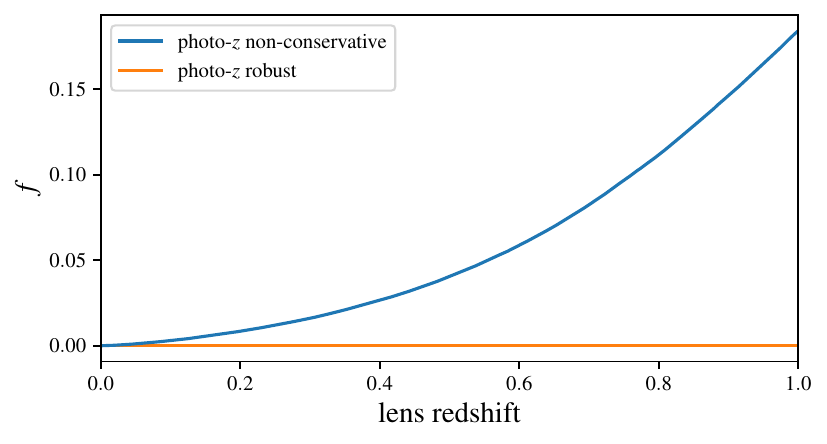}
    \caption{Photo-$z$ contamination fraction as a function of the lens redshift of non-conservative or robust photo-$z$ selected sources.}
    \label{fig: photoz_contamination}
\end{figure}

Finally, we select galaxies according to the robust criterion \citep{2017MNRAS.472.1946S}
\begin{equation}
z_{\rm s, min} > z_{\rm d} + \Delta z \, ,
\label{eq: robust selection}
\end{equation}
where $z_{\rm s, min}$ is the $2\sigma_{\tilde{\Delta} z}$ lower limit of the galaxy photo-$z$ distribution. Figure~\ref{fig: photoz_contamination} shows that the robust selection significantly reduces the foreground contamination of the source sample. 

\subsubsection{Impact of photo-\texorpdfstring{$z$}{z} uncertainties} \label{sec: Accuracy and precision with photo-z}

\begin{figure}
	\includegraphics[width=\columnwidth]{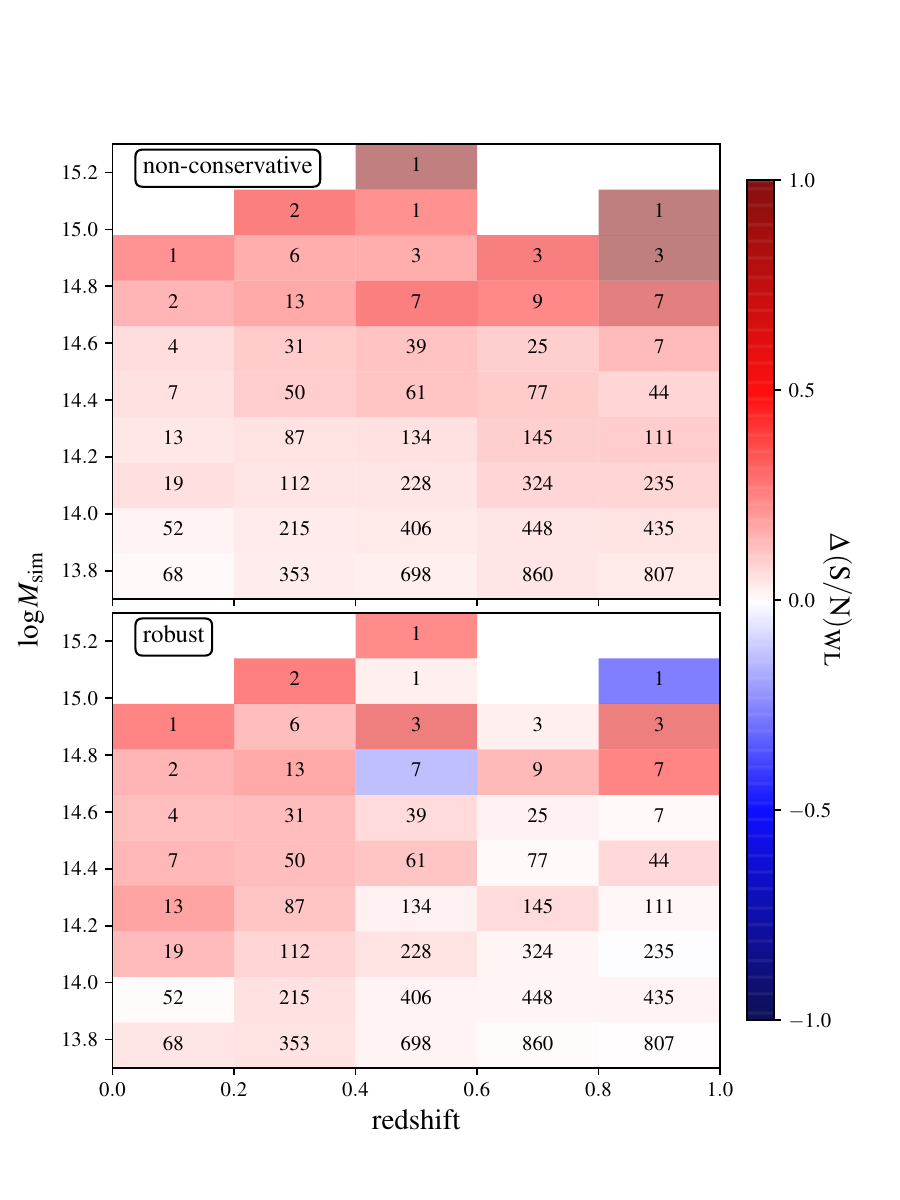}
    \caption{Difference of the $\rm (S/N)_\textsc{wl}$ as measured either for lensing with true redshifts, i.e., the `LSS + shape noise' case, or non-conservative (top) or robust (bottom) photo-$z$ selected samples, computed as the CBI of the values binned in mass and redshift. The number of clusters is displayed in each bin.}
    \label{fig: SNR_scatter_photoz}
\end{figure}

In Fig.~\ref{fig: SNR_scatter_photoz}, we plot the difference between the $\rm (S/N)_\textsc{wl}$ for the `LSS + shape noise' source catalogue with true galaxy redshifts, see Fig.~\ref{fig: SNR_distribution}, and that derived from a source catalogue with photo-$z$ noise. Photo-$z$ noise generally reduces the lensing $\rm (S/N)_\textsc{wl}$, particularly in the higher-mass bins. The reduction is larger for the non-conservative photo-$z$ selected sample, with $\rm (S/N)_\textsc{wl}$ lower by $15 \, \%$ on average, than for the robust one, $\rm (S/N)_\textsc{wl}$ lower by $5 \, \%$ on average. 
We notice a larger $\rm (S/N)_\textsc{wl}$ using the robust selection in some high-mass, high-redshift bins. However, their low statistics  do not allow for a significant conclusion about deviations from the `LSS + shape noise' case.
As summarised in Table~\ref{tab: SNR threshold}, photo-$z$ noise generally decreases the $\rm (S/N)_\textsc{wl}$. The decrease is slightly smaller in the robust case because the removal of foreground contaminants strengthens the lensing signal. However, at low mass and low redshift, the robust photo-$z$ selection decreases the $\rm (S/N)_\textsc{wl}$ slightly more than in the non-conservative case. This is because the robust photo-$z$ cut reduces the overall number of sources and shape noise dominates in this mass regime.

Figure~\ref{fig: mass_bias_systematics} shows the weighted average of the mass bias and scatter in different mass bins 
considering the ML point estimator. 
In the non-conservative case, the mass bias can be as large as $-40 \, \%$ and it exceeds the $-20 \, \%$ limit for $\log M_{\rm sim} \la 14.75$. On the other hand, for the robustly selected background sample, the mass bias does not significantly change from the fiducial case. 
This trend is also seen for the relative mass change in Fig.~\ref{fig: mass_change_systematics}, where the WL mass for the non-conservative selection differs from the WL mass measured on true redshift data. On the other hand, the robust photo-$z$ selection does not show a significant relative change.

The mass bias and scatter averaged over the full sample mass range are summarised in Table~\ref{tab: mass_bias}. The measured WL mass bias for the non-conservative photo-$z$ selection is on average $-34.4 \pm 1.6 \, \%$. On the other hand, the mass bias for the robust photo-$z$ selection is very close to the unbiased measurement, $\langle b_M \rangle = -14.6 \pm 1.9 \, \%$, but has a larger scatter due to the lower number of sources selected.

These results are consistent with the linear regression analysis, see Table~\ref{tab: linear_regression}. The slope and intercept of the linear regression for the non-conservative selection are lower than in the unbiased sample, whereas the scatter is not significantly different.
On the other hand, results for the robust photo-$z$ selection are very similar to the unbiased case except for the scatter that is larger, in agreement with results from the mass bias.


The cluster average density profiles plotted in Fig.~\ref{fig: stacked_shears} further support our finding that non-conservative source selection significantly biases cluster WL measurements.
On the other hand, we find that the average shear profiles of sources selected in the robust photo-$z$ case are consistent with the `LSS + shape noise' case.

Masses recovered from the average density profiles are given in Table~\ref{tab: stacked_mass} and the corresponding relative mass change is reported in Table~\ref{tab: relative_change}. The mass is $22.3 \pm 1.5 \, \%$ lower than the `LSS + shape noise' case for the the non-conservative selection, but differs by just $0.7 \pm 1.7 \, \%$ for the robust photo-$z$ selection. 

The mass change as derived from the ensemble average of individual WL masses is $-8.0 \pm 0.2 \, \%$ for the non-conservative selection. This is due to the effect of the mass prior on noisy individual estimates.


To summarise, we find that the WL mass measurement can be significantly impacted by source redshift uncertainties without further calibration. However, this effect can be significantly mitigated with a careful source selection. A robust selection in photo-$z$ to reduce foreground contamination can be coupled with a colour-colour selection to increase source completeness while preserving purity \citep{PaperIV}. Another possibility is to use calibrated redshift distributions from tomographic bins for cosmic shear along with an empirical  correction for cluster member contamination \citep{2023arXiv231012213B}. 

\subsection{Cluster detection} \label{sec: Euclid-like clusters}

\begin{figure}
	\includegraphics[width=\columnwidth]{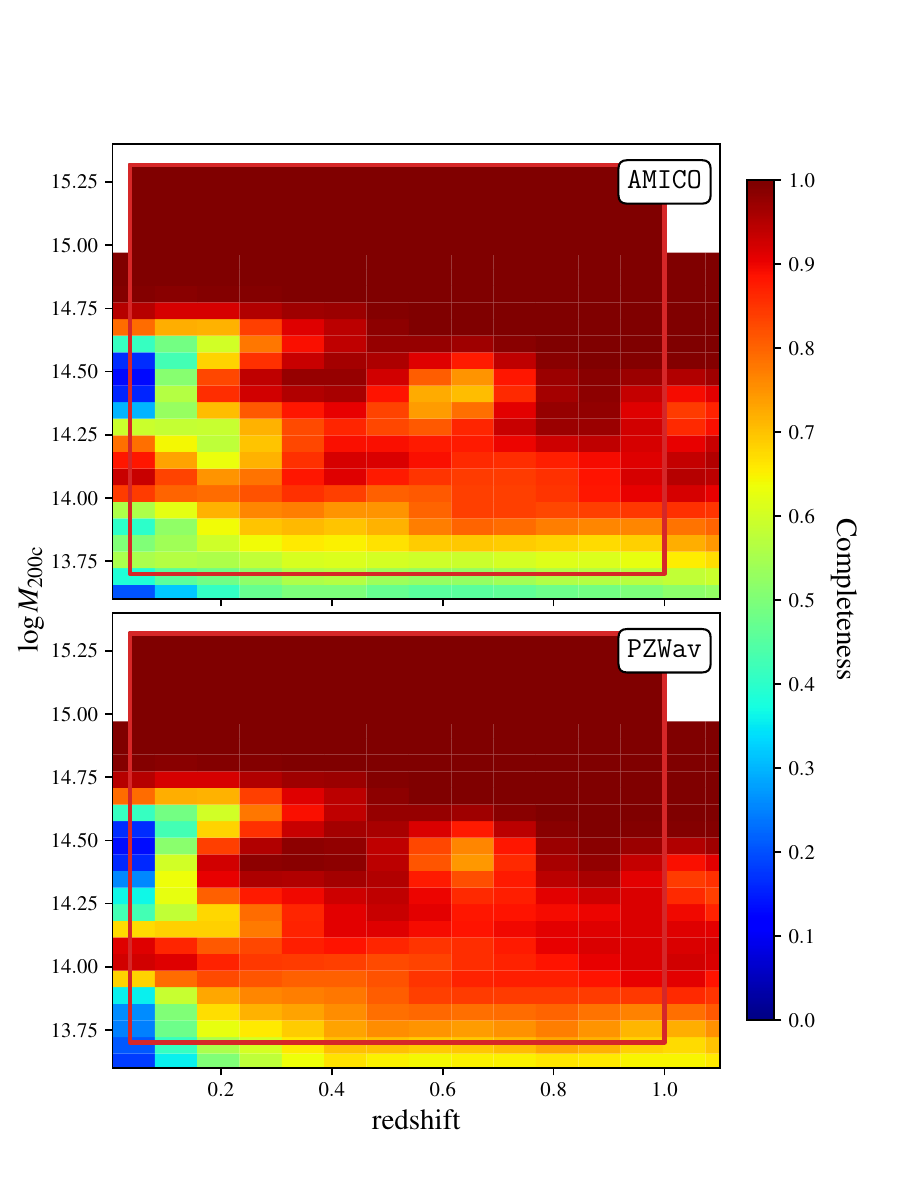}
    \caption{Expected completeness for the \AMICO and \PZWav algorithms as a function of mass and redshift. The light red box encloses the parameter space sampled by the DEMNUni-Cov haloes.}
    \label{fig: completeness_histogram}
\end{figure}

Cluster detection can be another source of WL mass systematic effects, due to, e.g., selection effects or miscentring. In this section, we study the potential impact of the optical cluster detection algorithms that will be used in the \Euclid data analysis pipeline: the Adaptive Matched Identifier of Clustered Object algorithm \AMICO \citep{2011MNRAS.413.1145B, 2018MNRAS.473.5221B, 2019A&A...627A..23E}, and the wavelet-based algorithm \PZWav \citep{2014becs.confE...7G, 2019A&A...627A..23E, 2024ApJ...967..123T}. These two optical cluster finding algorithms were selected to run on \Euclid data based on their performance in a comparative challenge of state-of-the art detection algorithms \citep{2019A&A...627A..23E}. 

To construct simulated lens samples reflecting the expected performance of each algorithm in the \Euclid survey, we first account for the cluster detection completeness. Euclid Collaboration: Cabanac et al. (in prep.) evaluates the performance of \AMICO and \PZWav by running them on the realistic simulated \Euclid galaxy catalogue \texttt{Flagship} version 2.1.10 \citep{2017ComAC...4....2P, EuclidSkyFlagship}. The completeness is calculated for galaxy clusters detected with $\rm (S/N)_{\rm det} > 3$. In Fig.~\ref{fig: completeness_histogram}, we show the mass-redshift completeness of \AMICO and \PZWav. In each mass-redshift bin, we randomly pick DEMNUni-Cov galaxy clusters with an associated probability equal to the completeness of the bin. Following this procedure, we obtain two \Euclid-like cluster samples, whose distributions are shown in Fig.~\ref{fig: clusters_distribution}. The sizes of the \AMICO and \PZWav-like catalogues are similar, 4557 and 4954 clusters, respectively.

\begin{figure}
	\includegraphics[width=\columnwidth]{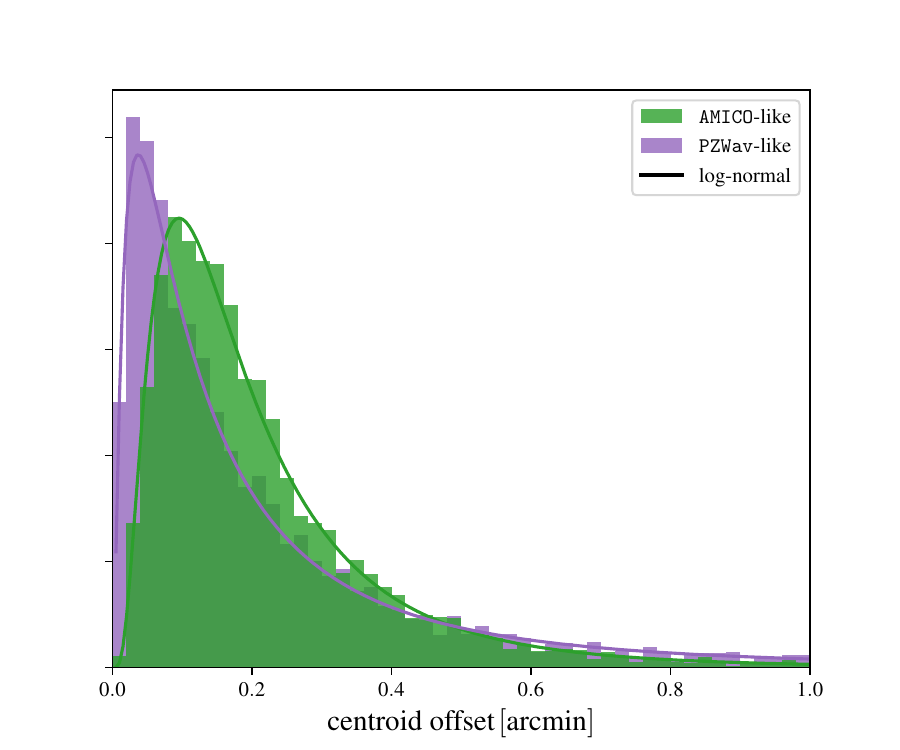}
    \caption{Centroid offsets of \AMICO-like and \PZWav-like clusters selected from the DEMNUni-Cov catalogue.}
    \label{fig: centroid_offset}
\end{figure}

To simulate realistic cluster catalogues, we take into account the centroid offset, i.e., the angular separation between the true cluster centre and the location of the detection, and realistic cluster redshifts based on the performance of the algorithms \citep{2019A&A...627A..23E}. We randomly shift the centre positions of \AMICO-like (\PZWav-like) clusters on the perimeter of a circle centred on the true position using a separation following a log-normal distribution with mean of $\ang{;0.22;}$ ($\ang{;0.27;}$) and RMS $\ang{;0.19;}$ ($\ang{;0.45;}$) \citep{2019A&A...627A..23E}. Figure~\ref{fig: centroid_offset} displays the centroid offset distribution generated for each cluster detection algorithm. 
Redshifts of detected clusters are scattered as $\sigma_{\tilde{\Delta} z} = 0.017 (M_{\rm 200c} / 10^{14} \, M_\odot)^{-1/2}$ for masses lower than $10^{14} \, M_\odot$, and $\sigma_{\tilde{\Delta} z} = 0.017$ for masses larger than $10^{14} \, M_\odot$.

We show the weighted average of the mass bias of \Euclid-like clusters in Fig.~\ref{fig: mass_bias_systematics}. For both samples, the mass bias is generally larger than for the full DEMNUni-Cov catalogue, which is well-centred and complete. The impact on the mass bias seems larger in the low-mass range, where \AMICO-like and \PZWav-like samples are less complete. The mass bias of \PZWav-like clusters is larger than for \AMICO-like clusters due to the larger dispersion of the centroid offset. 

Table~\ref{tab: mass_bias} presents the results of the ensemble average mass bias and scatter.
For the two detection algorithms and in comparison with results presented in Sect.~\ref{sec: Accuracy and Precision}, WL mass bias values are slightly more severe. \AMICO (\PZWav) clusters have a WL mass that is lower than the true mass on average by $18.2 \pm 2.0 \, \%$ ($21.0 \pm 2.0 \, \%$). The results of the linear regression presented in Table~\ref{tab: linear_regression} agree. For both \Euclid-like samples, intercept and slope parameters are lower than parameters for the unbiased cluster sample, whereas the intrinsic scatter slightly increases. Uncertainties in cluster redshift and position lowers the precision of the lensing mass.

Table~\ref{tab: relative_change} shows the relative change in WL mass. The ensemble WL mass change is not significantly impacted by systematic effects. The relative change given in WL mass as inferred from the averaged lensing signal, as shown in Fig.~\ref{fig: stacked_shears}, is $5.3 \pm 1.7 \, \%$ ($0.9 \pm 1.9 \, \%$) larger for \AMICO-like (\PZWav-like) samples.

Cluster samples discussed here are selected from the DM haloes in the DEMNUni-Cov simulation given the expected performance of the \Euclid detection algorithms, and we could not account for correlation between WL mass and optical proxies. Nevertheless, it is also known that optical cluster selection leads to WL selection biases \citep{2022MNRAS.515.4471W, 2022MNRAS.511L..30Z}. This is due to secondary selection effects of optical clusters, like a preference for concentrated haloes, haloes aligned along the line of sight, and haloes with more structure along the line of sight, leading to projection effects in the measured richness.

An alternative approach would be to perform an end-to-end mass bias analysis with the large \Euclid-like \texttt{Flagship} simulation \citep{2017ComAC...4....2P, EuclidSkyFlagship}, including galaxy clusters directly detected with \AMICO and \PZWav. However, no shear products were available in \texttt{Flagship} at the time that the analysis presented in this paper was carried out.

\subsection{Masks effects} \label{sec: Masks effects}

\begin{figure}
	\includegraphics[width=\columnwidth]{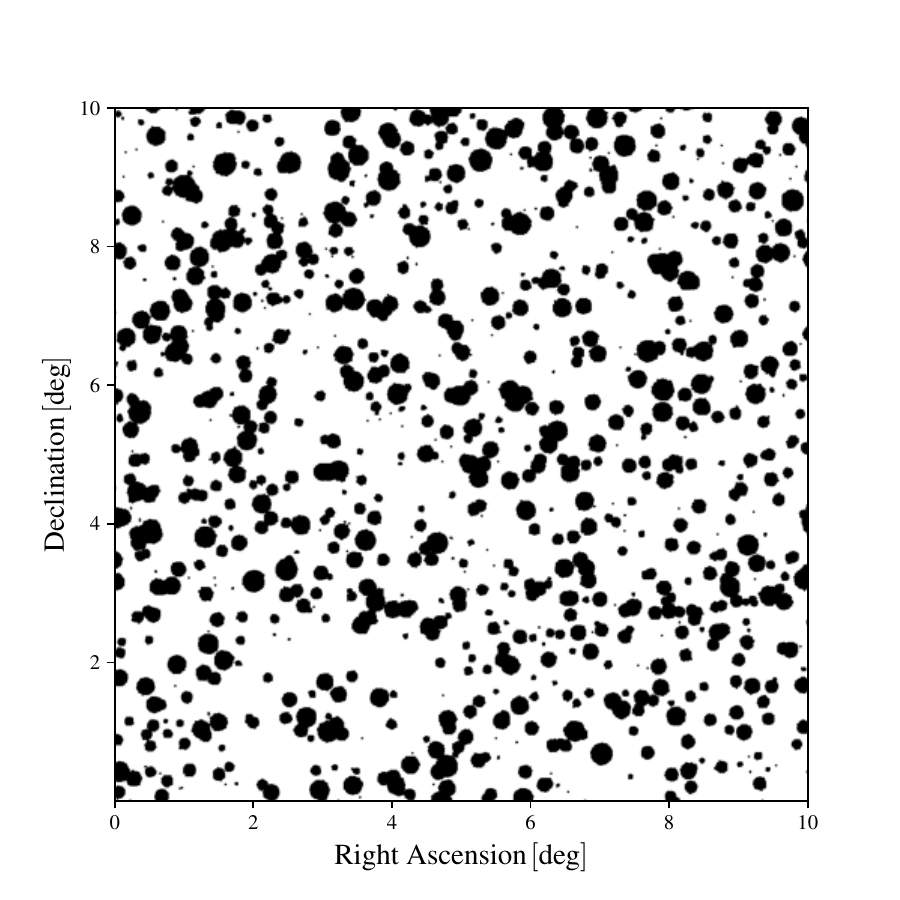}
    \caption{Mask map considered for the analysis \citep{2020A&A...638A.141P}.}
    \label{fig: masks}
\end{figure}

Masking of data plays a major role in large surveys. Regions of the sky must be masked due to, e.g., foreground bright stars. Masks lower the number of sources useful for WL, and increase the noise \citep{2014ApJ...784...31L}. To consider these effects, we apply the binary mask of \citet{2020A&A...638A.141P}, shown in Fig.~\ref{fig: masks}, to our catalogue. The mask was derived from \Euclid simulated catalogues \citep{2020A&A...644A..31E} using the code \texttt{FLASK} \citep{2016MNRAS.459.3693X}. The fraction of the masked sky is $22.4 \, \%$, both on source and lens catalogues.

The resulting mass bias is shown in Fig.\ref{fig: mass_bias_systematics} and Table~\ref{tab: mass_bias}. On the overall mass range, we do not see a significant difference with respect to the full-sky case with no masks, but we see a slightly larger scatter.
The parameters of the linear regression, see Table~\ref{tab: linear_regression}, are not significantly impacted. The relative mass change with respect to unmasked case, see Table~\ref{tab: relative_change}, is not statistically significant. 


Here clusters were selected within the unmasked regions. On real data sets, cluster detection algorithms run on fields with masks. One way to assess WL mass precision and accuracy more realistically would be to detect clusters directly on masked mocks. In this way, we would also account for galaxy clusters detected at the edges of the mask.

\section{Halo density model and priors} \label{sec: Model and prior}

Here, we discuss the bias related to the fitting procedure adopted to derive the mass, in particular, the assumed model of the halo density profile and the role of priors. 

\subsection{BMO profile} \label{sec: BMO profile}

The NFW profile has a non-physical divergence of its total mass \citep{2003MNRAS.340..580T}. The BMO profile, a smoothly truncated version of the NFW profile, see Eq.~\eqref{eq: BMO}, circumvents this problem, but it still suffers from the well-known degeneracy between mass and concentration. A halo with WL mass biased low can still fit the data if the concentration is biased high.

The BMO modelling gives results similar to NFW in terms of mass point estimators with respect to the true mass, see Table~\ref{tab: mass_bias} and Fig.~\ref{fig: mass_bias_systematics}. However, the WL mass and concentration inferred by fitting the averaged lensing signal are impacted. The BMO profile provides a better fit to the data at larger radii than the NFW profile, see Fig.~\ref{fig: stacked_shears}, with a smaller reduced $\chi^2$, see Table \ref{tab: stacked_mass}. As shown in Table~\ref{tab: relative_change}, the mass fitted to the averaged lensing profile is $5.1 \pm 1.6 \, \%$ higher, while the concentration is $42.8 \, \%$ lower, which echoes the mass-concentration degeneracy existing for the NFW profile. 


An additional step to further this analysis would be to consider the two-halo term. On larger scales than those reached in this study, i.e., $R \ga 6 \, \rm Mpc$, the lensing signal of the halo is dominated by correlated matter, e.g., neighbouring haloes or filaments \citep{2005ApJ...631...41T, 2010ApJ...724..878T, 2022MNRAS.511.1484I}. 

\subsection{Gaussian mass prior} \label{sec: Gaussian mass prior}

To explore the role of the model, we consider an informative prior on the mass. We repeat the Bayesian analysis, and we fit the WL signal to a BMO profile with a Gaussian prior on the mass, which, combined with the prior limits set in Sect.~\ref{sec: Fitting configuration}, mimics the distribution of the DEMNUni-Cov clusters, see Fig.~\ref{fig: clusters_distribution}. 
The mean of the Gaussian prior is set to the mode of the cluster mass distribution. The dispersion is twice the standard deviation, so that the tail in the distribution towards high masses is accounted for.
The prior is then truncated at the parameter limits.

The slope, intercept, and scatter of the linear regression between WL and true masses are reported in Table~\ref{tab: linear_regression}. By definition, the prior does not affect the ML analysis, but it affects results for point-estimators based on the posterior probability distribution. For the median, CBI, and mean point estimates, the slope and intercept parameters are lower than for the fiducial setting with a uniform prior. The intrinsic scatter is also two times lower. The informative prior leads to an increased precision but also to a larger bias.

Results of the analysis of the weighted mass bias, see Table~\ref{tab: mass_bias}, are consistent with those of the linear regression. The mass bias is more severe but the scatter is lower for the median, CBI, and mean WL mass leading to a decreased accuracy and increased precision.
Other prior functions could also be explored, in particular priors based on mass-concentration relations \citep[e.g.,][]{2014ApJ...797...34M, 2015ApJ...799..108D, 2019ApJ...871..168D, giocoli24}.

\section{Accuracy and precision in an \Euclid-like survey} \label{sec: Combined systematics}

Finally, to provide a preliminary overview of the accuracy and precision we can expect from WL mass measurements in \Euclid, we measure the mass bias and the relative mass change considering at the same time all the systematic effects previously discussed. To mimic a realistic scenario \citep{PaperIII}, we consider the following setting:
\begin{itemize}
    \item robust photo-$z$ source selection (Sect.~\ref{sec: Robust selection});
    \item \AMICO-like clusters (Sect.~\ref{sec: Euclid-like clusters});
    \item masked catalogues (Sect.~\ref{sec: Masks effects});
    \item BMO halo profile and Gaussian prior (Sect.~\ref{sec: Model and prior}).
\end{itemize}

In terms of precision of the lensing signal, we find 140 (about $3.9 \, \%$) clusters with $\rm (S/N)_\textsc{wl} > 3$, similar to the fraction of clusters with $\rm (S/N)_\textsc{wl} > 3$ measured from catalogues in the `LSS + shape nosie' case (about $3.7 \, \%$). 

In terms of precision of the recovered WL mass, the scatter as inferred either from the linear regression of the ML mass or the weighted average analysis is larger than the scatter of the `LSS + shape noise' case if we consider the ML point estimate.
However, looking at the median, CBI, or mean point estimates, the scatter is smaller, i.e., the precision is increased, mainly due to the Gaussian mass prior.



In terms of accuracy of the recovered WL mass, the slope and intercept values of the ML mass linear regression are statistically consistent with those of the `LSS + shape noise' case, with intercept $\alpha = -0.097 \pm 0.010$ and slope $\beta = 0.924 \pm 0.043$, see Fig.~\ref{fig: scatter_point_estimate}. For the median, CBI and mean WL mass, the recovered accuracy is lower.

The weighted average mass bias is summarised in Table~\ref{tab: mass_bias}. The ML mass is larger than $80 \, \%$ of the true mass in a significant mass range, see Fig.~\ref{fig: mass_bias_systematics}. The bias of the ML mass is $\langle b_M \rangle = -15.4 \pm 2.5 \, \%$ for the full sample. The median, CBI, and mean estimators exhibit a larger increase in mass bias with respect to the `LSS + shape noise' case measurement than the ML estimator. Looking at the relative mass change, see Table~\ref{tab: relative_change}, we find that the ensemble average ML mass is consistent with the `LSS + shape noise' case, suggesting that systematic effects can be well controlled. 

The difference in the average lensing profile is due to the effect of the selection function of \Euclid detected clusters. In the low-mass regime, clusters are not detected and the ensemble average of the cluster surface mass density profile increases compared to the profile of the full cluster population.
A residual difference in the fitted mass is due to the BMO halo density model compared to NFW.




\section{Summary} \label{sec: Summary}

To exploit the full cosmological constraining power of galaxy clusters, their WL masses must be well characterised. This requires testing the accuracy and precision of algorithms for WL mass measurement on simulated data. We can then use the resulting mass bias measurements to understand the impact of systematic effects on cluster WL masses.

In view of the first data from the \Euclid survey, we investigated the mass bias with \COMBCL, the \Euclid pipeline for cluster WL mass measurements. Using \COMBCL on the DEMNUni-Cov $N$-body simulations, we studied the WL mass of 6155 clusters with true mass $M_{\rm sim} > 5 \times 10^{13} \, M_\odot$ (or $\log M_{\rm sim} \ga 13.7$) and redshift $z_{\rm d} < 1$. 

We first assessed the precision and accuracy of the measured shear profile and WL mass in an ideal scenario with no measurement errors or uncertainties, but where the measured mass can differ from the true one due to, e.g.,
shape noise, LSS, triaxiality, and cluster orientation, mergings or irregular morphology \citep{2023A&A...675A.120E}. This allowed us to quantify the intrinsic bias and scatter of the WL mass with respect to the true cluster mass.
We summarise the results discussed in Sect.~\ref{sec: Accuracy and Precision} as follows.
\begin{itemize}
    \item The lensing signal of low-mass clusters is expected to be low. Only $3.7 \, \%$ of the clusters are recovered with $\rm (S/N)_\textsc{wl} > 3$, most of which lie at the sparsely populated, high-mass end. We found agreement with the semi-analytical forecasting presented in \citet{PaperIII}.
    \item The mass bias significantly depends on the mass point-estimator of choice. A summary statistic with only a location and a scale might be not enough to properly describe the mass probability function of a halo with low $\rm (S/N)_\textsc{wl}$. We measured the average mass bias via four statistical estimators: median, biweight location (CBI), mean, and maximum likelihood (ML). We found that the ML point estimate yields a WL mass almost two times more accurate than those of the median, CBI, and mean. In fact, at low masses, the mass bias can be as large as $\langle b_M \rangle \sim - 40\,\%$. The ML point estimate is more accurate, with $\langle b_M \rangle \ga -20 \, \%$ in the full mass range. For the full sample, the WL mass is on average biased low by about $23.5$--$26.7 \pm 1.2 \, \%$ for median, CBI, and mean measurements, and $14.6 \pm 1.7 \, \%$ for ML. However, the intrinsic scatter of the ML masses is about two times larger than other mass estimates.
    \item At the massive end of the halo mass function, $\log M_{\rm sim} \geq 14.7$, we found a mass bias in agreement with that measured with the massive Three Hundred clusters \citep{giocoli24}. Our sample extends to lower masses and larger redshifts, as expected for \Euclid-like clusters. We found that the bias over the full mass range is three to five times larger than that measured for massive clusters. The larger scatter in the DEMNUni-Cov sample is due to noise by correlated matter or LSS.
\end{itemize}

We went beyond the simple scenario with only intrinsic scatter and noise and checked how various systematic effects or uncertainties, e.g., photo-$z$ uncertainties, sample incompleteness, or masks, affect the WL mass measurements. We mostly focused on the ML mass estimator. Results are discussed in Sect.~\ref{sec: Systematics Assessment} and summarised below.
\begin{itemize}
    \item Following \citet{2023MNRAS.520.3529E}, we simulated observed galaxy photometric redshifts in two different scenarios, a robust photo-$z$ sample, with a conservative source selection, and a non-conservative photo-$z$ sample, which suffers from significant contamination or photo-$z$ outliers. For the non-conservative selection, which is similar to WL analyses in tomographic bins, we found that the lensing signal is biased low by $15 \, \%$, and that the WL mass bias is $-33.4 \pm 1.6 \, \%$ on the overall cluster mass range. For the robust selection, the lensing signal and the WL mass are well recovered over the full cluster mass range with a mass bias of $-14.7 \pm 1.9 \, \%$.
    \item We simulated \Euclid-like cluster samples considering the two cluster detection algorithms selected to run in the \Euclid analysis pipeline, \AMICO and \PZWav \citep{2019A&A...627A..23E}. Each sample was simulated with the corresponding completeness, centroid offset distribution, and redshift uncertainty. Thanks to the high completeness, \Euclid samples will recover the full halo population even in the low-mass regime, even though miscentering and redshift uncertainty could dilute the lensing signal. The relative change of the measured WL mass with respect to the true WL mass is on average $-1.5 (-2.5) \pm 0.3 \, \%$ for \AMICO(\PZWav)-like clusters. 
    \item We applied masks to the simulated data and found that the resulting WL masses remained consistent with the ideal scenario, even though some constraining power is lost due to the lower number of selected background galaxies.
    \end{itemize}

We additionally considered the effects of halo profile modelling and priors.
\begin{itemize}
    \item We switched the NFW model of the halo density profile to a BMO profile. On average, individual WL masses are consistent with mass measured with a NFW profile. We noticed that the BMO model better describes the averaged lensing profile, with a lower reduced $\chi^2$. On averaged density profiles, the BMO profile brings the WL mass higher, with a relative change with respect to the NFW WL mass of $5.1 \pm 1.6 \, \%$. 
    \item The prior affects the median, CBI, and mean mass point estimates. With a Gaussian prior on the mass, the scatter on the mass is two times lower and the bias more severe than for a uniform prior, leading to a higher precision and a lower accuracy.
    \end{itemize}
    
A final test was performed using a combination of the aforementioned effects including masks and realistic photo-$z$s.
\begin{itemize}
    \item We considered \AMICO-like clusters, detected in the presence of masks, with robustly selected source galaxies. We fitted a BMO model with a Gaussian mass prior. From the linear regression of the WL mass - true mass relation, we found a higher scatter than for the fiducial case with only instinsic scatter and noise, but a similar accuracy as measured with the slope and intercept parameters. Thanks to the high completeness of the lens sample and the robust source selection, the average mass bias, $\langle b_M \rangle = -15.5 \pm 2.4 \, \%$, agrees with that expected in an ideal scenario.
    \end{itemize}

\section{Discussion} \label{sec: Discussion}

Our analysis considered a broad selection of systematic effects, including triaxiality and cluster orientation, LSS and shape noise, source redshift uncertainties, and detection uncertainties from cluster-finding algorithms. However, in this work, we did not consider calibration of the spurious lensing signals induced by improper shear measurements or blended objects. Shear correction calibrated for cosmic shear may not be appropriate in the cluster regime since the shear response can become non-linear for stronger shears \citep{2020A&A...640A.117H}. Further improvements to the methodology and a full calibration of the cluster mass bias would require realistic cluster image simulations. Moreover, we do not account for intrinsic alignment and the clustering of sources, which are subdominant for single cluster WL mass analyses \citep[e.g.][]{2014MNRAS.445..726C}.

The analysis presented in this paper assumes that the miscentring of galaxy clusters follows an isotropic distribution. This approximation accounts for random deviations from the gravitational centre but does not consider inherent (directional) correlations between the centre proxy and WL shear measurements. Such correlations might arise, for example, when cluster centres are identified based on properties of the intracluster medium, as discussed in \citet{2024MNRAS.532.3359S}. 

We characterise WL mass bias using point estimates, simplifying the treatment by neglecting posterior shapes. As \citet{2018MNRAS.479..890L} show, bias and scatter can be derived using the full posterior of the WL mass measurement. This approach provides a close approximation ($\sim 4\%$ relative difference) to our mean point estimate.

Cluster WL mass bias could be further reduced with improvements to the halo density model. We found that fitting improves if we consider a truncation radius. Proper modelling for miscentring, concentration, or correlated or infalling matter could further improve the fit \citep{2019MNRAS.484.1598B, 2021A&A...653A..19G, 2022MNRAS.511.1484I}.

We underline that this study is part of a larger series of work to characterise cluster WL masses with \Euclid. \citet{PaperIII} tests the robustness of \Euclid cluster mass measurement algorithms on an extensive data set of Stage-III surveys or precursors. \citet{PaperIV} derives colour selections from \Euclid and ground-based photometry to optimise the quality of the WL signal. Ongoing testing and future works will exploit the \texttt{Flagship} simulation \citep{2017ComAC...4....2P, EuclidSkyFlagship} to perform a complete end-to-end cluster WL analyses within \Euclid. 



\begin{acknowledgements}
\AckEC.

LI and MS acknowledge financial contributions from INAF Theory Grant 2023: Gravitational lensing detection of matter distribution at galaxy cluster boundaries and beyond (1.05.23.06.17).

LI acknowledges financial support from the INAF grant 2023: Testing the origin of giant radio halos with joint LOFAR (1.05.23.05.11).

MS acknowledges financial contributions from contract ASI-INAF n.2017-14-H.0 and contract INAF mainstream project 1.05.01.86.10. 

LI and LM acknowledges the grants ASI n.2018- 23-HH.0 and ASI n. 2024-10-HH.0 ``Attività scientifiche per la missione Euclid – fase E''.

LM acknowledges the grant ASI n.I/023/12/0.

GC thanks the support from INAF theory Grant 2022: Illuminating Dark Matter using Weak Lensing by Cluster Satellites, PI: Carlo Giocoli.
GC and LM thank Prin-MUR 2022 supported by Next Generation EU (n.20227RNLY3 {\it The concordance cosmological model: stress-tests with galaxy clusters}).
The DEMNUni-cov simulations were carried out in the framework of ``The Dark Energy and Massive Neutrino Universe covariances" project, using the Tier-0 Intel OmniPath Cluster Marconi-A1 of the Centro Interuniversitario del Nord-Est per il Calcolo Elettronico (CINECA). We acknowledge a generous CPU and storage allocation by the Italian Super-Computing Resource Allocation (ISCRA) as well as from the coordination of the ``Accordo Quadro MoU per lo svolgimento di attività congiunta di ricerca Nuove frontiere in Astrofisica: HPC e Data Exploration di nuova generazione'', together with storage from INFN-CNAF and INAF-IA2.

GD acknowledges the funding by the European Union - $NextGenerationEU$, in the framework of the HPC project – ``National Centre for HPC, Big Data and Quantum Computing'' (PNRR - M4C2 - I1.4 - CN00000013 – CUP J33C22001170001).

LM acknowledges the financial contribution from the grant PRIN-MUR 2022 20227RNLY3 ``The concordance cosmological model: stress-tests with galaxy clusters'' supported by Next Generation EU and from the grants ASI n.2018-23-HH.0 and n. 2024-10-HH.0 ``Attività scientifiche per la missione Euclid – fase E''.

\end{acknowledgements}





\citestyle{aa}
\bibliographystyle{aa}
\bibliography{biblio, Euclid}


\appendix
\onecolumn

\section{Samples properties}

\begin{table}[h!]
    \caption{Properties of the different samples of the analysis.}
    \resizebox{!}{\vsize-2cm}{
    \rotatebox{90}{
    \begin{tabular}{l|cccc|cccc|cccc}
        \toprule
        \multirow{2}{*}{data set} & \multicolumn{4}{c|}{lens catalogue} & \multicolumn{4}{c|}{source catalogue} & \multirow{2}{*}{covariance} & \multirow{2}{*}{mask fraction} & \multirow{2}{*}{model} & \multirow{2}{*}{prior} \\
         & selection & size & position & redshift & shape noise & selection & density & redshift & & & & \\
        \midrule
        LSS + shape nosie (\ref{sec: Accuracy and Precision}) & \splitcelltwo{$z_{\rm d} < 1$}{$\log M_{\rm sim} \gtrsim 13.7$} & 6155 & true & true & $\sigma_{\epsilon}=0.26$ & $z_{\rm s} > z_{\rm d} + 0.1$ & $30 \, \rm arcmin^{-2}$ & true & stat + LSS & none & NFW & \splitcellthree{uniform}{$\log M_{\rm 200c} \in [13, 16]$}{$\log c_{\rm 200c} \in [0, 1]$} \\
        
        photo-$z$ non-conservative (\ref{sec: Basic selection}) & \splitcelltwo{$z_{\rm d} < 1$}{$\log M_{\rm sim} \gtrsim 13.7$} & 6155 & true & true & $\sigma_{\epsilon}=0.26$ & $z_{\rm s} > z_{\rm d} + 0.1$ & $30 \, \rm arcmin^{-2}$ & \splitcellfour{normal}{$\mu_{\tilde{\Delta} z} = -0.002$}{$\sigma_{\tilde{\Delta} z}=0.45$}{12.7\% outliers} & stat + LSS & none & NFW & \splitcellthree{uniform}{$\log M_{\rm 200c} \in [13, 16]$}{$\log c_{\rm 200c} \in [0, 1]$} \\
        
        photo-$z$ robust (\ref{sec: Robust selection}) & \splitcelltwo{$z_{\rm d} < 1$}{$\log M_{\rm sim} \gtrsim 13.7$} & 6155 & true & true & $\sigma_{\epsilon}=0.26$ & \splitcelltwo{$z_{\rm s} > z_{\rm d} + 2\sigma_{\tilde{\Delta} z} + 0.1$}{30\% removed} & $21 \, \rm arcmin^{-2}$ & \splitcellthree{normal}{$\mu_{\tilde{\Delta} z} = -0.003$}{$\sigma_{\tilde{\Delta} z}=0.29$} & stat + LSS & none & NFW & \splitcellthree{uniform}{$\log M_{\rm 200c} \in [13, 16]$}{$\log c_{\rm 200c} \in [0, 1]$} \\
        
        \AMICO-like (\ref{sec: Euclid-like clusters}) & \splitcellthree{$z_{\rm d} < 1$}{$\log M_{\rm sim} \gtrsim 13.7$}{\AMICO detection} & 4557 &  \splitcellthree{log-normal}{$\mu = 0.22$}{$\sigma=0.19$} & $\sigma_{\tilde{\Delta} z} = 0.03 / \sqrt{M}$ & $\sigma_{\epsilon}=0.26$ & $z_{\rm s} > z_{\rm d} + 0.1$ & $30 \, \rm arcmin^{-2}$ & true & stat + LSS & none & NFW & \splitcellthree{uniform}{$\log M_{\rm 200c} \in [13, 16]$}{$\log c_{\rm 200c} \in [0, 1]$} \\
        
        \PZWav-like (\ref{sec: Euclid-like clusters}) & \splitcellthree{$z_{\rm d} < 1$}{$\log M_{\rm sim} \gtrsim 13.7$}{\PZWav detection} & 4954 & \splitcellthree{log-normal}{$\mu = 0.27$}{$\sigma=0.45$} & $\sigma_{\tilde{\Delta} z} = 0.03 / \sqrt{M}$ & $\sigma_{\epsilon}=0.26$ & $z_{\rm s} > z_{\rm d} + 0.1$ & $30 \, \rm arcmin^{-2}$ & true & stat + LSS & none & NFW & \splitcellthree{uniform}{$\log M_{\rm 200c} \in [13, 16]$}{$\log c_{\rm 200c} \in [0, 1]$} \\
        
        masks (\ref{sec: Masks effects}) & \splitcelltwo{$z_{\rm d} < 1$}{$\log M_{\rm sim} \gtrsim 13.7$} & 4756 & true & true & $\sigma_{\epsilon}=0.26$ & $z_{\rm s} > z_{\rm d} + 0.1$ & $23 \, \rm arcmin^{-2}$ & true & stat + LSS & 22.4\% & NFW & \splitcellthree{uniform}{$\log M_{\rm 200c} \in [13, 16]$}{$\log c_{\rm 200c} \in [0, 1]$} \\
        
        BMO + prior (\ref{sec: Model and prior}) & \splitcelltwo{$z_{\rm d} < 1$}{$\log M_{\rm sim} \gtrsim 13.7$} & 6155 & true & true & $\sigma_{\epsilon}=0.26$ & $z_{\rm s} > z_{\rm d} + 0.1$ & $30 \, \rm arcmin^{-2}$ & true & stat + LSS & none & BMO & \splitcellthree{Gaussian + uniform}{$\log M_{\rm 200c} \in [13, 16]$}{$\log c_{\rm 200c} \in [0, 1]$} \\
        
        \splitcellbis{photo-$z$ robust +}{\AMICO-like +}{masks +}{BMO + prior} (\ref{sec: Combined systematics}) & \splitcellthree{$z_{\rm d} < 1$}{$\log M_{\rm sim} \gtrsim 13.7$}{\AMICO detection} & 3547 & \splitcellthree{log-normal}{$\mu = 0.22$}{$\sigma=0.19$} & $\sigma_{\tilde{\Delta} z} = 0.03 / \sqrt{M}$ & $\sigma_{\epsilon}=0.26$ & \splitcelltwo{$z_{\rm s} > z_{\rm d} + 2\sigma_{\tilde{\Delta} z} + 0.1$}{30\% removed} & $16 \, \rm arcmin^{-2}$ & \splitcellthree{normal}{$\mu_{\tilde{\Delta} z} = -0.003$}{$\sigma_{\tilde{\Delta} z}=0.29$} & stat + LSS & 22.4\% & BMO & \splitcellthree{Gaussian + uniform}{$\log M_{\rm 200c} \in [13, 16]$}{$\log c_{\rm 200c} \in [0, 1]$} \\
        \bottomrule
    \end{tabular}
    }}
    \label{tab: specifics_samples}
\end{table}

\label{lastpage}
\end{document}